\documentclass[fleqn,usenatbib]{mnras}

\usepackage[T1]{fontenc}
\usepackage{ae,aecompl}
\usepackage{graphicx}
\usepackage{float}
\usepackage{natbib}
\usepackage{amsmath}    
\usepackage{amssymb}    
\usepackage{color}
\usepackage{aas_macros}
\usepackage{array}

\newcolumntype{L}[1]{>{\raggedright\let\newline\\\arraybackslash\hspace{0pt}}m{#1}}
\newcolumntype{C}[1]{>{\centering\let\newline\\\arraybackslash\hspace{0pt}}m{#1}}
\newcolumntype{R}[1]{>{\raggedleft\let\newline\\\arraybackslash\hspace{0pt}}m{#1}}

\newcommand*\mathinhead[2]{\texorpdfstring{$\boldsymbol{#1}$}{#2}}

\def\Re{\mbox{$R_{\rm e}$}}

\def\mst{\mbox{$M_{\star}$}}

\def\lsim{\mathrel{\rlap{\lower3.5pt\hbox{\hskip0.5pt$\sim$}}
    \raise0.5pt\hbox{$<$}}}                
\def\gsim{~\rlap{$>$}{\lower 1.0ex\hbox{$\sim$}}}

\def\SN{\mbox{$S/N$}}

\def\zphot{\mbox{$z_{\rm phot}$}}

\def\MErrauto{\mbox{{\tt MAGERR\_AUTO}}}

\def\photoz{\mbox{photo-$z$}}

\def\Log{\mbox{Log}}
\def\Zsun{\mbox{$Z_{\rm \odot}$}}

\def\Fig{\mbox{Fig.~}}
\def\Figs{\mbox{Figs.~}}
\def\Tab{\mbox{Tab.~}}
\def\Sec{\mbox{Sect.~}}
\def\Secs{\mbox{Sects.~}}

\def\Eq{\mbox{Eq.~}}

\title[Structural parameters in KiDS]{Evolution of galaxy size--stellar mass relation from the Kilo Degree Survey}

\author[Roy et al.]{
N.~Roy$^{1,2}$, N.R.~Napolitano$^{1}$,  F.~La Barbera$^{1}$,
C.~Tortora$^{4}$, F.~Getman$^{1}$, \and M.~Radovich$^{3}$,
M.~Capaccioli$^{2}$, M.~Brescia$^{1}$, S.~Cavuoti$^{1,2,6}$,
G.~Longo$^{2}$, M.A.~Raj$^{1,2}$, \and E.~Puddu$^{1}$,
G.~Covone$^{2,6}$, V.~Amaro$^{2}$, C.~Vellucci$^{2}$,
A.~Grado$^{1}$, K.~Kuijken$^{5}$, \and G.~Verdoes ~Kleijn$^{3}$, E.~Valentijn$^{3}$\\~\\
$^1$ INAF -- Osservatorio Astronomico di
Capodimonte, Salita Moiariello, 16, 80131 - Napoli, Italy\\
$^2$ Dipartimento di Fisica "E. Pancini", Universit\`{a} di Napoli
Federico II, Compl. Univ. Monte S. Angelo, 80126 - Napoli, Italy\\
$^3$ Kapteyn Astronomical Institute, University of Groningen, P.O.
Box 800, 9700 AV Groningen, the Netherlands\\
$^4$ INAF -- Osservatorio Astronomico di Padova, Via Ekar, 36012
Asiago VI 0424 462032\\
$^5$Leiden Observatory, Leiden University, P.O. Box 9513, 2300 RA
Leiden, the Netherlands\\
$^6$Istituto Nazionale di Fisica Nucleare, Sezione di Napoli,
Complesso Universitario di Monte S. Angelo, Via Cintia Edificio 6,
80126 \\Napoli, Italy\\
}

\date{Accepted XXX. Received YYY; in original form ZZZ}
\pubyear{2017}

\begin{document}
\label{firstpage}

\pagerange{\pageref{firstpage}--\pageref{lastpage}}

\maketitle

\begin{abstract}
We have obtained structural parameters of about $340,000$ galaxies
from the Kilo Degree Survey (KiDS) in 153 square degrees of data
release 1, 2 and 3. We have performed a seeing convolved 2D single
S\'ersic fit to the galaxy images in the 4 photometric bands ($u$,
$g$, $r$, $i$) observed by KiDS, by selecting high signal-to-noise
ratio ($\SN > 50$) systems in every bands.

We have classified galaxies as spheroids and disc-dominated by
combining their spectral energy distribution properties and their
S\'ersic index. Using photometric redshifts derived from a machine
learning technique, we have determined the evolution of the
effective radius, \Re\ and stellar mass, \mst, versus redshift,
for both mass complete samples of spheroids and disc-dominated
galaxies up to z$\sim0.6$.

Our results show a significant evolution of the
structural quantities at intermediate redshift for the massive
spheroids ($\Log\ M_*/M_\odot>11$, Chabrier IMF), while almost no
evolution has found for less massive ones ($\Log\
M_*/M_\odot < 11$). On the other hand, disc dominated systems show
a milder evolution in the less massive systems ($\Log\
M_*/M_\odot < 11$) and possibly no evolution of the more massive
systems. These trends are generally consistent with predictions
from hydrodynamical simulations and independent datasets out to
redshift $z \sim 0.6$, although in some cases the scatter of the
data is large to drive final conclusions.

These results, based on 1/10 of the expected KiDS area,
reinforce precedent finding based on smaller statistical samples
and show the route toward more accurate results, expected with the
the next survey releases.

\end{abstract}

\section{Introduction}
Spheroids play an important role in the observational studies of
galaxy formation and evolution as their structure reveals clear
traces of evolution from past to present.
They are known to follow well-defined empirical scaling laws that
relate their global or local observational properties: the
Faber-Jackson (FJ; \citealt{Faber_Jackson+76}), the $\mu_e-R_e$
relation (\citealt{Kormendy+77}, \citealt{Capaccioli+92a}),
fundamental plane (\citealt{Dressler+87_FP};
\citealt{DOnofrio+97_distance}), size vs. mass (\citealt{Shen+03},
\citealt{HB09_curv}), colour vs. mass (\citealt{Strateva+01}),
colour vs. velocity dispersion, $\sigma$ (\citealt{Bower+92_CM}),
Mg2 vs. $\sigma$ (e.g., \citealt{Guzman+92_Mg2};
\citealt{Bernardi+03_Mg2}), colour gradient vs. mass
(\citealt{Tortora+10CG}; \citealt{LaBarbera+11_CG}), black hole
mass vs. galaxy mass and $\sigma$, i.e., $M_{\rm BH}-\mst$ and
$M_{\rm BH}-\sigma$ (\citealt{deZeeuw+01_Blk_HOLe};
\citealt{Magorrian+98_BH}; \citealt{Ferrarese_Merritt+00_BH};
\citealt{Gebhardt+00_BH_mass_velo};
\citealt{Tremaine+02_BH_velo}), total vs. stellar mass
(\citealt{Moster+10}), dynamical vs. stellar mass in the galaxy
centers (\citealt{Tortora+09,SPIDER-VI}), Initial mass function
(IMF) vs. $\sigma$ (e.g., \citealt{Treu+10};
\citealt{Conroy_vanDokkum12b}; \citealt{Cappellari+12};
\citealt{LaBarbera+13_SPIDERVIII_IMF};
\citealt{Tortora+13_SPIDER_IMF,Tortora+14_DMslope,Tortora+14_MOND}).

Late-type galaxies show also similar scaling relations, in
particular a size-mass relation, which has a different slope with
respect to the one of early-type galaxies (\citealt{Shen+03};
\citealt{vanderwel+14_SM}). Closely related to that, there is also
the size-velocity relation (\citealt{Courteau+007}), which shows
that discs with faster rotations are also larger in size
(\citealt{Mo+1998}). Another fundamental scaling relation is the
Tully-Fisher relation between the mass or intrinsic luminosity and
angular velocity or emission line width of a spiral galaxy
(\citealt{TF77}), with the variant accounting for the stellar
mass-velocity relation (\citealt{Dutton+07} and reference therein)
and the baryonic mass-velocity relation (\citealt{Lelli+016}).

Scaling relations provide invaluable information about the
formation and evolution of galaxies, setting stringent constraints
to their formation models. In particular, studying the structural
and mass properties of galaxies at different redshifts can give
more insights into the mechanisms that have driven their assembly
over time.

For instance, spheroidal systems (e.g. early-type galaxies, ETGs)
follow a steep relation between their size and the stellar mass,
the so called, size-mass relation. Most of the ETGs are found to
be much more compact in the past with respect to local
counterparts (\citealt{Daddi+05}; \citealt{Trujillo+06};
\citealt{Trujillo+07}; \citealt{Saglia+10}; \citealt{Trujillo+11},
etc.). A simple monolithic-like scenario, where the bulk of the
stars is formed in a single dissipative event, followed by a
passive evolution, is inconsistent with these observations, at
least under the assumption that  that most of the high-z compact
galaxies are the progenitors of nowadays ETGs (see
\citealt{delarosa+16_highz_compact}, for a different prospective).
Thus, several explanations have been offered for the dramatic size
difference between local massive galaxies and quiescent galaxies
at high redshift. The simplest one is related to the presence of
systematic effects, most notably an under-(over)-estimate of
galaxy sizes (masses). However, recent studies suggest that it is
difficult to change the sizes and the masses by more than a factor
of 1.5, unless the initial mass function (IMF) is strongly altered
(e.g., \citealt{Muzzin+09_Mass_of_gal}; \citealt{Cassata+10};
\citealt{Szomoru+10}). Other explanations include extreme mass
loss due to a quasar-driven wind (\citealt{Fan+08}), strong radial
age gradients leading to large differences between mass-weighted
and luminosity-weighted ages (\citealt{Hopkins+09_DELGN_III};
\citealt{LaBarbera_deCarvalho+09}), star formation due to gas
accretion (\citealt{Franx+08_surf_dens}), and selection effects
(e.g., \citealt{vandokkum+09_compactness};
\citealt{vanderwel+09_size_mass}).

The best candidate mechanism to explain the size evolution of
spheroids is represented by galaxy merging. As cosmic time
proceeds the high-z ``red nuggets" are thought to merge and evolve
into the present-day massive and extended galaxies. Spheroids
undergo mergings at different epochs, becoming massive and red in
colour (\citealt{Kauffmann+96_merging}). Rather than major
mergers, the most plausible mechanism to explain this size and
mass accretion is minor merging (e.g.,
\citealt{Bezanson+09_minor_merge}; \citealt{Naab+09};
\citealt{vanDokkum+10}; \citealt{Hilz+13};
\citealt{Tortora+14_DMevol,Tortora+18}). Numerical simulations
predict that such mergers are frequent
(\citealt{Guo_White+08_merge}; \citealt{Naab+09}) leading to
observed stronger size growth than mass growth
(\citealt{Bezanson+09_minor_merge}). The minor merging scenario
can also explain the joint observed evolution of size and central
dark matter (\citealt{Cardone+11SIM};
\citealt{Tortora+14_DMevol,Tortora+18}). However, recently it has
been found that a tiny fraction of the high-z red nuggets might
survive intact till the present epoch, without any merging
experience, resulting in compact, relic systems in the nearby
Universe (\citealt{Trujillo+12_compacts};
\citealt{Damjanov+15_compacts};
\citealt{Tortora+16_compact_KiDS}).

Late--Type galaxies (LTGs) or disc-dominated galaxies shows a
shallower trend in size and stellar masses compared to ETGs
(\citealt{Shen+03}). Furthermore, the size and stellar mass of
LTGs evolve mildly with lookback time (e.g.
\citealt{vanderwel+14_SM}) while the evolution is stronger for the
ETGs.

In the recent years, the size evolution of ETGs and LTGs has been
studied based on different survey data such as DEEP2 (galaxies
within the redshift range 0.75 < z < 1.4:
\citealt{Davis+003_DEEP2}); GAMA (250 square degrees with galaxies
up to redshift 0.4: \citealt{Driver+11_GAMA}); 2dFGRS (measuring
redshifts for 250000 galaxies; \citealt{Colless+001_2dfGRS}), and
SDSS (10000 square degrees in northern sky in $u$, $g$, $r$, $i$
and z bands; \citealt{York+000_SDSS_tech}). The latter has been
the most successful survey in the field of galaxy evolution
studies (\citealt{Kauffmann+03}) in the recent years with pioneer
results showing the size evolution of both passive galaxies and
active, disc-dominated systems (see, e.g. \citealt{Shen+03},
 \citealt{HB09_curv}, \citealt{Baldry+12_gama_mass}, \citealt{Kelvin+012},
\citealt{Mosleh+13_size_evol}, \citealt{Lange+015}).

However, other ground based instrumentations and telescopes are
providing, and will provide in the future, higher data quality and
we are currently in the position to improve our understanding of
structural evolution of galaxies over larger datasets. The Kilo
Degree Survey (KiDS) is one of the latest survey aimed at
gathering best data quality from the ground, and expand the SDSS
results to larger redshifts and lower masses. KiDS is a large sky
optical imaging survey, which will cover 1500 square degrees over
$u$, $g$, $r$, and $i$ bands, using VLT Survey telescope (VST,
\citealt{Capaccioli_Schipani11}) equipped with the 1 deg$^2$
camera OmegaCAM (\citealt{deJong+15_KiDS_paperI,
deJong+017_KIDS_DR3}). KiDS has been designed to perform extensive
weak lensing studies (\citealt{Kuijken+15_weak_lensing_kids},
\citealt{Hildebrandt+017}) taking advantage of the high spatial
resolution of VST (0.2"/pixel) and the optimal seeing conditions
of Cerro Paranal. However, with a depth $\sim 2$ magnitudes deeper
than SDSS, KiDS is suitable to perform detailed galaxy evolution
studies and to be a unique "rarity seeker". In particular, KiDS
has proven to be very efficient to perform the census of
particular classes of objects, as the ultra-compact massive
galaxies (UCMGs,
\citealt{Tortora+16_compact_KiDS},\citealt{Tortora+18_UCMG}),
galaxy clusters (\citealt{Radovich+17_KiDS}) and strong
gravitational lenses (\citealt{Napolitano+15_proc_lensing_KiDS},
\citealt{Petrillo+17_CNN}, \citealt{Spiniello+18_quads}).

Based on the number of galaxies analyzed in the present work, we
estimate that KiDS, after completion, will allow us to measure
structural parameters, in $ugri$, for about 4 million galaxies, up
to refshift $z\lsim0.7$ (\citealt{Tortora+16_compact_KiDS}).
With the help of high quality data obtained with KiDS and
the use of machine learning techniques to determine photometric redshifts
(\citealt{Cavuoti+15_KIDS_I,Cavuoti+17_KiDS}), we
are intended to study the size evolution of galaxies up to
redshift $z \lsim 0.7$.

The paper is organized as follows.
Sample selection is presented in \Sec\ref{sec:sample}, while
\Sec\ref{sec:surf_phot} is devoted to the description of the
structural parameter measurement, the derivation of the
measurement errors and the analysis of the impact of various
systematics. The galaxy classification, the size-mass relation and
its evolution in terms of redshifts are shown in
\Sec\ref{sec:Resu}. Finally, a discussion of the results,
conclusions and future prospects is provided in
\Sec\ref{sec:disk_concl}. We will adopt the following cosmology:
$H_{0}=75$ km/s/Mpc, $\Omega_m = 0.29$ and $\Omega_\Lambda = 0.71$
(e.g., \citealt{Komatsu+11_WMAP7}).

\begin{figure}
\centering
\includegraphics[width=\columnwidth]{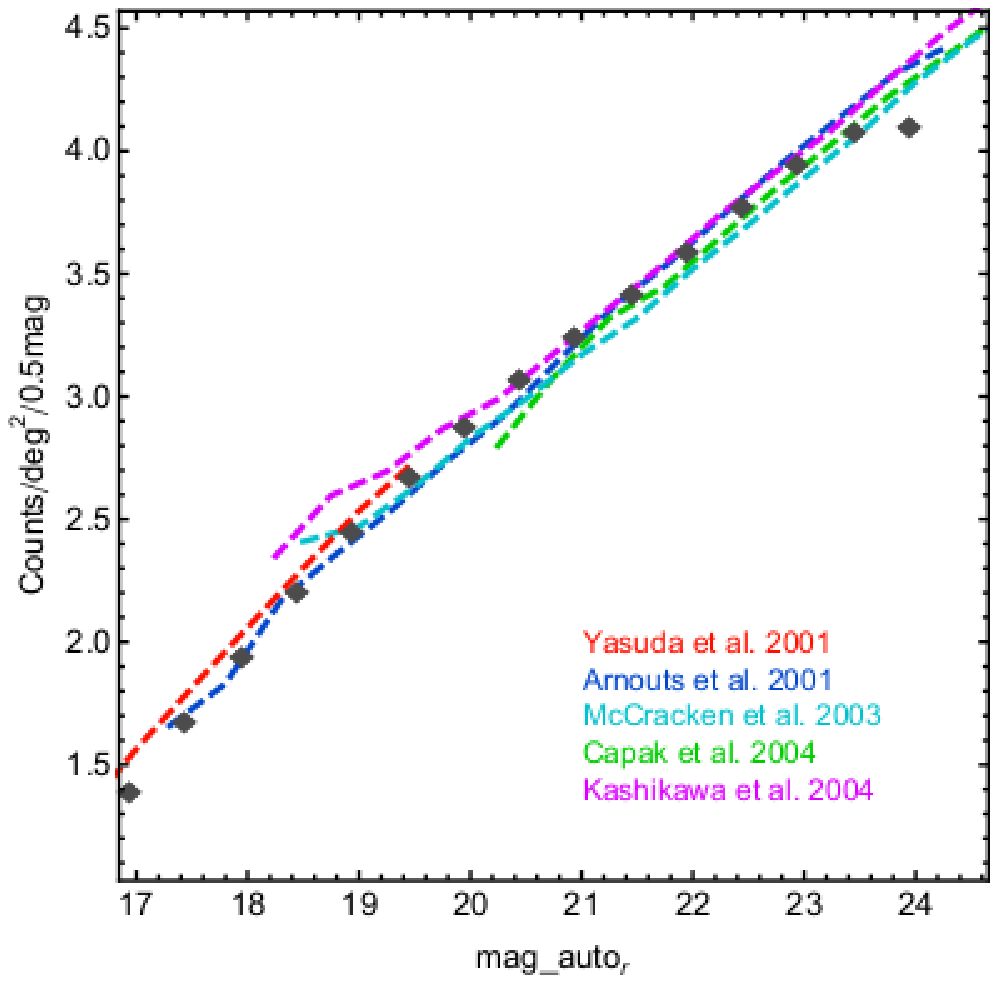}
\includegraphics[width=\columnwidth]{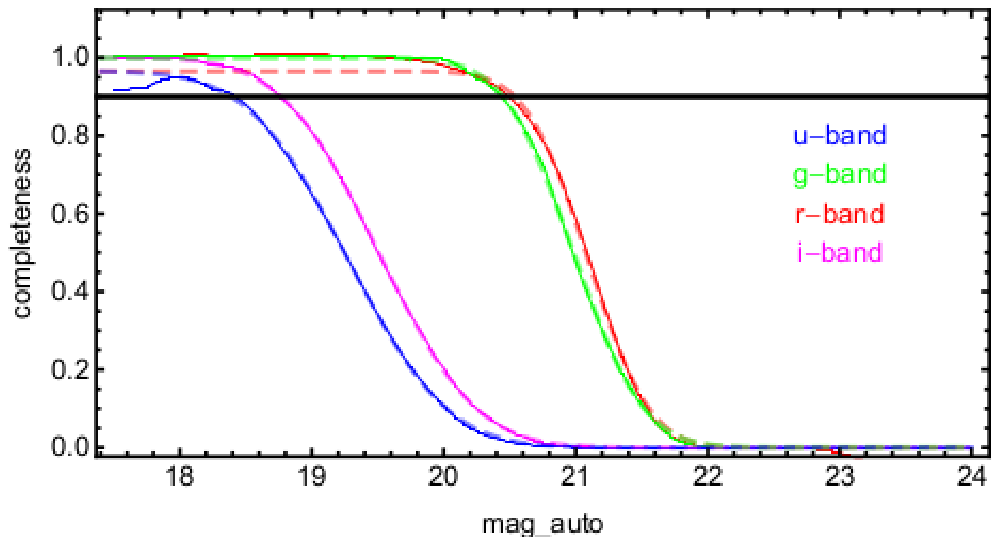}
\vspace{0.2cm} \caption{{\it Top}: Galaxy counts (grey boxes) as a
function of their $\texttt{MAG\_AUTO}$ in $r-$band are compared
with other literature estimates (as in the legend). The match with
previous literature is very good at fainter magnitudes while is
not perfect at the brightest ones due to the limited area covered.
See also the discussion in the text. {\it Bottom}: completeness of
the "high \SN" sample in $u,~ g, ~r$, and $i$-band with colour
code as in the legend. The completeness has been computed with
respect to the 6 million sample. The derived completeness from
data are shown as solid lines, while the best fit using Eq.
\ref{eq:compl_erf} are plotted as dashed lines}
\label{completeness}
\end{figure}

\section{Sample Selection}\label{sec:sample}
The sample adopted in this analysis consists of galaxies extracted
from 153 square degree of the KiDS survey
(\citealt{deJong+15_KiDS_paperI}) which have been already
presented in \citet{Tortora+16_compact_KiDS}. Details about the
data reduction and calibration can be found in
\citet{deJong+15_KiDS_paperI}. In the following we give a brief
summary of the way the galaxy sample has been selected.

Single band source lists for the observed tiles are extracted
using a stand-alone procedure named \texttt{KiDS-CAT}, which uses
Sextractor (\citealt{Bertin_Arnouts+96}) for the source detection,
star galaxy separation and the catalog extraction. In particular,
the star/galaxy (S/G) separation is based on the
$\texttt{CLASS\_STAR}$ parameter from S-Extractor measured on the
$r$-band images, the deepest and best seeing ones for KiDS,
following the procedure described in \citet[Sect.
4.5.1]{deJong+15_KiDS_paperI}.

While the S/G separation is mainly based on the single $r-$band
shape information, source colours are measured based on multi-band
source catalogs, which have been obtained using S-Extractor in
dual image mode by taking the $r-$band images as reference for
source extraction and then measuring the source fluxes in the
registered images from the other bands, at the sky position of the
$r-$band detection. The fluxes from the multi-band catalog have
been used to perform the stellar population synthesis as described
in \Sec\ref{sec:SED}. Among the sources selected as galaxies
($\sim$ 11 millions), we have retained those sources which were
marked as being out of critical areas from our masking procedure
(see \citealt{deJong+15_KiDS_paperI}, Sect. 4.4). The effective
uncritical area has been found to be  $103$ square deg, which
finally contains $\sim$6 million galaxies. This latter sample
turned out to be complete out to $\sim24$ mag in $r-$band by
comparing the galaxy counts as a function of extinction-corrected
$\texttt{MAG\_AUTO}$ (used as robust proxy of the total magnitude)
with previous literature (e.g. \citealt{Yasuda+01_gal_counts},
\citealt{Arnouts+01_gal_counts},
\citealt{McCracken+03_gal_counts}, \citealt{Capak+04_gal_counts},
\citealt{Kashikawa+04_gal_counts}), as shown in
\Fig\ref{completeness}.

Finally, in order to perform accurate structural parameter
measurement for these systems, we have selected galaxies with
``high'' signal-to-noise (\SN), defined as $1/\MErrauto$
(\citealt{Bertin_Arnouts+96}). Specifically, we have used $\SN >
50$ as initial guess for reliable structural parameters
(\citealt{LaBarbera_08_2DPHOT}). This choice of \SN\ will be fully
checked by applying the 2D surface brightness fitting procedure
(see \Sec\ref{sec:stru_par}) to mock galaxies in
\Sec\ref{sec:simul}. We refer to the samples resulting from the
\SN\ selection, as the ``high-\SN'' samples, consisting of  4240,
128906, 348025, and 129061 galaxies, in the $u$, $g$, $r$, and $i$
bands, respectively. These represent the galaxy samples used for
the model fitting procedure in the different bands as described in
\S\ref{sec:surf_phot}. The final output sample to be used for
structure parameter analysis will be discussed in
\S\ref{sec:etg_ltg}

\subsection{Magnitude completeness}\label{sec:mag_compl}
The difference in counts among the different bands is due to their
intrinsic depth, being the latter a combination of exposure time
and seeing, with the $u$-band the shallowest band and the $r$-band
the deepest in the KiDS survey plan (see
\citealt{deJong+15_KiDS_paperI}).

In order to evaluate the completeness magnitude of our sample in
different bands, we have computed the fraction of the detected
galaxies of the high--\SN\ sample in bin of $\tt MAG\_AUTO$ with
respect to number of galaxies in the same bins of a deeper and
complete samples and finally fit the binned fractions with a
standard error function model (see e.g. \citealt{Rykoff+015}).
\begin{equation}
{\rm comp}=(1/2)\left[ 1-erf\left( \frac{m-m_{50}}{\sqrt{2w}} \right) \right],
\label{eq:compl_erf}
\end{equation}
where $m_{50}$ is the magnitude at which the completeness is 50\%,
and $w$ is the (Gaussian) width of the rollover. The magnitude at
which the sample is 90\% complete has been extrapolated by the
best fit function. As shown in Fig. \ref{completeness} (top
panel), the full sample of 6 million galaxies detected in the KiDS
area has counts consistent with other literature samples and can
be used as a reference counts to obtain the fraction of galaxies
of the high--\SN\ sample as shown in the bottom panel of Fig.
\ref{completeness}. In this latter plot, we show the interpolated
completeness function from the data as solid lines and the best
fit curves as dashed lines. The derived 90\% completeness limit
are 18.4, 20.4, 20.5, and 18.8 for $u$, $g$, $r$, and $i$-band
respectively.

\subsection{Photometric Redshifts} \label{sec:photo_z}
\begin{figure}
    \centering
    \includegraphics[width=\columnwidth]{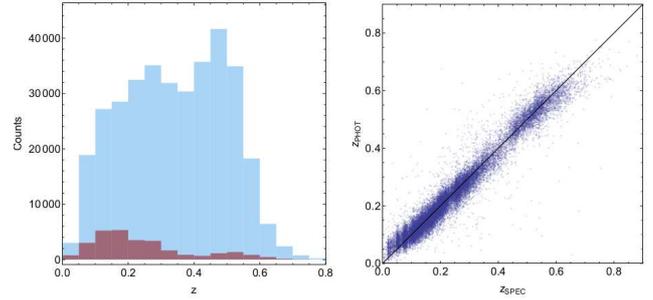}
    \caption{{\it Left}: The distribution of the spectroscopic sample adopted as knowledge base for
     the MLPQNA method (in red) and the \photoz\ distribution of the ``high \SN'' sample (in light blue).
     {\it Right}: Comparison between spectroscopic and photometric redshifts for the blind test
     set. See the text for more details.
    }\label{zspec_zphot}
\end{figure}

Photometric redshifts have been derived from Multi Layer
Perceptron with Quasi Newton Algorithm (MLPQNA) method (see
\citealt{Brescia+13}; \citealt{Brescia+14},
\citealt{Cavuoti+15_photoz}), and fully presented in
\cite{Cavuoti+15_KIDS_I}, which we address the interested reader
for all details. This method makes use of an input knowledge base
(KB) consisting of a galaxy sample with both spectroscopic
redshifts and multi-band integrated photometry to perform the best
mapping between colours and redshift. In particular, we have used
4$''$ and 6$''$ diameter apertures to compute the magnitudes to be
used to best perform such a mapping on the training set (see
\citealt{Cavuoti+15_KIDS_I} for more details). While the
spectroscopic redshifts for the KB are given by the Sloan Digital
Sky Survey data release 9 (SDSS-DR9; \citealt{Ahn+12_SDSS_DR9})
and Galaxy And Mass Assembly data release 2 (GAMA-DR2;
\citealt{Driver+11_GAMA}). This sample consists of $\sim 60,000$
galaxies with spectroscopic redshifts out to $z\lsim 0.8$, as
shown in \Fig\ref{zspec_zphot}. $60$ per cent of the sample is
used as {\it training set}, to train the network, looking at the
hidden correlation between colours and redshifts. While the rest
of the galaxies in the KB are collected in the {\it blind test
set}, needed to evaluate the overall performances of the network
with a data sample never submitted to the network previously (see
right panel in \Fig\ref{zspec_zphot}). The scatter in the
measurement, defined as $(z_{spec}-z_{phot})/(1+z_{spec})$, is
$\sim 0.03$ (see \citealt{Cavuoti+15_KIDS_I}). The advantage of
the machine learning techniques resides in the possibility of
optimizing the mapping between the photometry and the spectroscopy
regardless the accuracy in the photometric calibration, but the
disadvantage consists in the limited applicability of the method
only to the volume in the parameter space covered by the KB sample
(see \citealt{Cavuoti+15_KIDS_I}). In our case, for instance, of
the 6 millions starting systems, accurate photo-z have been
derived for systems down to $r\sim21$, i.e. $\sim$1.1 million
galaxies. This sample is still deeper than the high--\SN\ sample
(see \Sec\ref{sec:mag_compl}). After completing the analysis
presented in this paper, new set of machine learning photo-z were
made available to the KiDS collaboration (see
\citealt{Bilicki+017_photoz_arxiiv} for details). This will be
used for the forthcoming analysis of the next KiDS data releases.

\subsection{Stellar Mass and galaxy classification}\label{sec:SED}
Stellar masses, rest-frame luminosities from stellar population
synthesis (SPS) models and a galaxy spectral-type classification
are obtained by means of the SED fitting with {\tt Le Phare}
software (\citealt{Arnouts+99}; \citealt{Ilbert+06}), where the
galaxy redshifts have been fixed to the \zphot\ obtained with
MLPQNA. We adopt the observed $ugri$ magnitudes (and related $1\,
\sigma$ uncertainties) within a $6''$ aperture of diameter, which
are corrected for Galactic extinction using the map in
\cite{Schlafly_Finkbeiner11}.

To determine stellar masses and rest-frame luminosities, we have
used single burst SPS models from \cite{BC03} with a
\cite{Chabrier01} IMF. We use a broad set of models with different
metallicities ($0.005 \leq Z/\Zsun \leq 2.5$) and ages ($age \leq
\rm age_{\rm max}$), the maximum age, $\emph{$\rm age_{\rm
max}$}$, is set by the age of the Universe at the redshift of the
galaxy, with a maximum value at $z=0$ of $13\, \rm Gyr$. Total
magnitudes derived from the S\'ersic fitting, $m_{S}$, (see
\Sec\ref{sec:stru_par}) are used to correct the outcomes of {\tt
Le Phare}, i.e. stellar masses and rest-frame luminosities, for
missing flux. Typical uncertainties on the stellar masses are of
the order of 0.2 dex (maximum errors reaching 0.3 dex).

We have finally used the spectrophotometric classes from {\tt Le
Phare} to derive a classification of our galaxies. As template set
for this aim, we adopted the 66 SEDs used for the CFHTLS in
\cite{Ilbert+06}. The set is based on the four basic templates
(Ell, Sbc, Scd, Irr) in \cite{CWW80}, and starburst models from
\cite{Kinney+96}. Synthetic models from \cite{BC03} are used to
linearly extrapolate this set of templates into ultraviolet and
near-infrared. The final set of 66 templates (22 for ellipticals,
17 for Sbc, 12 for Scd, 11 for Im, and 4 for starburst) is
obtained by linearly interpolating the original templates, in
order to improve the sampling of the redshift-colour space and
therefore the accuracy of the SED fitting. We did not account for
internal extinction, to limit the number of free parameters.

This fitting procedure provided us with a photometrical galaxy
classification, which allows us to separate ETGs (spheroids) from
LTGs (disc-dominated galaxies).

\subsection{Mass completeness as a function of the redshift}\label{sec:mass_compl}
In the following we will study the behaviour of the galaxy
properties as a function of the redshift. It is well known that
some of the galaxy physical quantities (e.g. size, S\'ersic index,
colour, etc.) correlate with mass. Hence it is important to define
a mass complete sample in each redshift bins.

\begin{figure}
    \centering
    \includegraphics[scale=.9]{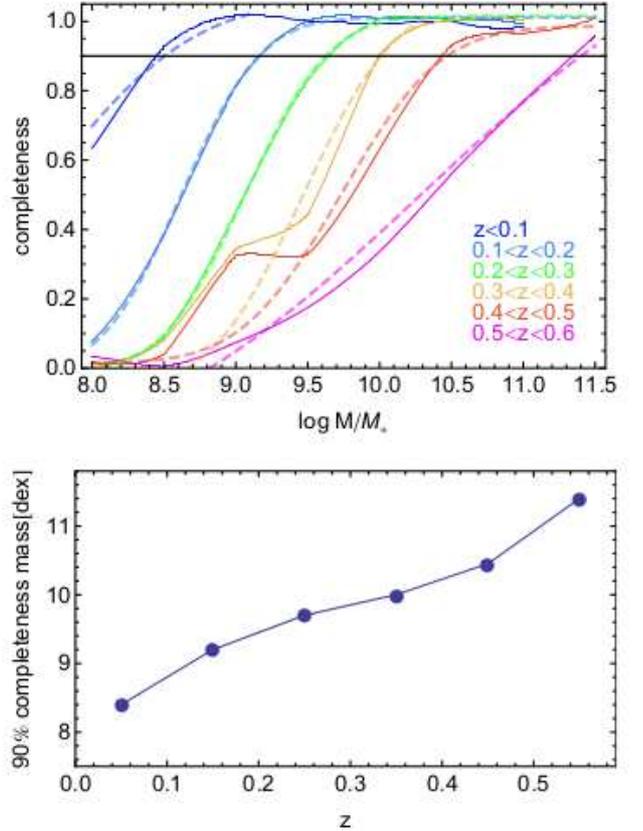}
    \caption{Mass completeness as a function of redshift: the ratio of the
    high \SN\ sample and the \photoz\ sample for galaxies separated in different redshift bins are shown.
    In the bottom panel the derived completeness from data are shown
    as dashed lines, while the best fit using Eq. \ref{eq:compl_erf} are
    plotted as solid lines (except for the most massive bin where there was
    not convergence due to the poor sampling above 90\% completeness).
    The numerical values are reported in
    \Tab\ref{tab:mass_compl}. See text for details.
    }
    \label{mass_compl}
\end{figure}

To do that, we have proceeded in the same way we have computed the
completeness magnitudes in \Sec\ref{sec:mag_compl}, i.e. by
comparing the high--\SN\ galaxy counts against the \photoz\
sample, once galaxies have been separated in different \photoz\
bins. Results are shown in \Fig\ref{mass_compl} and completeness
masses are reported in \Tab\ref{tab:mass_compl}. The table stops
at $z=0.6$ because the high--\SN sample starts to be fully
incomplete in mass above that redshift.

\begin{table}
\centering \caption{90\% completeness mass as a function of the
photometric redshift for the high \SN\ sample.}
\label{tab:mass_compl}       
%
%
\begin{tabular}{cc}
\hline\noalign{\smallskip}
\photoz\ bin & 90\% compl. $\Log\ M_*/M_\odot$ \\
\noalign{\smallskip}\hline\noalign{\smallskip}
$\leq0.1 $& 8.5\\
$0.1<z\leq0.2$ & 9.2 \\
$0.2<z\leq0.3$ & 9.6 \\
$0.3<z\leq0.4$ & 10.0 \\
$0.4<z\leq0.5$ & 10.5 \\
$0.5<z\leq0.6$ & 11.4 \\
\noalign{\smallskip}\hline\noalign{\smallskip}
\end{tabular}
\end{table}

\section{Surface Photometry}\label{sec:surf_phot}
In this section we present the measurement of structural
parameters for the galaxy sample described above, using 2DPHOT
(\citealt{LaBarbera_08_2DPHOT}). We evaluate parameter
uncertainties and determine the reliability of the fitting
procedure using mock galaxy images, with same characteristics as
the KiDS images (see \Sec\ref{sec:simul}). We finally compare the
results obtained with KiDS for galaxies in common with an external
catalog from SDSS data (i.e. \citealt{SPIDER-I},
\citealt{Kelvin+012}).

\begin{figure}
    \includegraphics[width=\columnwidth]{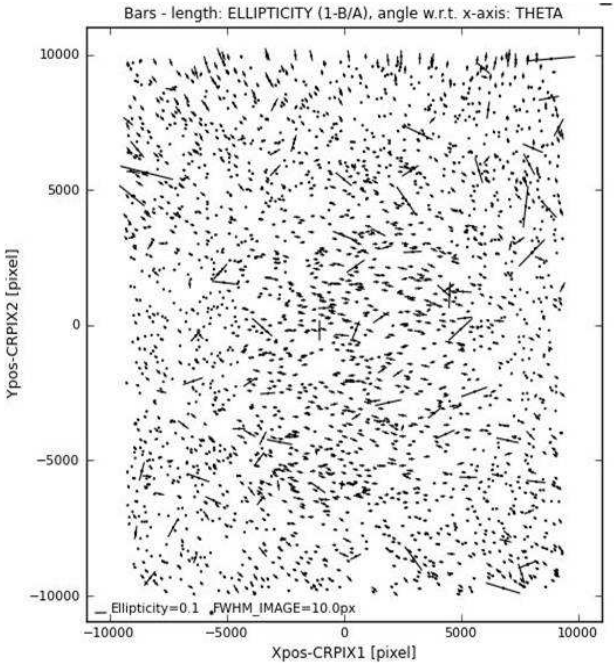}
    \vspace{-0.5cm}
    \caption{PSF anisotropy within the coadd KIDS\_129.0\_-0.5 in $r$-band.
    The elongation is aligned in a specific direction on the borders but random in the middle of the image.} \label{PSF}
\end{figure}

\subsection{Structural Parameters}\label{sec:stru_par}
Surface photometry of the high--\SN\ sample has been performed
using 2DPHOT (\citealt{LaBarbera_08_2DPHOT}), an automated
software environment that allows 2D fitting of the light
distribution of galaxies on astronomical images.
\begin{figure*}
    \includegraphics[width=\columnwidth]{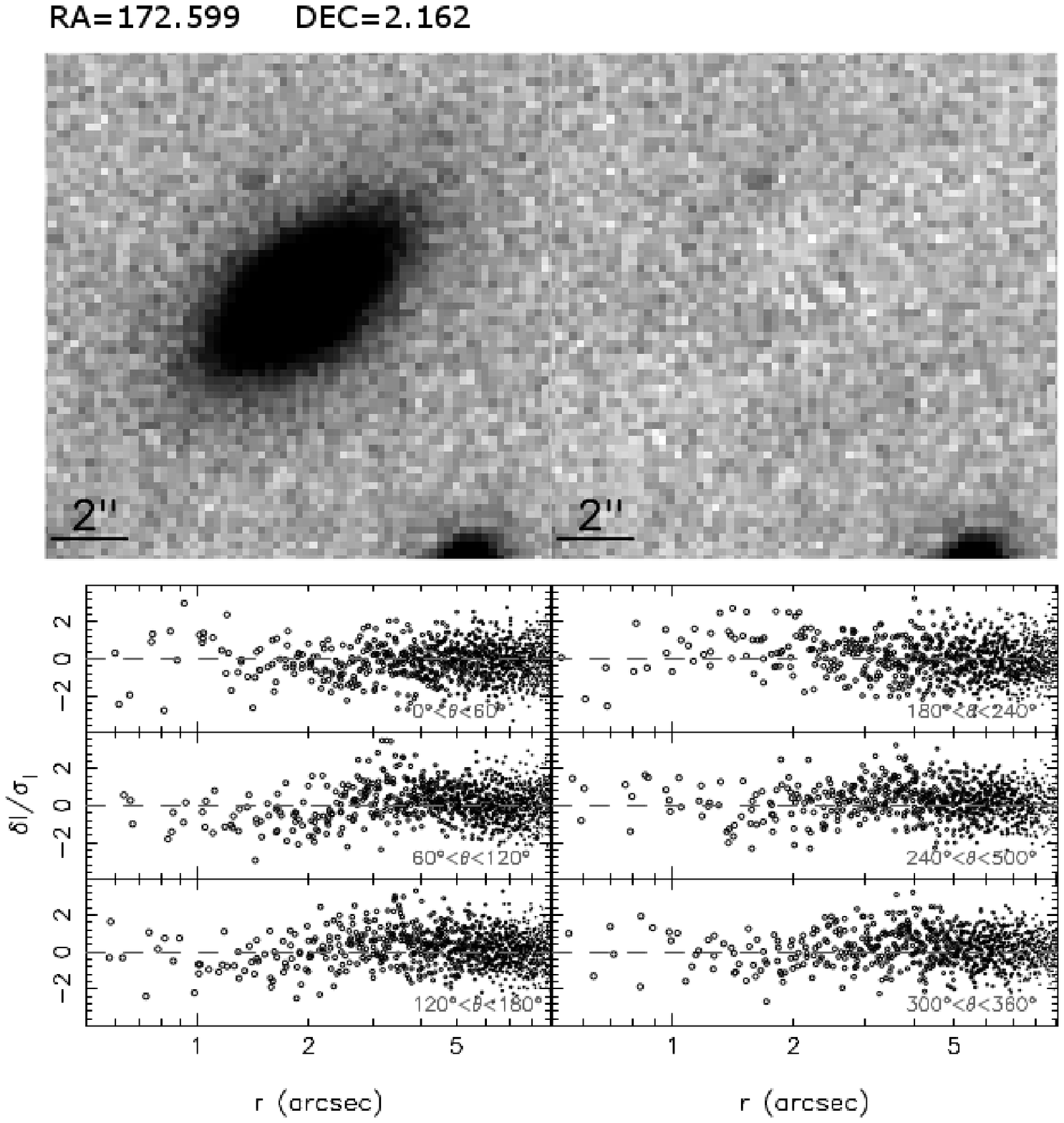}
    \includegraphics[width=\columnwidth]{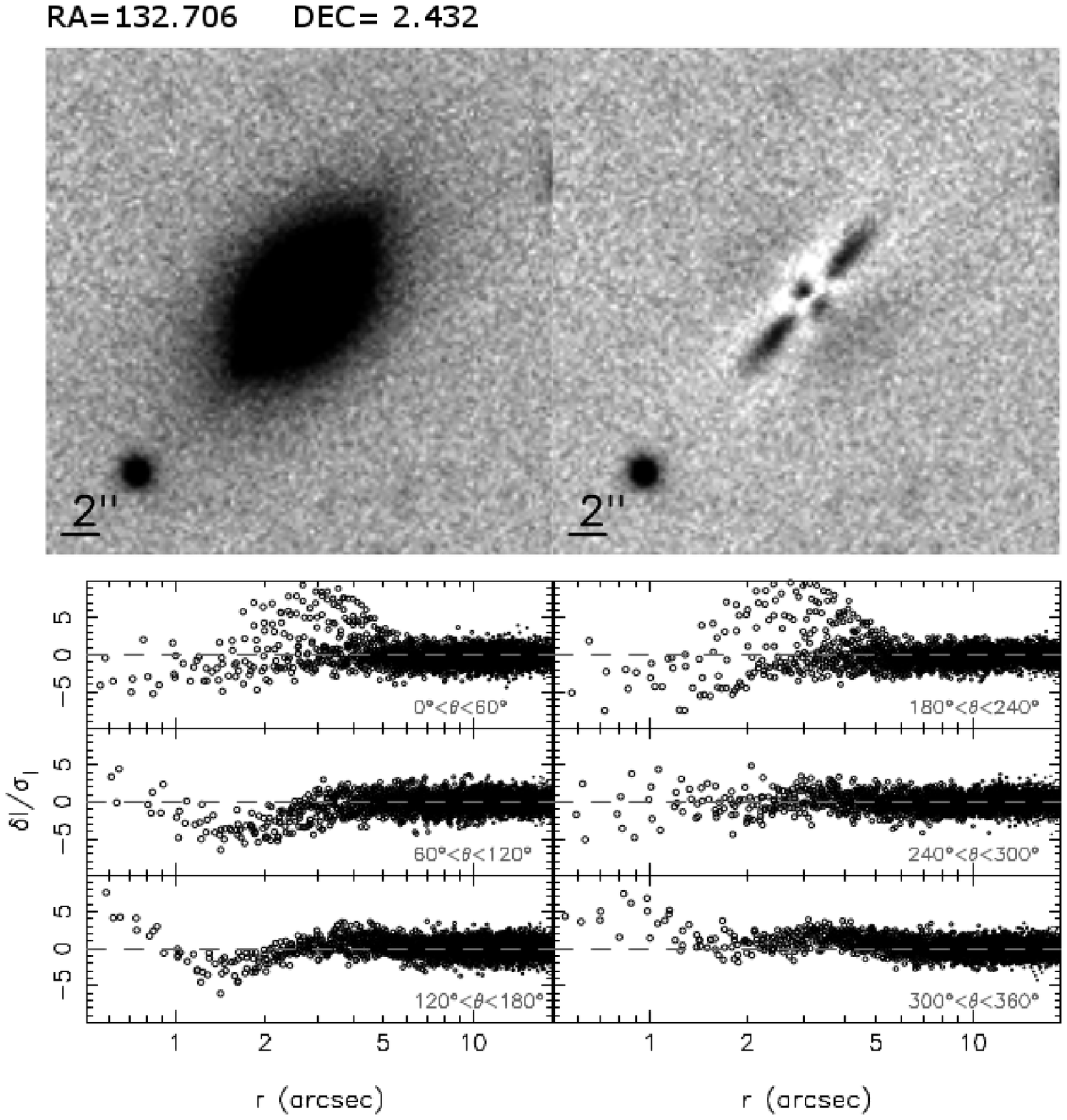}
    \caption{2DPHOT fitting in $r$-band for two example
    galaxies with $\chi\prime^2<1.3$ (left) and $\chi\prime^2>1.3$ (right). In each panel we show the galaxy image (left) and model
    subtracted image (residual, right). In the six bottom panels, residuals of the galaxy flux per pixel, after the model
    subtraction, are shown as a function of the distance to the galaxy center, in different bins of the polar angle.
    See also the text for details.}\label{2dphot_out}
\end{figure*}

In particular, 2DPHOT has been optimized to perform a Point Spread
Function (PSF) convolved S\'ersic modelling of galaxies down to
subarcsec scales (\citealt{SPIDER-I}). Typical FWHM of KiDS
observations are $1.0''\pm0.1''$ in $u$-band, $0.9''\pm0.1''$ in
$g$-band, $0.7''\pm0.1''$ in $r$-band, and $0.8''\pm0.2''$ in
$i$-band (see \citealt{deJong+15_KiDS_paperI,
deJong+017_KIDS_DR3}). As usual in large field detectors, the PSF
is somehow a strong function of the position across the
field-of-view: in \Fig\ref{PSF} we show a typical PSF pattern in
VST/OmegaCAM, images where the solid lines show the amplitude of
the elongation and orientation (anisotropy) of the PSF. Especially
in the image borders, the orientation of PSFs is strongly aligned,
while in the center the PSF tend to be more randomly oriented
(isotropic), with smaller elongations. The PSF strongly affects
the measurement of the surface brightness profile of galaxies by
anisotropically redistributing the light from the inner brighter
regions to the outer haloes (see e.g. \citealt{deJong+08_PSF}),
hence altering the inferred galaxy structural parameters (e.g.
effective radius, axis ratio, slope of the light profile, etc.).
For each source, 2DPHOT automatically selects nearby sure stars
and produces average modelled 2D PSF from two or three of them
(depending on the distance of the closest stars). The PSF is
modelled with two Moffat profiles (see
\citealt{LaBarbera_08_2DPHOT}). The best-fit parameters are found
by $\chi^2$ minimization where the function to match with the 2D
distribution of the surface brightness values is the convolved
function given by
\begin{equation}\label{Eq:fitting}
    \centering
    M(BG,\{p_k\}) = BG + B(\{p_k\}) ~\rm{o}~ S
\end{equation}
where B is the galaxy brightness distribution, which is described
by a set of parameters $\{p_k\}$; S is the PSF model; BG is the
value of the local background; and the symbol o denotes
convolution. The modelled PSF is convolved with a 2D S\'ersic
profiles with the form
\begin{equation}\label{sersic}
    \centering
   B(r, R_{m}, n) = I_{0} + \frac{2.5 b_{n}}{ln(10)} [(R/R_{m})^{1/n} - 1]
\end{equation}
For the S\'ersic models, the parameters $\{p_k\}$ are the
effective major semiaxis $R_m$, the central surface brightness
$I_{0}$, the S\'ersic index n, the axial ratio $b/a$, the position
angle PA, the coordinates of the photometric center, and the local
value of the background. In \Fig\ref{2dphot_out} two illustrative
examples of two-dimensional fit results for galaxies in $r$-band
are given.
More in details, \Re\ is computed as the circularized radius of the
ellipse that encloses half of the total galaxy light, i.e., $\Re =
(b/a)^{1/2}R_{m}$. The total (apparent) magnitude, $m_{T}$, is, by
the definition,
\begin{equation}\label{mag}
    \centering
    m_{T} = -2.5 \Log\ (2\pi) - 5 \Log\ (R_{e}) + <\mu>_{e}.
    \end{equation}

\begin{figure*}
    \includegraphics[scale=0.42]{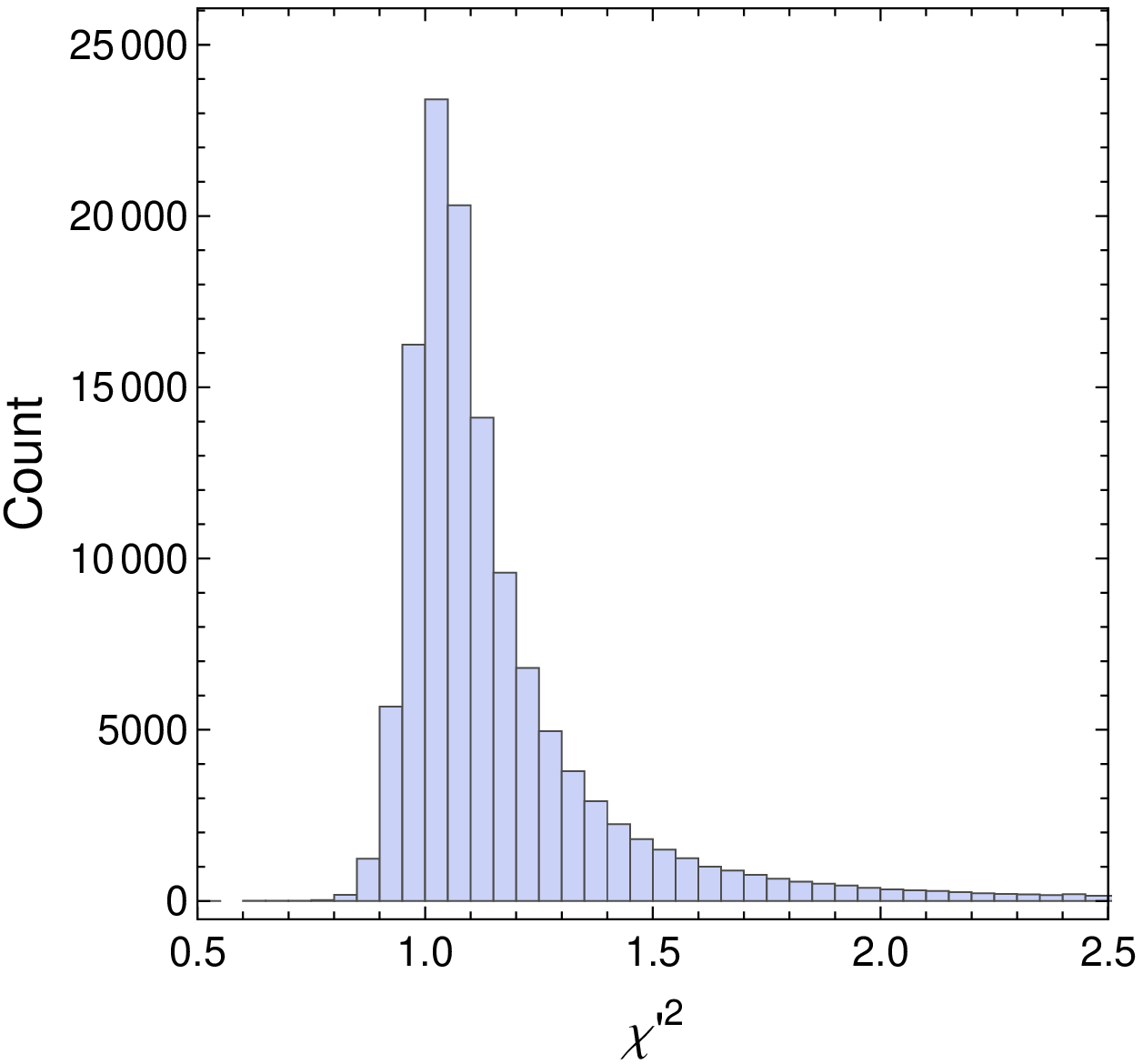}
    \includegraphics[scale=0.42]{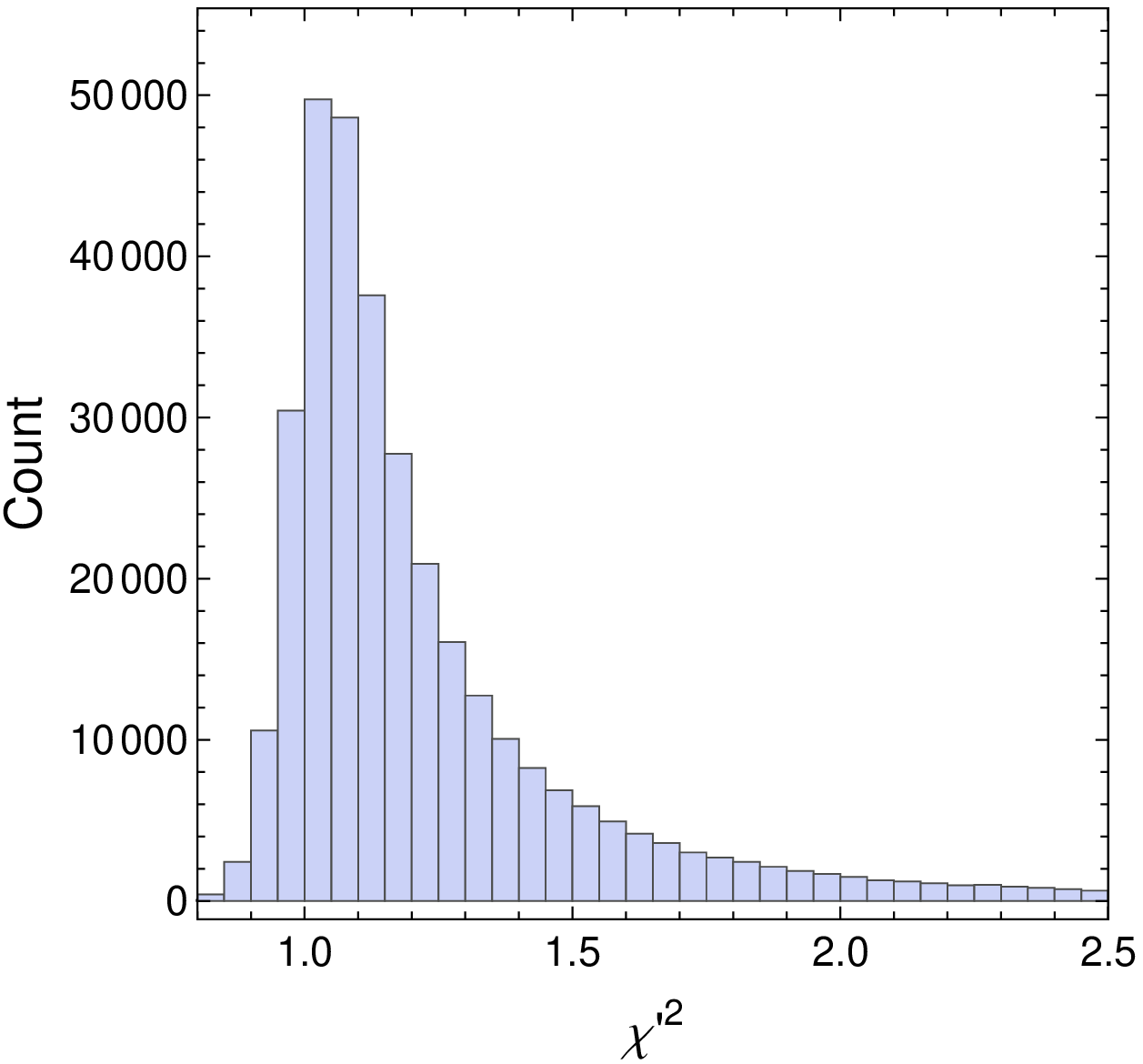}
    \includegraphics[scale=0.42]{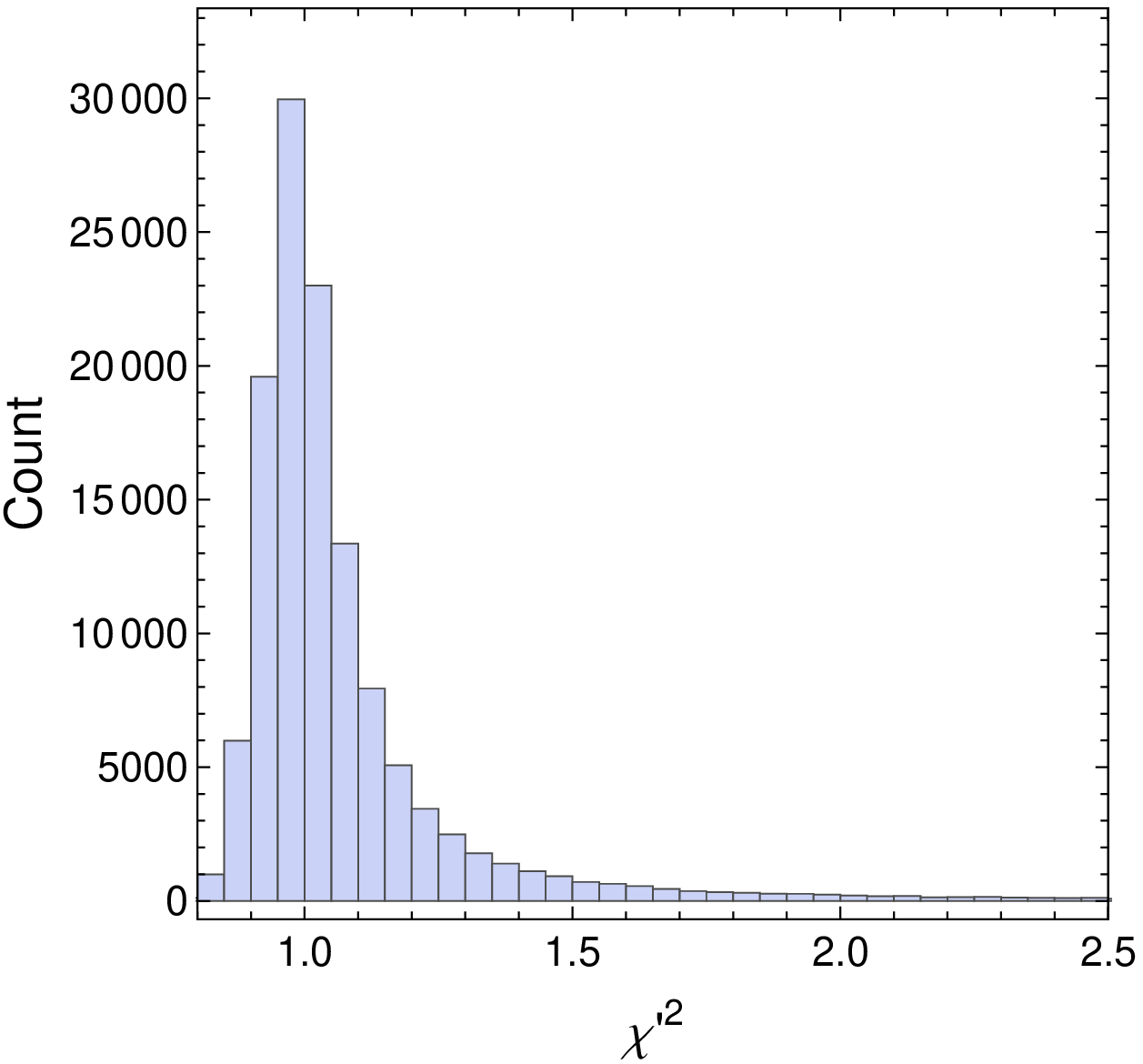}
    \caption{ $\chi\prime^2$ distribution of galaxies in g, r, and i bands, from left to right.}\label{fig:chi_g_r_i}
\end{figure*}

\subsection{Selection of best--fitted data}
In order to select the galaxies with most reliable parameters, we
defined a further $\chi\prime^2$, including in the calculation
only the pixels in the central regions. This procedure is
different from the standard $\chi^2$ definition where the sum of square
residuals over all the galaxy stamp image is minimized. The new
quantity will provide a better metric to select the galaxies with
best fitted parameters as it relies only on pixels with higher \SN,
while it is not used in the best fitting procedure itself.

\begin{figure*}
    \centering
    \includegraphics[width=.8\textwidth]{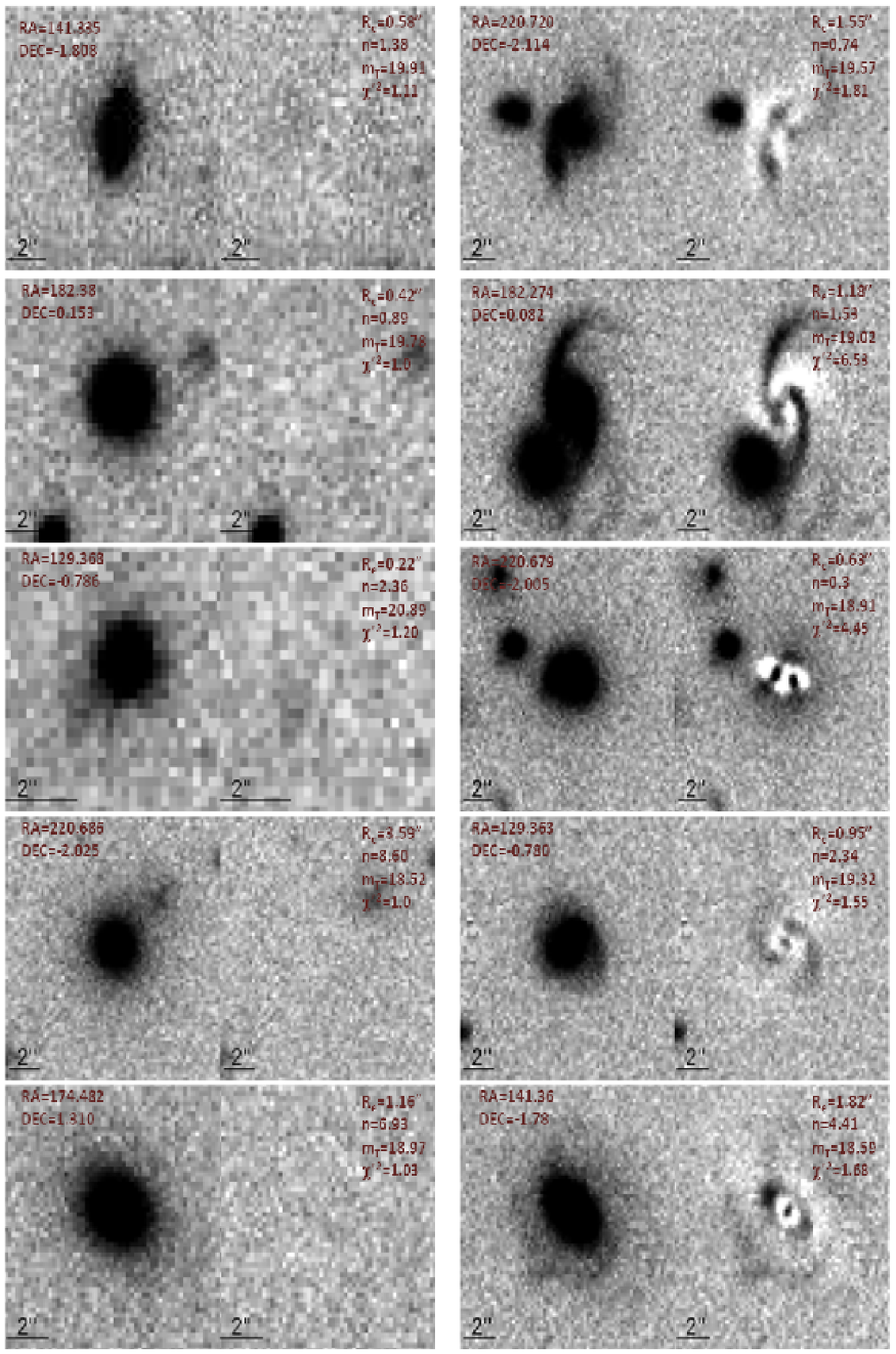}
   \vspace{0.5cm}
    \caption{More examples of the 2D fit results for galaxies in r band. The left panels show
    the results for galaxies with good fits ($\chi\prime^2 < 1.3$) and the
    right panels those with bad fits ($\chi\prime^2 > 1.3$). In each panel the source
    and the model subtracted residual maps are shown.}\label{fit_examples}
\end{figure*}

\begin{figure}
    \centering
    \includegraphics[width=\columnwidth]{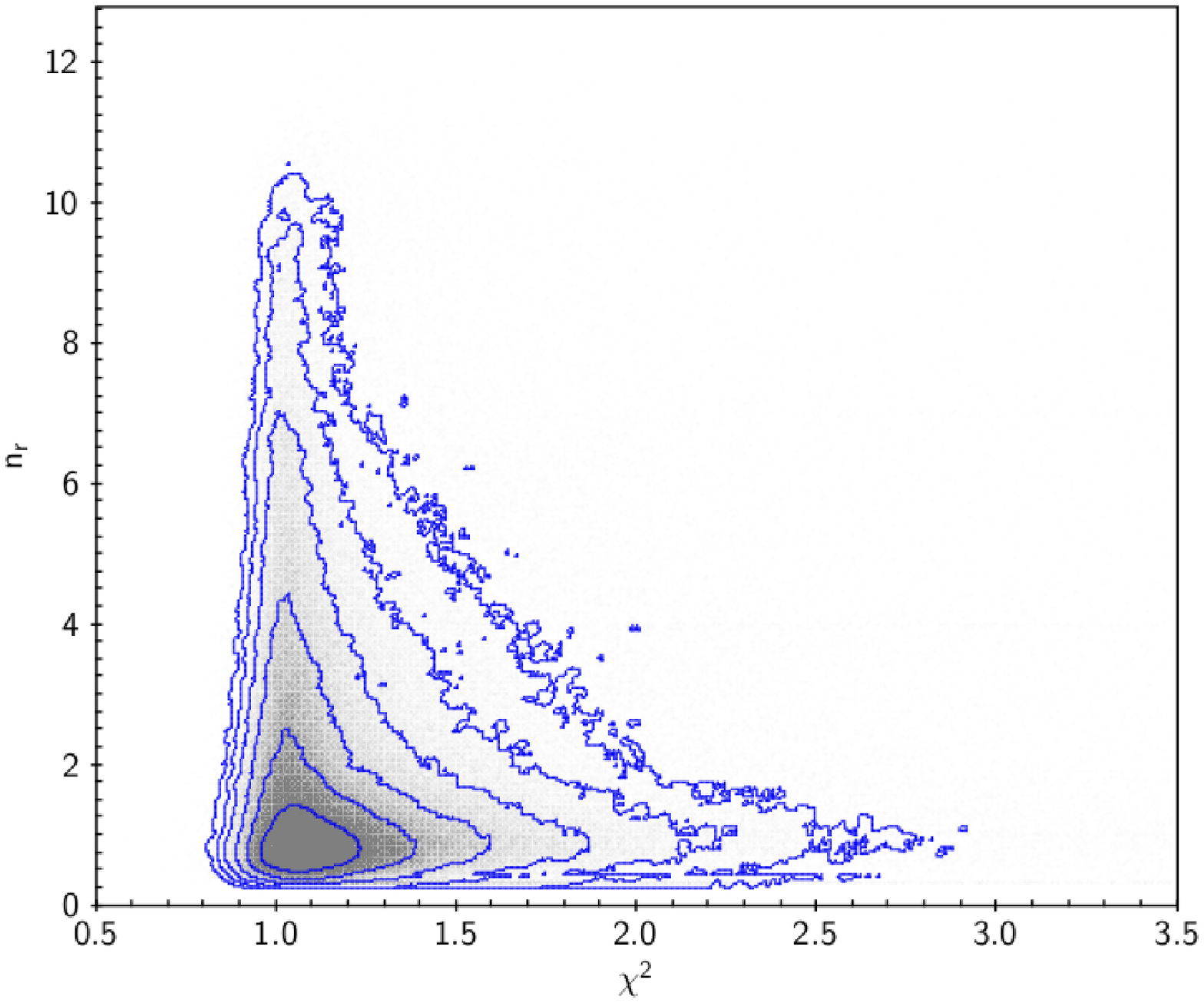}
    \caption{The plot shows the S\'ersic index vs. $\chi\prime^2$ in $r-$band. We note that at lower $n$ ($\lsim 2.5$)
    there is an excess of large $\chi\prime^2$ ($>1.3$), due to the presence of substructures
    in the residuals, demonstrating that these $n$ values are a good proxy of
    later morphological types. Log-spaced isodensity contours show that the tails
    of high-$\chi\prime^2$ become dominant in the $\chi\prime^2$ distribution of the best-fit
    at the smaller S\'ersic index  index (i.e. $\chi\prime^2 \lsim 2$).} \label{fig:chi2-n}
\end{figure}

To compute the $\chi\prime^{2}$ for each galaxy, all pixels
1$\sigma$ above the local sky value background value are selected
and the 2D model intensity value of each pixel is computed from
the two dimensional seeing convolved S\'ersic model as in
\Eq\ref{Eq:fitting}. For the selected pixels, the $\chi\prime^{2}$
is computed as the rms of residuals between the galaxy image and
the model. The distribution of the $\chi'^2$ for the whole high
$\SN$ sample in the  $g$, $r$, and $i$ bands are given in
\Fig\ref{fig:chi_g_r_i}. As shown in the right panel of
\Fig\ref{2dphot_out} we have galaxies with larger $\chi\prime^{2}$
(e.g. $\chi\prime^{2}>1.3$), which corresponds to lower quality
models. This is clearly shown in \Fig\ref{fit_examples}, which
displays more examples of galaxy images and residual maps in the
r-band. Here, galaxies with $\chi\prime^2 < 1.3$ are shown on the
left two columns and examples of $\chi\prime^2 > 1.3$ galaxies and
residuals are on the right two columns. In the first group the
S\'ersic fit performs very good with almost null residuals, while
in the second group substructures like spiral arms, rings, double
central peaks from ongoing mergers, etc. show up in the residuals.
We substantiate our argument using \Fig\ref{fig:chi2-n} where we
plot the $n$-index vs. $\chi\prime^2$, which shows that for lower
S\'ersic index ($n<2.5$) there is an excess of large
$\chi\prime^2$, i.e. worse fit, due the fact that at these low-$n$
late-type systems are predominant (\citealt{Ravindranath+02},
\citealt{Trujillo+07}, \citealt{LaBarbera+02_2dphot}) and tend to
have significant substructures. Indeed, the fraction of high
$\chi\prime^2$ is larger in bluer bands, which is probably
affected by star forming regions generally populating
substructures of regular discs in late-type systems.

This is a relevant result which show that the good KiDS image
quality, combined with an accurate surface photometry analysis,
can allow us to correlate the structural properties of the
galaxies, as the S\'ersic index, with the residuals in the
subtracted images, e.g. the typical late-type features. This could
provide further parameters for galaxy classification, which we
plan to investigate further in future analyses.

The use of a single S\'ersic profile is not the more general
choice we could make, as it is well known that galaxies generally
host more than one photometric component (see e.g.
\citealt{Kormendy+09}). This is not only true for late-type
systems, showing a bulge+disc structure, but also for some large
ellipticals, now systematically found to have extended
(exponential) haloes (e.g. \citealt{Iodice+16}). Looking at the
$\chi\prime^2$ distribution in \Fig\ref{fig:chi_g_r_i}, the
fraction of galaxies with $\chi\prime^2>1.3$ is not negligible,
and amounts to $\sim40\%$ in $r$-band.

However, the adoption of multi-component models has two main
disadvantages: the degeneracies among parameters and the higher
computing time due to the higher dimensionality of the parameter
space. In particular, the amount and the quality of the
information (e.g. the number of pixels across which typically
high$-z$ galaxies are distributed on CCDs of the order of few
tens) makes very hard to obtain reliable modelling of
multi-component features in galaxies, especially when the ratio
between the two components is unbalanced toward one (see e.g. the
case in the right panel of \Fig\ref{2dphot_out}, where the inner
disc represents a minor component of the dominant bulge).

\begin{figure*}
  \centering
    \includegraphics[scale=0.55]{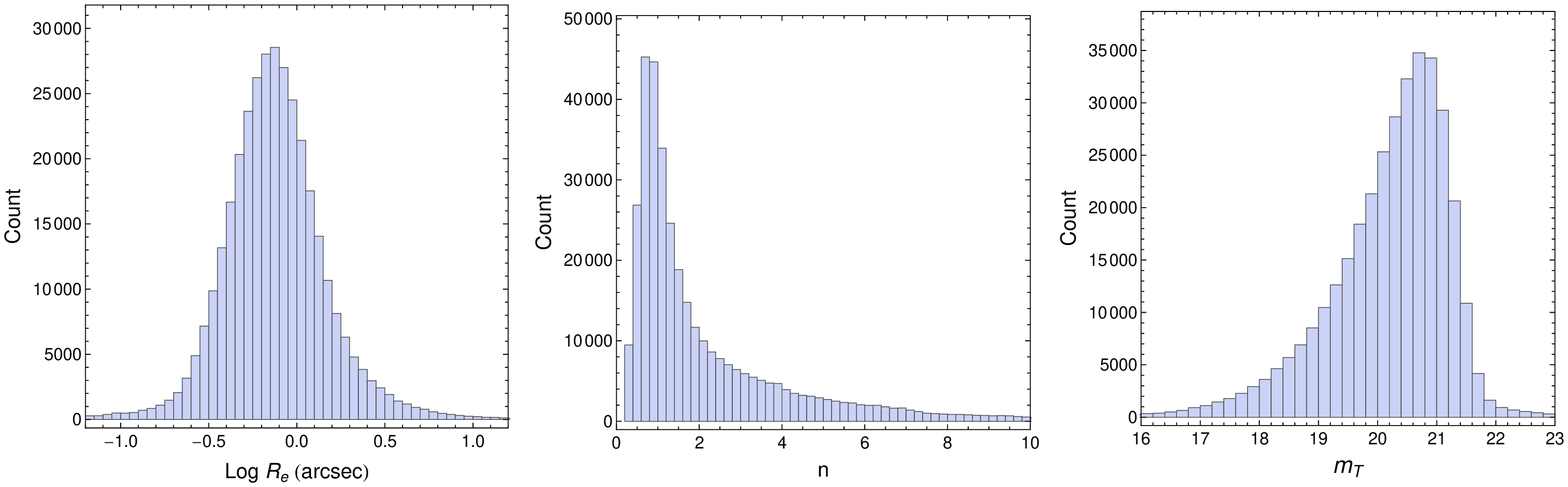}
    \caption{Distribution of structural parameters in r-band: for $\Log\ \Re$,
    $n$, $m_T$, from left to right panel.}\label{param_distribution}
\end{figure*}

For our analysis we have adopted image stamps centered on each
galaxy of $\sim$ 100 arcsec by side, i.e. 500 pixels given the
resolution of telescope of 0.2 arcsec/pix. This stamp size has
been chosen as best compromise between computational speed and
area covered. We have excluded from our analysis galaxies with
$\Re>50''$, as these might be (i) galaxies for which the 2D light
distribution is poorly sampled, resulting into overly large \Re\
values or (ii) galaxies with a second extended component, that is
modelled as a single component with large $n$, resulting into
large \Re.
We conclude this section by showing the distribution of the
best-fit structural parameters obtained in $r$-band to give a
perspective of the parameter space covered by the sample. In
\Fig\ref{param_distribution} this is given for the effective
(half-light) radius, \Re, the S\'ersic index, $n$, and the total
magnitude, $m_T$. The median effective radius of the sample is 5.4
arcsec, while the median of the S\'ersic index is 1.3 and the
median of the total mag is 20.4 in $r$-band. The distribution of
the $R_e$ is quite symmetric and show that we can reach galaxy
sizes of the order of the tenths of the arcsec for the smallest
systems, while the largest galaxies measured can be as large as 10
arcsec and more. The S\'ersic index distribution shows a large
tail toward the larger n-index, i.e. at $n>2$. This shows that the
spheroidal-like systems are not the dominant class of galaxies in
our sample. The total magnitude distribution also shows the effect
of the sample completeness as the median almost corresponds to the
completeness magnitude (see \S\ref{sec:mag_compl}).

\subsection{Uncertainties on structural parameters}\label{sec:uncert}
We have estimated the statistical errors on the estimated structural
parameters using two approaches: 1) internal: by comparing estimates
obtained by or best fit in contiguous bands; 2) simulations: by applying
our procedure on mock galaxies mimicking KiDS observations and checking
how the estimated parameters compare to the know input ones of the
simulated galaxies.
\subsubsection{Internal check}
We first estimate the uncertainties on structural parameters by
comparing the differences in $\Log\ \Re$, $\langle \mu_{e}
\rangle$, and $\Log\ n$ between contiguous wavebands, in our case
we have adopted $r$ and $i$ bands. The basic assumption is that
these two bands are close enough that the variation of the galaxy
properties from one band to other is dominated by the measurement
errors (\citealt{SPIDER-I}). Therefore, this approach provides an
upper limit to the uncertainty on structural parameters.

\begin{figure*}
     \centering
    \includegraphics[scale=0.55]{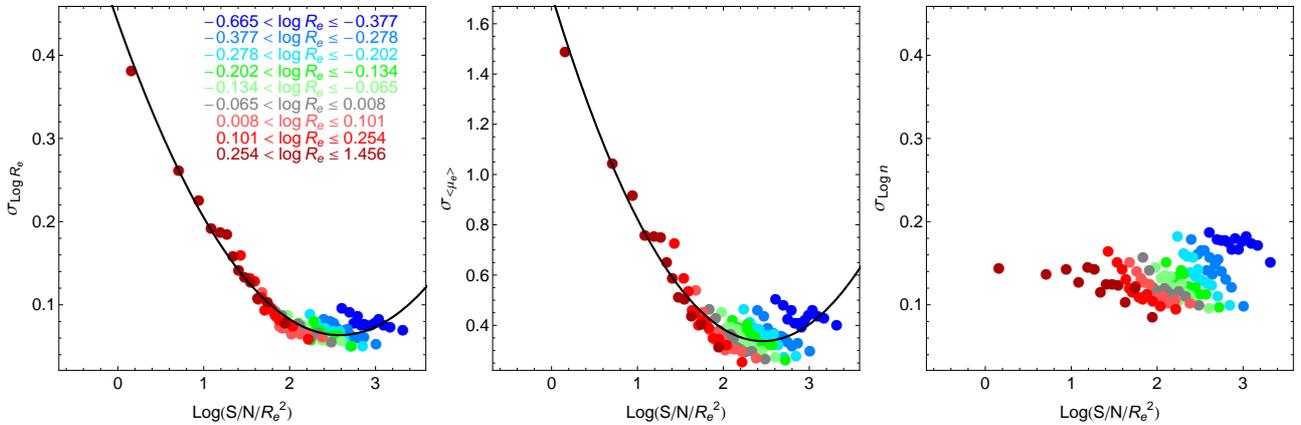}
    \caption{Uncertainties in the parameters $\Log\ \Re$, $\langle\mu_{e}\rangle$, and
    $\Log\ n$ as a function of the Logarithm of the \SN\ per unit area. Different colours show different bins of $\Log\ \Re$,
    where \Re\ is in arcsec. For a given colour the points are the uncertainties in different bins of Logarithm of
    $S/N/\Re^2$. The black curve is the best fitting functional form used to model the dependence of the uncertainties
    on S/N. This fit is not performed for $\Log\ n$ as it does not shows any correlation with S/N$/\Re^2$.}\label{uncertainty}
\end{figure*}

For the uncertainty calculation we follow the method explained in
\cite{SPIDER-I}. We bin the differences in the $\Log\ \Re$,
$\langle \mu_{e} \rangle$, and $\Log\ n$ between $r$ and $i$ bands
with respect to the Logarithm of the mean effective radius $\Log\
\Re$ and \SN\ per unit area of the galaxy image,
$\SN/{R_{e}}^{2}$. In this case the \SN\ is defined as the mean
value of the inverse of $\MErrauto$, between the two bands. Bins
are made such that the number of galaxies in each bin is same.
Measurement errors on $\Log\ \Re$, $\langle \mu_{e} \rangle$, and
$\Log\ n$ are computed from the mean absolute deviation of the
corresponding differences in that bin. The results are shown in
Fig \ref{uncertainty}.

The errors on the parameters show a dependency on the S/N per unit
area: as the value of S/N per area decreases ($\Log\ (S/N/\Re^{2})
< 2$), the errors tends to increase. This is due to the combined
effect of the S/N and the number of pixels where the signal is
distributed. At low $S/N/{\Re}^{2}$, there are sources with large
\Re\ and small S/N, whereas high $S/N/{R_{e}}^{2}$ are systems
that might have large S/N, but due to the small number of pixels
induces the uncertainty on parameters. Most of the galaxies have
\Re\ in the range $-0.5 < \Log\ \Re <0.2$, where the errors on the
parameters are less than 0.1 dex for \Re\ and less than 0.4 dex
for $\langle \mu_{e} \rangle$, but the errors on $n$ are more
randomly distributed and do not show particular trends. However,
also in this case, they stay remarkably contained below 0.2 dex.

\subsubsection{Simulated galaxies}\label{sec:simul}
\begin{figure*}
\centering
\includegraphics[width=2\columnwidth]{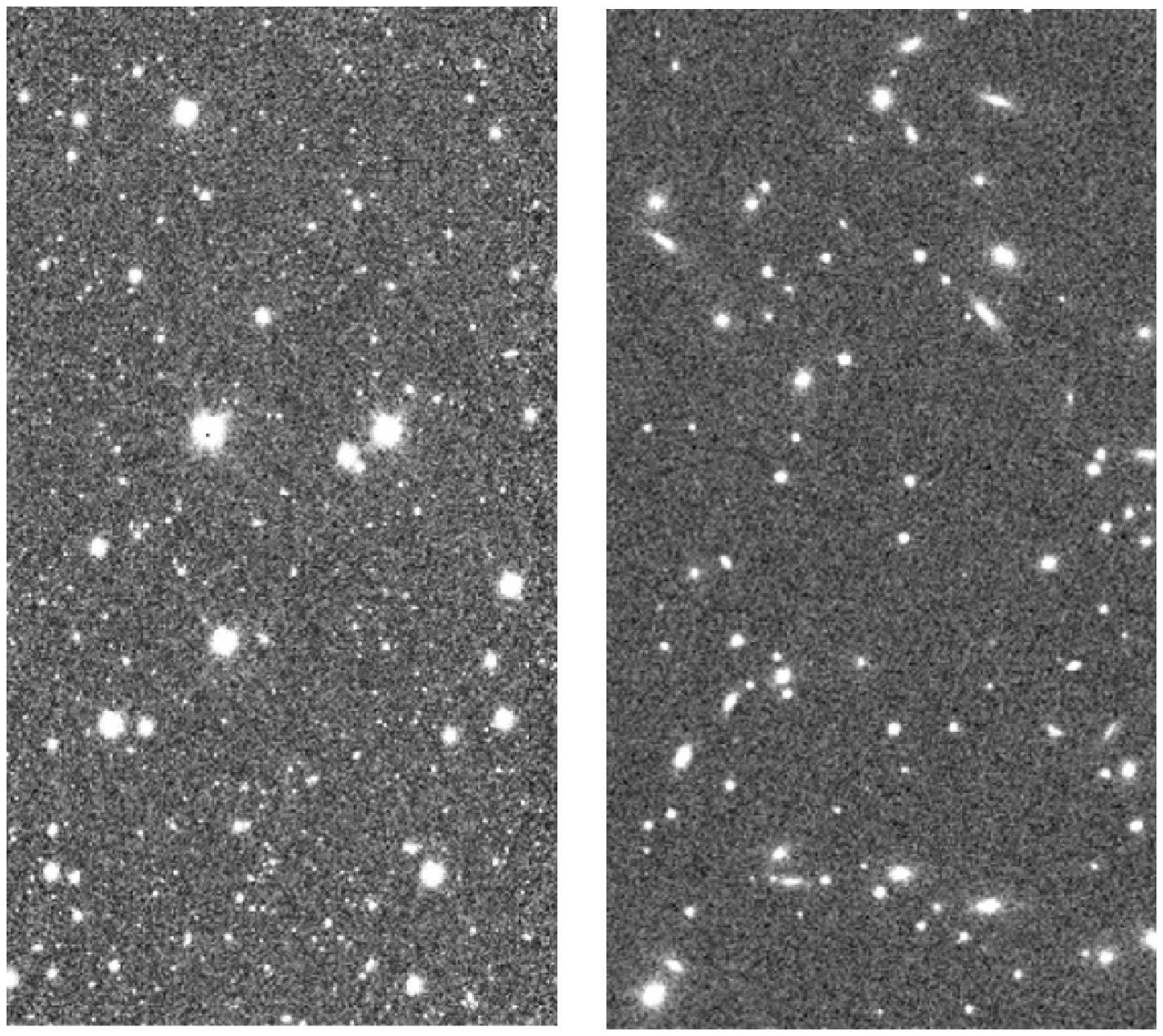}
\vspace{0.5cm} \caption{A real KiDS image (left) vs. a mock image
with simulated galaxies (right). Seeing FWHM are 0.69 and 0.66 for
real and mock images respectively. }\label{fig:mock}
\end{figure*}

A further approach to assess the reliability of the parameters
obtained from the fitting procedure and estimate their intrinsic
statistical errors, is based on mock galaxy images generated on
top of a gaussian background noise, given by the background rms
measured for the KiDS images. The artificial galaxies have
physical parameters, i.e., magnitude, S\'ersic index, effective
radius, and axis ratio, which are assigned based on a grid of
values. For each parameter, the grid of values was chosen based on
the range of values for the observed galaxies.  In particular we
have uniformly sampled the parameters in the following intervals:
$0.2 \leq \Re \leq 20$ arcsec, $0.6\leq n \leq 10$, $0.5\leq
b/a\leq 1$, and $16 \leq m_T\leq 24$ mag.  About the choice of
using a uniform distribution in total magnitude, instead of using
a realistic luminosity function, we stress here that we are not
interested in producing realistic images, but rather realistic
individual systems which we want to analyse to assess the
robustness of our procedures. This causes a lack of faint systems
in our simulated images with respect to real images as seen in
\Fig\ref{fig:mock}. As this does not impact the local background
of the brighter systems, representative our complete sample, the
overall results of the analysis are not affected. We have
simulated about 1800 galaxies on image chunks of 3000 pixels by
side in order to reproduce the same galaxy density observed in
KiDS images. We have generated such mock observations in different
bands and in different seeing conditions. In \Fig\ref{fig:mock} we
show an example of simulated $r$-band image, compared with a real
one.

\begin{figure*}
\includegraphics[scale=.55]{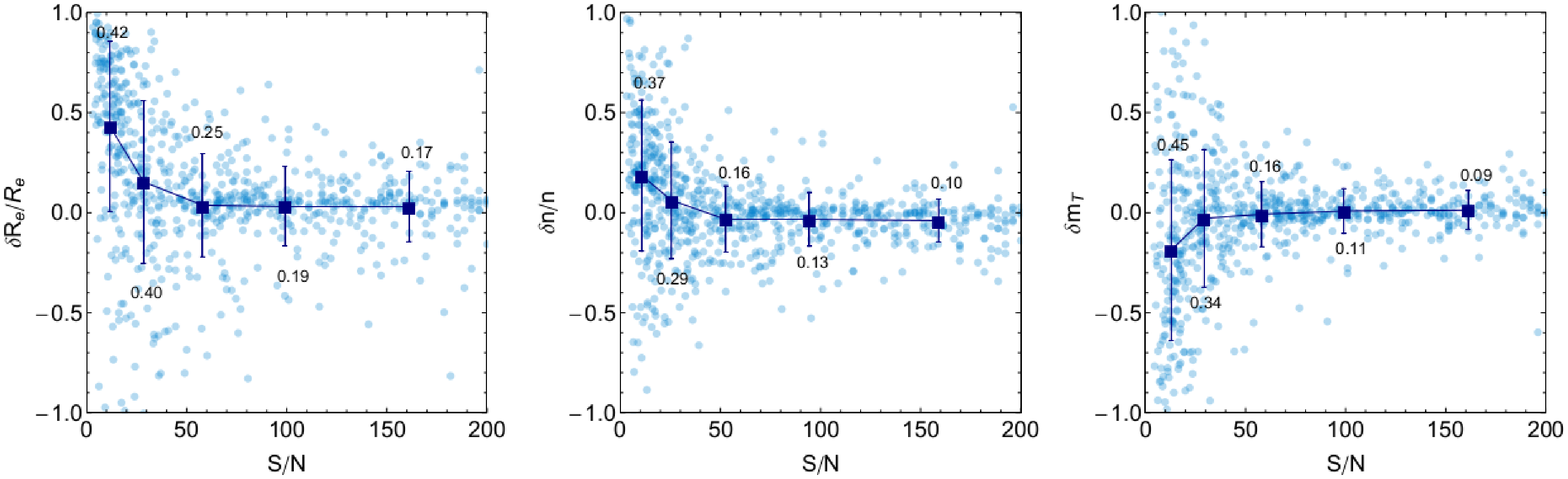}
\caption{Figure shows differences between the input and output
parameters for \Re\, n, and $m_T$, with respect to \SN. We define
the quantity $\delta p_k = (p_k^{\rm in}-p_k^{\rm out})$, with
$p_k=\Re\,~n,~m_T$. We plot $\delta \Re / \Re^{in}$, $\delta n /
n^{in}$ and $\delta m_{T}$ in terms of \SN. Datapoints for single
galaxies are plotted as blue points. Mean values are plotted as
filled squares and error bars show the standard deviation in bins
of \SN. The numbers given are the standard deviations in each
bin.} \label{simulation}
\end{figure*}

We have then applied 2DPHOT to the mock images with the same setup
used for the real images (see \Sec\ref{sec:surf_phot}). The
relative differences between the measured quantities and the input
ones adopted to generate the simulated galaxies are shown in
\Fig\ref{simulation} as a function of the \SN.

The figure shows that the input and output values are well in
agreement with each another, except in the low-\SN\ regime (i.e.,
$S/N\lsim50$), where we start observing a systematic deviation of
the measured values from the input ones. This is an {\it a
posteriori} confirmation that our choice of $\SN
> 50$ for robust structural parameter studies was correct.

In the same Figure we show the relative differences of the same
observables against the input values (bottom row): in this case
there is no trend in the derived quantities and statistical errors
stay always below 10\%. We have found that these good accuracies
are independent of the band and of the seeing, as long as we
restrict to galaxies with $\SN>50$ in any given bands.

\subsection{Check for systematics on the estimated parameters}
In this section we proceed with a series of validation tests to
check the presence of biases in the parameter estimates. To do that we
have selected literature samples having an overlap with our KiDS galaxy sample.
However, before going on with tests on external catalogs
we will start with a basic check on the effect of the background
evaluation on the parameter estimates.

\subsubsection{Effect of sky background}
We have discussed in \Sec\ref{sec:stru_par} that the background is
a free parameter in our fitting procedure (see e.g.
\Eq\ref{Eq:fitting}). However, it is well known that the
simultaneous fit of the background and the photometric laws can be
degenerate and produce some systematics.

In order to estimate the effect of background fitting on the
estimate of structural parameters, we have repeated the fitting of
galaxy image by keeping the background as a fixed parameter. We
measured the background value far from the galaxy (local
background value calculated from the galaxy stamp images, which is
1.5 times the S-Extractor $ISOAREA$ parameter, see
\citealt{LaBarbera_08_2DPHOT} for more details) and entered as the
initial guess in the fitting procedure. Here, we fix this value of
background for the modelling.

\begin{figure*}
\includegraphics[scale=0.55]{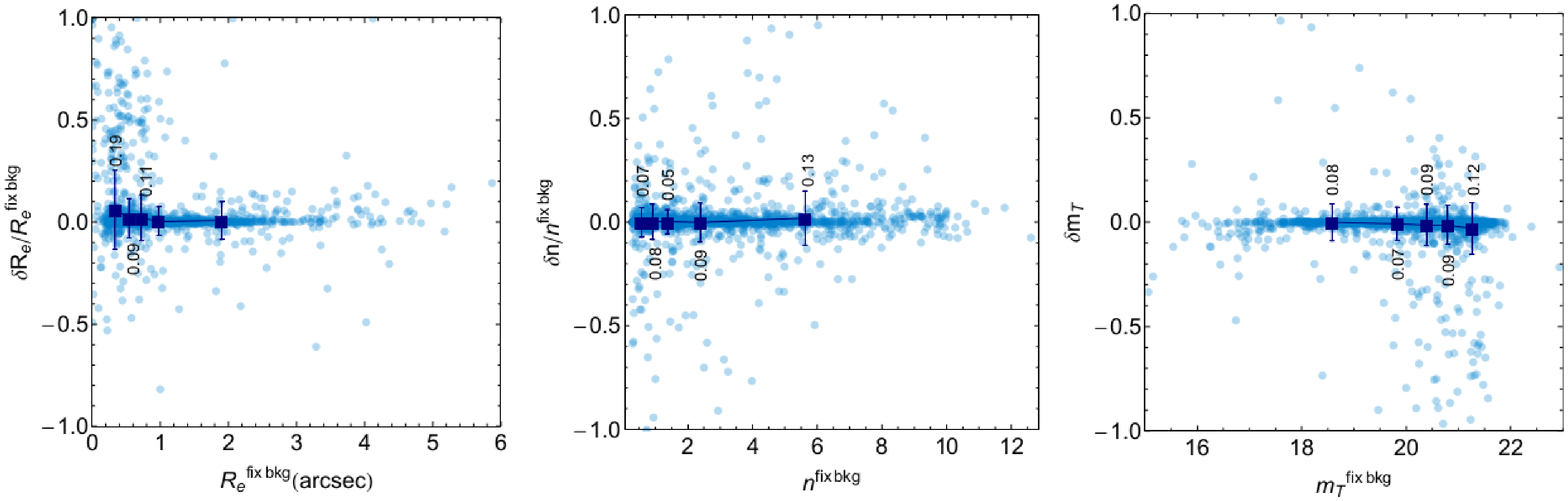}
\caption{ Differences in the $r$-band parameters \Re\, n, and
magnitude when background is kept constant with respect to the
value when background is subjected to change. We define the
quantity $\delta p_k = (p_k^{\rm fix}-p_k^{\rm var})$, with
$p_k=\Re\,~n,~m_T$. We plot $\delta \Re / \Re^{fix}$, $\delta n /
n^{fix}$ and $\delta m_{T}$ as a function of $\Re^{fix}$,
$n^{fix}$ and $m_{T}^{fix}$, respectively. Mean values are plotted
as filled squares and are given along with the single datapoints.
Error bars show the standard deviation in bins of parameter
plotted on x axis. The numbers given are the standard deviation in
each bin.} \label{background_figs}
\end{figure*}

We randomly selected $\sim$ 3000 galaxies from our high--S/N
galaxy sample and again extracted the structural parameters. We
compare the two sets of structural parameters we have obtained
with the standard procedure and the one with fixed background. The
differences in structural parameters are shown in
\Fig\ref{background_figs}.

Squares and error bars represent mean and standard deviation of
the scattered plot. For most of the selected galaxies the
differences between measured and input parameters are negligible.
The background fit does not introduce systematics and the error
associated to the background measurement is of the order of
10-20\% in \Re\, less than 10\% in $n$, and less then 20\% in the
total magnitude, which are in line with the estimates in
\Sec\ref{sec:uncert}.

\begin{figure*}
\includegraphics[scale=.55]{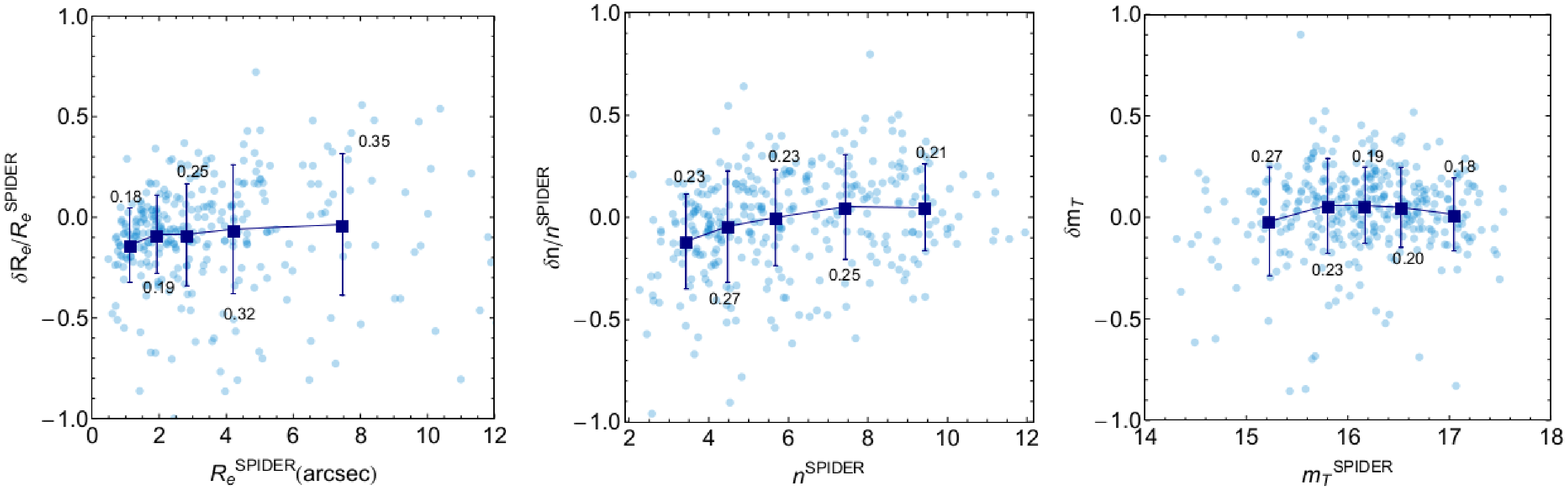}
\caption{Comparison of KiDS structural parameters with the ones
derived within the SPIDER survey. The SPIDER dataset consists of
spheroids with redshifts in the range $0.05 < z < 0.095$, selected
from SDSS; the structural parameters are derived using 2DPHOT. We
define the quantity $\delta p_k = (p_k^{\rm SPIDER}-p_k^{\rm
KiDS})$, with $p_k=\Re,~n,~m_T$. We plot $\delta \Re /
\Re^{SPIDER}$, $\delta n / n^{SPIDER}$ and $\delta m_{T}$ in terms
of \Re, n and $m_{T}$, respectively. Data are shown as points.
Mean values and standard deviations are plotted as filled squares
and error bars. The numbers are the standard deviations in each
bin.} \label{spider_kids}
\end{figure*}

\begin{figure*}
\includegraphics[scale=.46]{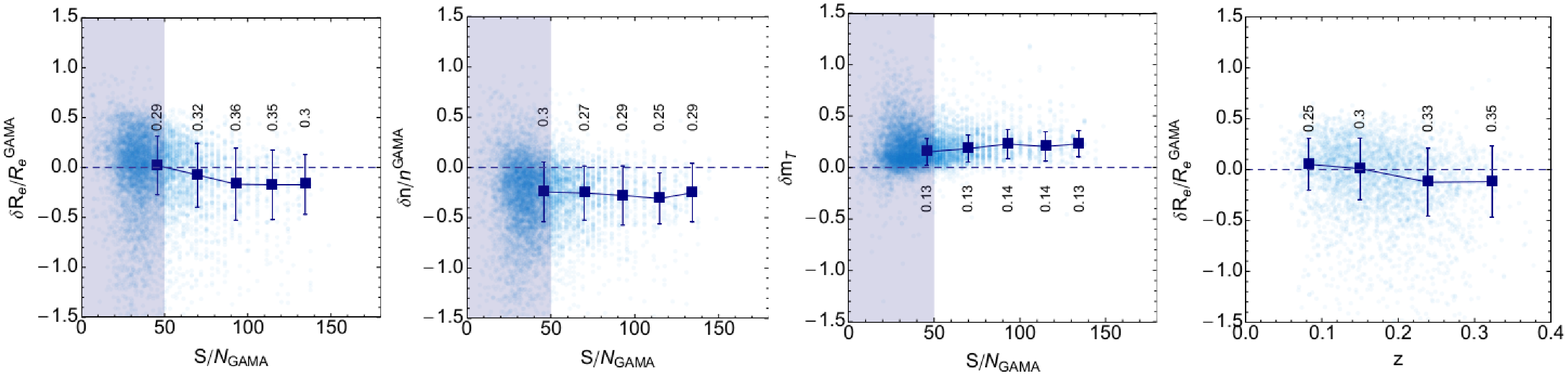}
\caption{Comparison of KiDS structural parameters with the ones
derived by GAMA using SDSS images (\citealt{Kelvin+012}). The
structural parameters are derived using GALFIT
(\citealt{Peng+002_GALFIT}). We define the quantity $\delta p_k =
(p_k^{\rm GAMA}-p_k^{\rm KiDS})$, with $p_k=\Re,~n,~m_T$. In the
first three panels we plot $\delta \Re / \Re^{GAMA}$, $\delta n /
n^{GAMA}$ and $\delta m_{T}$ as a function of \SN. In the fourth
panel to the right, we also plot the $\delta \Re / \Re^{GAMA}$ vs.
redshift, which shows that there is no significant systematics
between the GALFIT parameters and the ones obtained with 2DPHOT as
a function of the redshift. Median values and median deviations
divided by 0.675 (as equivalent to the standard deviation) are
plotted as filled squares and error bars. The numbers are the
standard deviations in each bin.} \label{fig:sdss_kids}
\end{figure*}

\subsubsection{Comparison of KiDS and SDSS structural parameters}\label{sec:sdss_comp}
We want now to compare our results with some external catalogs to
check the presence of biases. The accuracy of our structural
parameter estimates is compared with two samples which overlaps
with KiDS sky area.

First, the SPIDER galaxies (\citealt{SPIDER-I}),
which includes 39,993 spheroids with SDSS
optical imaging and UKIDSS Near Infra Red (NIR) imaging, with
redshifts in the range $0.05 \lsim z \lsim 0.1$.

This sample has structural parameters derived with the same
software (2DPHOT) used in this paper for KiDS, but applied on
SDSS images, which have a poorer image quality. This would
give us the effect of depth (KiDS is two
magnitudes deeper than SDSS) and  image quality (both pixel scale
and seeing are about twice smaller in KiDS) on the parameter
estimates being the analysis tool substantially the same for the
two datasets. By matching the KiDS data with SPIDER we found 344
galaxies in common for which we can have a direct comparison of
the derived parameters. This allows us to measure the relative
differences among the structural parameters. The results are shown
in \Fig\ref{spider_kids}, where we can see a good agreement among
the parameters from the two datasets with the scatter (measured by
the errorbars) in line with the statistical errors ($\sim$10\% or
below) discussed in \Sec\ref{sec:simul}.

Secondly, we have checked our structural parameters with the ones
obtained by the GAMA collaboration (\citealt{Kelvin+012}) using
GALFIT (\citealt{Peng+002_GALFIT}) on SDSS optical images. This
subsample consists of 7857 galaxies and the results are shown in
\Fig\ref{fig:sdss_kids}, where again we plot the relative
differences among the structural parameters. In this test, both
data and analysis methods are different, hence we can check
whether the combination of the image quality and the analysis
set-up can introduce some differences in the galaxy inferences.

The comparison with SDSS and KiDS data shows a clear offset
between the two sets of parameters of the order of 20\%. This was
already found when comparing the 2DPHOT estimates with GALFIT on
SDSS data (see e.g. \citealt{SPIDER-I} for details), hence this
has to be related to the different tools' performances. In
\Fig\ref{fig:sdss_kids} we plot the structural parameters against
the \SN\ defined as for the KiDS case. We can see that a large
part of the GAMA sample have a \SN$<50$, a region where the
scatter among the two analysis increases and results from SDSS
should be less robust. However, the offset shows-up at the higher
\SN\ which suggests that the differences are not due to the poorer
SDSS quality. In general, effective radii and S\'ersic indices
with GALFIT are smaller with respect to those of 2DPHOT by 15\%
and 25\% or less respectively, whereas the total magnitude from
2DPHOT is brighter by $\sim0.2$ mag compared to the SDSS. The
offset of the \Re\, in particular, seems consistent with zero
within the (albeit large) scatter.

There might be many reasons why the two software might have
brought to systematics (e.g. PSF sampling, convolution methods,
background estimate etc.) and a detailed discussion of the origin
of this is beyond the scope of this paper. Based on our test done
with mock galaxies in \Sec\ref{sec:simul}, corroborated by the
check vs. the SPIDER sample, we are confident that the 2DPHOT
estimates are fairly accurate. However, we will perform a
challenge of different surface photometry tools on an advanced
mock galaxy catalog on the next paper (Raj et al., in
preparation). We just remark here that there seems to be no trend
of the offset with the redshift, as shown on the last panel of
\Fig\ref{fig:sdss_kids}: since most of the focus of the paper is
on the galaxy size evolution with redshift, we believe that our
results should not suffer any severe systematics.

\begin{figure*}
\centering
\includegraphics[scale=0.57]{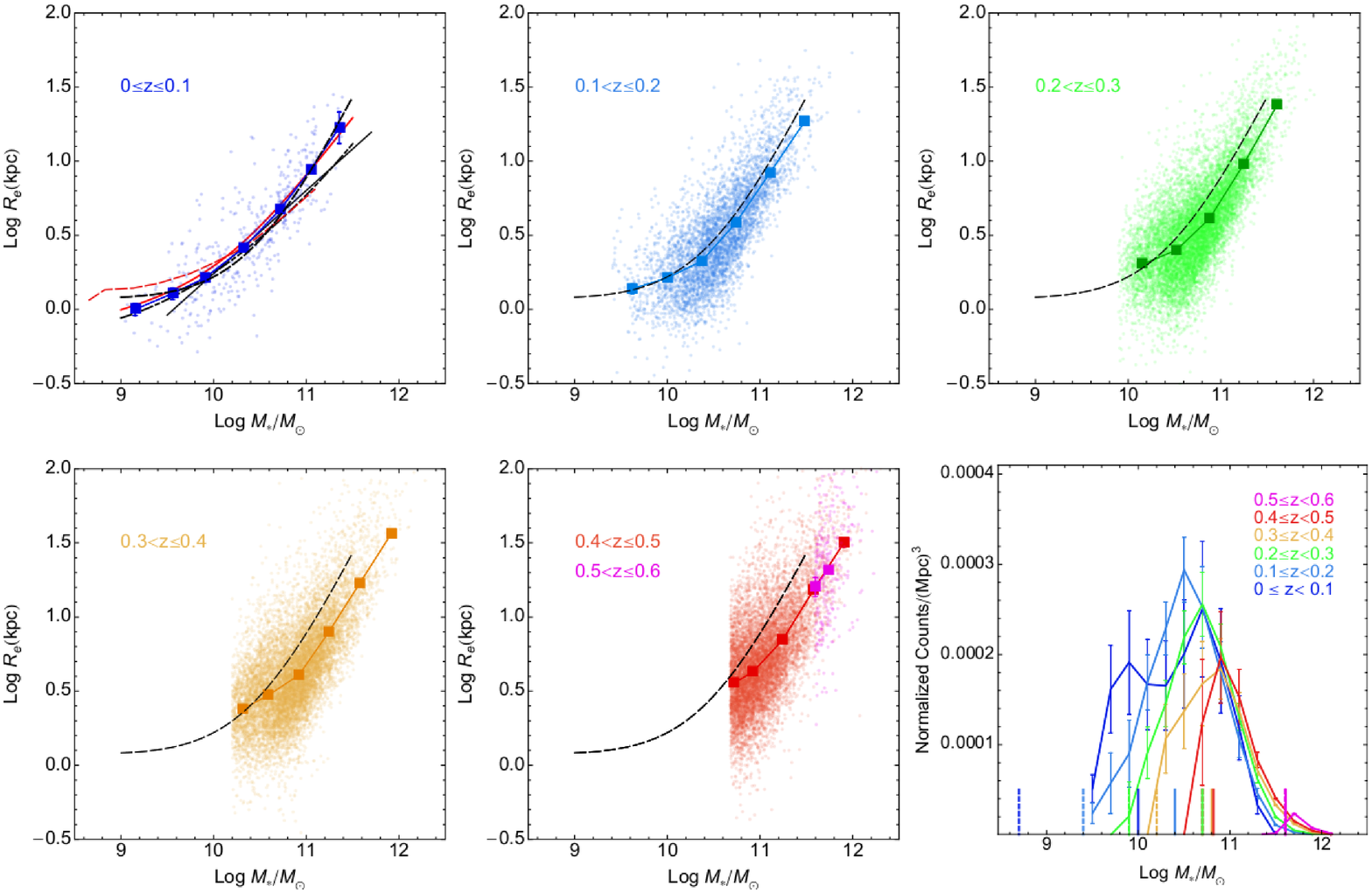}
\caption{Size--Mass relation for spheroids (top panels and left
and central bottom panels). Individual galaxy values are plotted
together with mean and standard deviation of the mean (boxes and
error bars). For the $0\leq z<0.1$ bin we overplot some local
relations from literature (solid line: \citealt{Shen+03};
dot-dashed line: \citealt{HB09_curv}; dashed black line:
\citealt{Mosleh+13_size_evol}; dashed red line:
\citealt{Baldry+12_gama_mass}; solid red line:
\citealt{Lange+015}). For all other $z$ bins we show the $z=0$
relation form \citet{Mosleh+13_size_evol} to visually appreciate
the deviation of the average relation from the local one. Bottom
right panel: the stellar mass distributions in different $z$ bins
normalized to the total covolume. The vertical coloured line at
the bottom of the bottom-right panel are the rough mass
completeness derived by the histogram shown in the same panel.
Here we took as fiducial completeness mass the mass roughly
corresponding to the peak of the distribution, except for the
lowest z bin where we also keep the second peak of the mass
distribution as a significant feature.} \label{fig:size_mass1}
\end{figure*}

\begin{figure}
\includegraphics[scale=0.6]{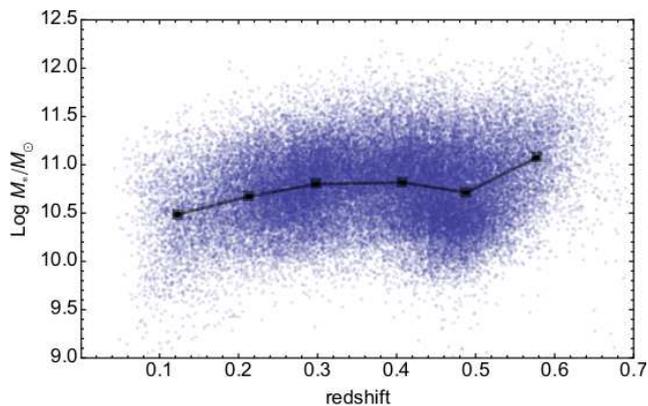}
\caption{Mass vs. redshift plot for the spheroids sample.
Overplotted to the individual galaxy values, we show the mean and
standard deviation of the mean of the sample (error bars are
comparable to the size of the boxes). Note that the steepening of
the $z\sim0.6$ bin is due to the mass incompleteness of this bin.
In the lowest $z$ bin ($z\sim0.1$), the sample suffers some volume
incompleteness (see discussion in \Sec\ref{sec:sample}), which
produces the mean mass in the bin to be biased toward less massive
systems.} \label{fig:mass_z}
\end{figure}

\begin{figure*}
\centering
\includegraphics[scale=0.54]{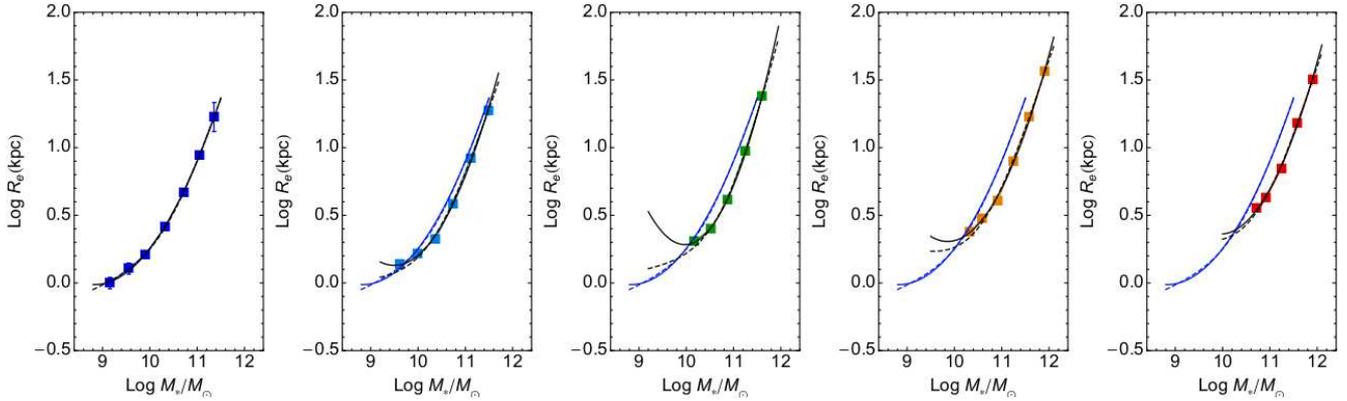}
\caption{Parametric fit to the Size--Mass relation for spheroids.
The average size mass in different bins (colour coded as in
\Fig\ref{fig:size_mass1}) is fitted with the parametric formulae
as in Eqs. \ref{size_M_eq1} (dotted lines) and \ref{size_M_eq2}
(solid lines). The $z=0$ fit has been reported in the subsequent
$z$ bins in blue, to visually check the difference of the $z>0.1$
relations. These curves are used to define the \Re\ corresponding
to different mass intercepts as shown in
\Fig\ref{fig:size_mass_interc}.} \label{fig:size_mass_fit}
\end{figure*}

\begin{figure}
\centering
\includegraphics[scale=0.69]{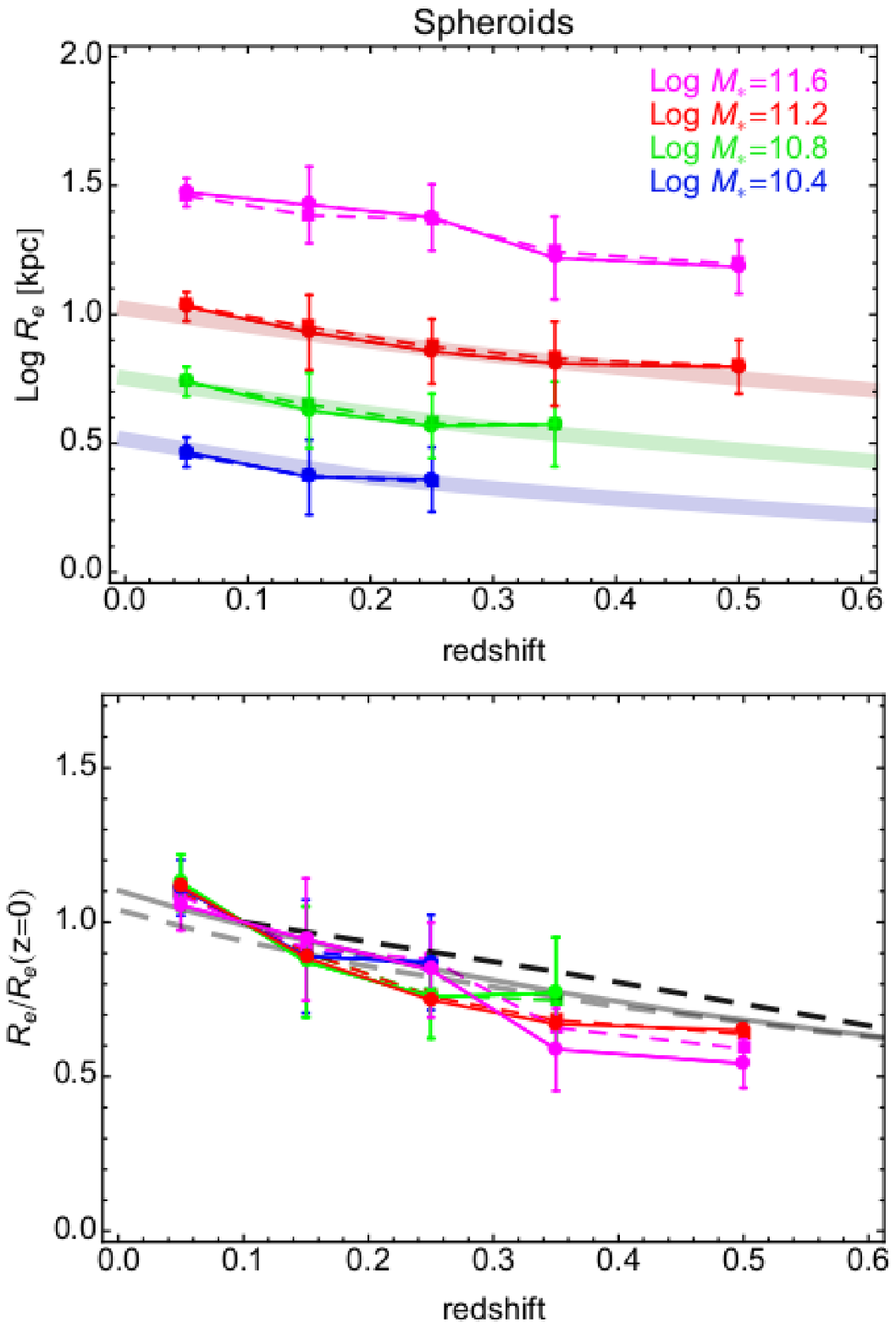}
\caption{Size vs. $z$ plots from the average size-mass parametric
fit of spheroids. {\it Top}. We plot the derived absolute
intercept of the best fit relations as in
\Fig\ref{fig:size_mass_fit} at mass values in the legends. Error
bars account for the $1\sigma$ errors in the best fit. Dotted and
solid lines show the results of fit from the M+13 relations as in
Eq. \ref{size_M_eq1} and HB+09 relation as in Eq. \ref{size_M_eq2}
respectively. We also overplot results from HST data in vdW+14,
corresponding to $\Log\ M_*/M_\odot=10.25,~10.75,~11.25$, from
bottom to the top, which well compare to our measurements in
similar mass bins. {\it Bottom}. We plot the size evolution with
respect to the local size at different mass intercepts. The
evolution of the size with redshift becomes increasingly
significant at larger masses. The black dashed line is the
relation as found by \citet{Trujillo+07} for $\Log\
M_*/M_\odot>11.0$. The grey dashed line is for ETGs with $10.5<
\Log\ M_*/M_\odot<11.0$ and the grey solid line for those with
$11.2< \Log\ M_*/M_\odot<12$ from \citet{Huertas+013_SM}.}
\label{fig:size_mass_interc}
\end{figure}

\section{Results}\label{sec:Resu}
In this section we present results about the evolution across
cosmic time of galaxy sizes and size--mass relations. The
evolution of the size--mass correlation is strictly related to the
way the galaxies have been assembled. It is known that the two
main classes of galaxies, spheroids and disc-dominated, show a
different dependency between size and stellar mass with
disc-dominated galaxies having a weak, if any, dependence on the
redshift, and spheroids showing a clear variation with the
redshift (see e.g. \citealt{Shen+03}, \citealt{vanderwel+14_SM}),
which suggest a different evolution pattern for the two
populations. In the following, we will refer to effective radii
derived in $r$-band if not otherwise specified.

\subsection{Spheroids and disc-dominated galaxy
classification}\label{sec:etg_ltg} We start by separating
spheroids and disc-dominated galaxies using two independent
criteria, based on: a) the S\'ersic index values
(\Sec\ref{sec:surf_phot}) and b) the SED fitting classification
using the spectrophotometric classes discussed in
\Sec\ref{sec:SED}. We define "spheroids" those systems with steep
light profiles, i.e. with $r$-band $n
> 2.5$ (\citealt{Trujillo+07}, \citealt{vanderwel+14_SM}), and with
photometry best-fitted by one of the 22 elliptical galaxy model
templates (see \Sec\ref{sec:SED};
\citealt{Tortora+16_compact_KiDS}). Instead, "disc-dominated"
galaxies are defined as systems with more extended and shallower
light profiles, i.e. with $r$-band $n < 2.5$, and with photometry
which is best-fitted by model templates of late-type galaxies
(i.e., Sbc and Scd types).

The final sample consists of $49\,972$ spheroids and $144\,859$
disc-dominated galaxies in $r$-band. We just remark that there are
a number of galaxies ($13\,403$) which turned out to be neither
spheroids nor disc-dominated (classified as star burst or
irregular systems), which we have excluded from our analysis.
Furthermore, in order to use with caution the warning of the
offset found with the GAMA estimate in \S\ref{sec:sdss_comp}, we
show that our results are insensitive to a more conservative
choice of the S\'ersic-index (e.g. adopting $n>3.5$) in Appendix
\ref{app:sersic}.

\subsection{Size--Mass as a function of redshift}
Once we have defined the two main galaxy classes interested by
this analysis, we can proceed to investigate the size--mass
relation as a function of the redshift and compare this with
previous literature data and simulations.

\subsubsection{Spheroids}\label{sec:size_mass_etg}
In \Fig\ref{fig:size_mass1} we show the size--mass relation of
spheroids in different redshift bins with overplotted the mean as
boxes and the standard deviation of the mean as errorbars. In
\Fig\ref{fig:size_mass1} only the 90\% complete sample is shown,
and this becomes clear in particular at $z>0.3$ where the sample
starts to be severely incomplete at $\Log\ M_*/M_\odot<10.2$. The
two bins at $z>0.4$ are shown together as the contribution of
galaxies in the bin $0.5<z\leq0.6$ is minimal and limited to the
very high mass end. The mean contour of the latter redshift bins
are fully consistent with the ones derived for the lower z bin,
$0.4<z\leq0.5$, hence we decided to cumulate the two samples.

In the figure we have also plotted some relevant literature trends
obtained at $z=0$ (i.e. \citealt[S+03 hereafter]{Shen+03};
\citealt[HB+09 hereafter]{HB09_curv}; \citealt[M+13
hereafter]{Mosleh+13_size_evol}; \citealt[B+12
hereafter]{Baldry+12_gama_mass}; \citealt[K+12
hereafter]{Kelvin+012}; \citealt[L+15 hereafter]{Lange+015}),
after having scaled all masses to the Chabrier IMF, which is our
reference choice. All the literature results used for comparison
had been obtained with a single S\'{e}rsic model (as for our
results) except for HB+09 which used a simple de Vaucoleurs
profile. Also, we had to take into account the different size
definitions as circularized radii (i.e. the ones adopted by us)
were used by Shen+03, HB+09, and M+13, while B+12, K+12 and L+15
adopted major axis effective radii and needed to be corrected by
the galaxy axis ratio (see \S\ref{sec:stru_par}). Since we did not
have information on the axis ratio of all literature samples, we
have adopted an average correction between the major axis and the
circularized radii as a function of the mass for the low-$z$ bin
obtained from our galaxy sample as discussed in Appendix
\ref{app:maj_circ} (and shown in Fig. \ref{fig:maj_circ}), which
we have applied to the datasets adopting major axis effective
radii (i.e. B+12 and K+12). This corresponds to have compared our
major axis estimates with the equivalent ones in B+12 and K+12,
and then re-arranged all back to some circularized radii
consistent with the same average ellipticity of the KiDS galaxies.

We first remark a very good agreement of our mean values (data
points with error bars) with the non-parametric estimates from
M+13 shown as dashed line in
\Fig\ref{fig:size_mass1}\footnote{Note that, the M+13 effective
radii are obtained from a non-parametric procedure, based on the
growth curve.}. In particular, we clearly see in our data a
flattening of the relation at masses below $\Log\
M_*/M_\odot\sim10.0$ in the lowest $z$ bin. The $z=0$ relation
from M+13 nicely matches also the average trend in our next $z$
bin ($0.1 < z \leq 0.2$), where the flattening of the relation is
even more evident.

Differently from M+13, S+03 use a single power-law to best fit
their data, i.e. $\Re \propto M^\alpha$, while HB+09 have
performed a parabolic fit in the Log-Log plane to reproduce the
curvature they have observed in their data too and that is also
seen in our \Fig\ref{fig:size_mass1}. Both S+03 and HB+09 show a
good agreement with our data at the intermediate mass scales,
while they diverge at the lower masses. In particular, S+03 does
not seem to catch the flattening of the average size-mass
relation, while HB+09 seems to over-predict the flattening we also
observe. We expect to better quantify this tension at lower mass
scales by using the larger dataset to be gathered with the third
data release. We note, though, that the sample is complete at this
mass bin according to \Tab\ref{tab:mass_compl}. At higher masses
($\Log\ M_*/M_\odot>10.8 $), the main issue of the S+03 relation,
is that they tend to underestimate \Re\ because of sky subtraction
in the SDSS Photo pipeline. To conclude our comparison with
previous literature, we also show the average relation obtained by
\cite{Baldry+12_gama_mass} with GAMA galaxies, where we see also a
flattening of the relation at $\Log\ M_*/M_\odot\sim10.0$, but the
overall relation seems tilted with respect to our average
relation.

\begin{figure*}
\centering
\includegraphics[scale=0.57]{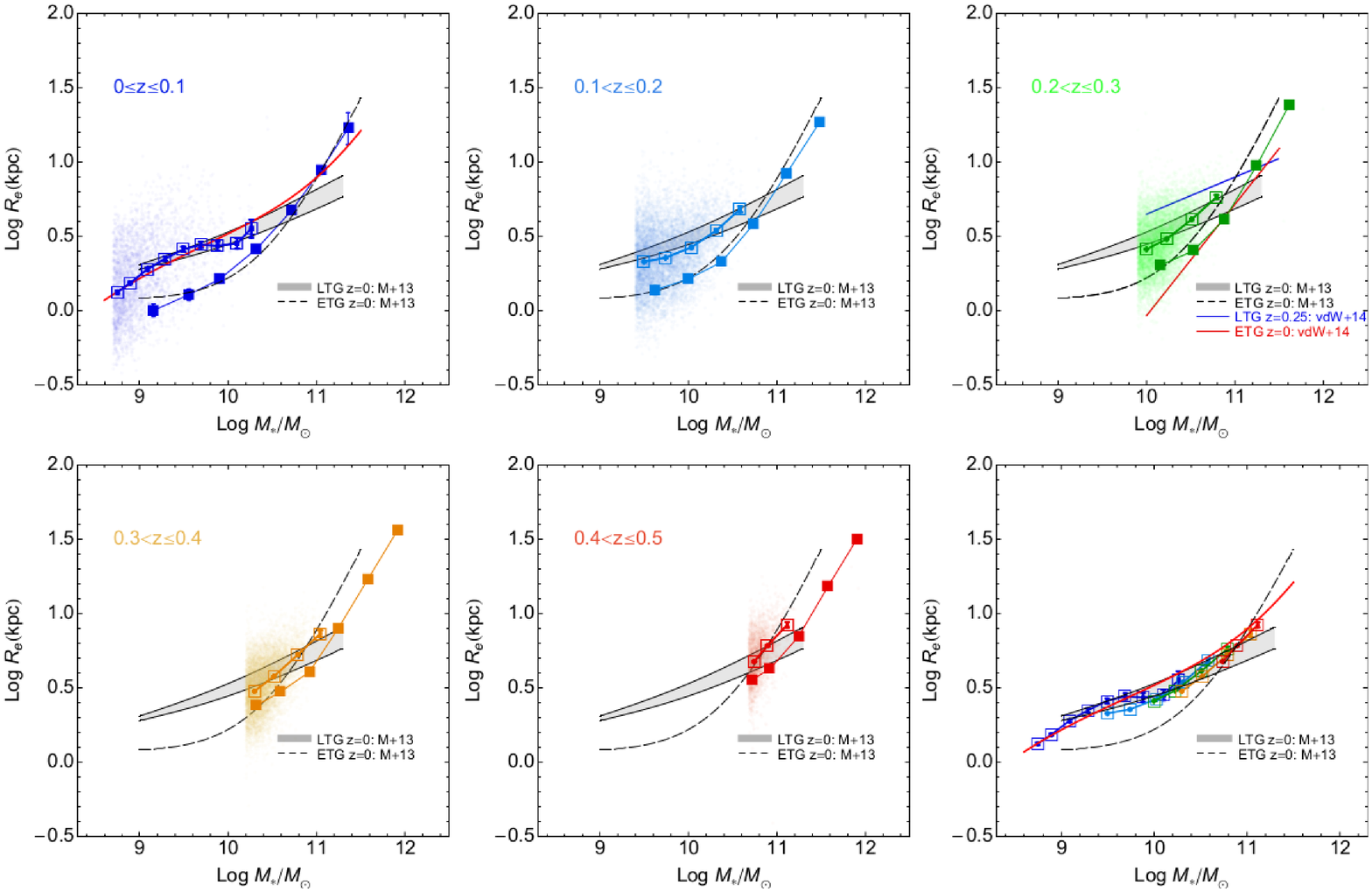}
\caption{Size--Mass relation for disc-dominated galaxies. Symbols
have the same meaning of the spheroids sample in
\Fig\ref{fig:size_mass1}, but now data are shown with open symbols
in contrast to the spheroids average relation also shown as full
symbols. The  local relation is given by a shaded area which show
the range spanned by the average relation from
\citet{Mosleh+13_size_evol} (i.e. their LTG, $n<2.5$, blue
samples) and solid red line from \citep{Lange+015}. In the bottom
right panel we summarize all results: the disc-dominated galaxies
have generally larger sizes at masses, especially for $\lsim
10^{11.0}M_\odot$ and show a trend with redshift (see
\Sec\ref{sec:size_mass_etg}) which seems weak or absent.}
\label{fig:size_mass_ltg}
\end{figure*}

We use the M+13 results as a $z=0$ reference to compare the
size-mass relations in the other redshift bins and visually
evaluate the evolution of the size--mass relation with lookback
time. Going toward higher $z$, in \Fig\ref{fig:size_mass1} we show
that the mean correlation (boxes connected by the solid lines)
starts to deviate from the $z=0$ relation after $z=0.2$ as
galaxies become more and more compact with respect to their low-z
counterparts. The difference is significant within the errors at
stellar masses $\Log\ M_*/M_\odot\gsim10.5$, while at lower masses
there is little evolution in size, or even, an opposite trend with
respect to that seen in the high mass regime, i.e. galaxy sizes
becoming larger. However, this might be due to the fact that we
are in a mass regime ($M \sim 10^{10}M_\odot$) close to the
completeness limit of the sample.

On the higher mass side, the sample does not suffer any particular
incompleteness, as shown by the mass distribution in the $z$-bins
in the bottom-right panel of the same \Fig\ref{fig:size_mass1}
(except possibly for the low-$z$ bin, see also below). Here, the
counts have been normalized to comoving volume and corrected by
the completeness function (i.e. the fraction of galaxy lost per
mass unit in different $z$-bins). The error bars mainly reflects
the propagation of \photoz\ errors on the determination of the
comoving volume in the different $z$-bins. The drop of the counts
after the first peak at $\Log\ M_*/M_\odot=10.5-11.0$ going
towards lower masses, is typical of the spheroids mass function
measured at all redshifts (see e.g. \citealt{Kelvin+014}) and does
not reflect an intrinsic incompleteness of the sample. We conclude
that the observed trend with $z$ relation, which moves the
spheroids sample progressively away from the $z=0$, is genuine and
has to be related to an evolution pattern in the galaxy structural
parameters. All these effects go in the sense of favoring more and
more massive (hence larger) galaxies at higher $z$, which goes in
the opposite direction as the trend of galaxy sizes decreasing
with z.

This should not be due to an evolution of the stellar mass, as the
average stellar mass of our sample does not show any significant
trend with the redshift. This is demonstrated in
\Fig\ref{fig:mass_z}, where the average masses stay almost
constant in the range $\Log\ M_*/M_\odot\sim10.7-10.8$ as a
function of $z$ although a steepening is observed only at the
$z\sim0.6$ bin, which is due to the mass incompleteness of this
bin (below $\Log\ M_*/M_\odot=11.5$). A possible selection effect
is also present in the lowest $z$ bin ($z\sim0.1$), due to the
volume incompleteness discussed in \Sec\ref{sec:sample}), which
causes the average mass in the bin to be biased toward the less
massive systems.

We conclude that the the driver of the evolution of the mass-size
relation is the  change of the galaxy size with $z$.
Visually, this means that galaxies more massive than $\Log\
M_*/M_\odot\sim10.5$ have sizes (i.e. \Re) that decrease with
increasing redshift at any given mass. To better quantify this
effect and to estimate also the amount of the size variation in
the different mass intervals, we have performed a fit to the
average size--mass at different redshifts and then evaluated the
\Re\ corresponding to different mass intercepts (see also
\citealt{vanderwel+14_SM}, hereafter vdW+14).

To fit the size--mass we have used the two fitting formula used in
M+13 and HB+09 (as showed in \Fig\ref{fig:size_mass1}), which we
report here below for clarity:

\begin{equation}\label{size_M_eq1}
\Re= \gamma (M_*)^\alpha(1 +M_*/M_0)^{(\beta-\alpha)}~ {\rm [from~
M+13]},
\end{equation}
where \Re\ is in kpc, $M_*$ in solar units, and $\alpha$, $\beta$,
$\gamma$, $M_0$ are free parameters, and

\begin{equation}\label{size_M_eq2}
Y=p_0+p_1 X +p_2 X^2~ {\rm [from ~HB+09]}
\end{equation}
where $Y=\Log\ \Re{\rm /kpc}$, $X=\Log\ M_*/M_\odot$ and
$p_0,~p_1,~p_3$ are free parameters to be adjusted to best fit the
data points. The best fit relations for both cases are shown in
\Fig\ref{fig:size_mass_fit}. The fit is generally very good for
both fitting function across the data points, however
\Eq\ref{size_M_eq2} seems to predict a very strong up-turn of the
trend at low masses, right outside the first datapoint, which we
cannot confirm with our current dataset.

In \Fig\ref{fig:size_mass_interc} we show the trend of the \Re,
obtained from \Eq\ref{size_M_eq1} and \ref{size_M_eq2}, for
different mass values, as a function of $z$, while errorbars show
errors from the  best fit at every mass bin for
\Eq\ref{size_M_eq1} only, for clarity (being the ones of
\Eq\ref{size_M_eq2} very similar).
The errors on the individual estimate take into account the
$1\sigma$ errors in the best fit. There is an evident trend of the
sizes to decrease with redshift in all mass values except $\Log\
M_*/M_\odot = 10.4$. This trend is nicely consistent with a
similar analysis performed by vdW+14 on HST data for CANDELS
(\citealt{CANDELS+11}) and shown in the same figure, where we show
their results for $\Log\ M_*/M_\odot= 10.25,~10.75,~11.25$, from
bottom to the top (see also the colour code, as in the legenda).
Our results are consistent CANDELS at higher-$z$ (>0.3) for the
lowest mass value for which our sample is complete out to
$z\sim0.5$ ($\Log\ M_*/M_\odot = 10.4$).

If we use the standard parametrization for the size evolution vs. redshift
of the form
\begin{equation}
\Re=B_z (1+z)^{\beta_z}
\label{eq:re-z}
\end{equation}
we note that the steepest variation of the sizes is found
in our highest mass intercept ($\Log\ M_*/M_\odot = 11.6$), for
which we measure a slope of $\beta_z=-2.0 \pm0.3$ as in
\Tab\ref{tab:Re-z_etg_ltg}, where we report the best fit to the
data point obtained from \Eq\ref{size_M_eq1} (but the use of
\Eq\ref{size_M_eq2} would not have changed the final results).
This is different from the ones of the lower mass bins which have
an average slope of -1.5, which is consistent with the one
reported by vdW+14 (i.e. -1.48). This corresponds to a reduction
of the size with respect to the value at $z = 0.1$ of galaxies
with mass $\Log\ M_*/M_\odot=11.6$ that reaches about 50\% at z
> 0.5 and that is larger than the 40\% of the galaxies of the
close mass bin ($\Log\ M_*/M_\odot=11.2$), as shown in the bottom
panel of \Fig\ref{fig:size_mass_interc}. We used the $z=0.1$ value
as normalization value, consistently with previous literature
(\citealt{Trujillo+07}, \citealt{Huertas+013_SM}) also shown in
the figure as comparison.

The evolution of the galaxy size over cosmic time becomes
increasingly significant at larger masses. We could not track back
these discrepancies in the $\Re(z)/\Re(z=0)$ in the original
samples from the two analyses mentioned above as the galaxy
selection are somehow different from ours (e.g.
\citealt{Trujillo+07} use systems with $n>2.5$,
\citealt{Huertas+013_SM} distinguish group and field galaxies) and
also the local values adopted by them are different.

\begin{figure}
\hspace{-0.3cm}
\includegraphics[width=\columnwidth]{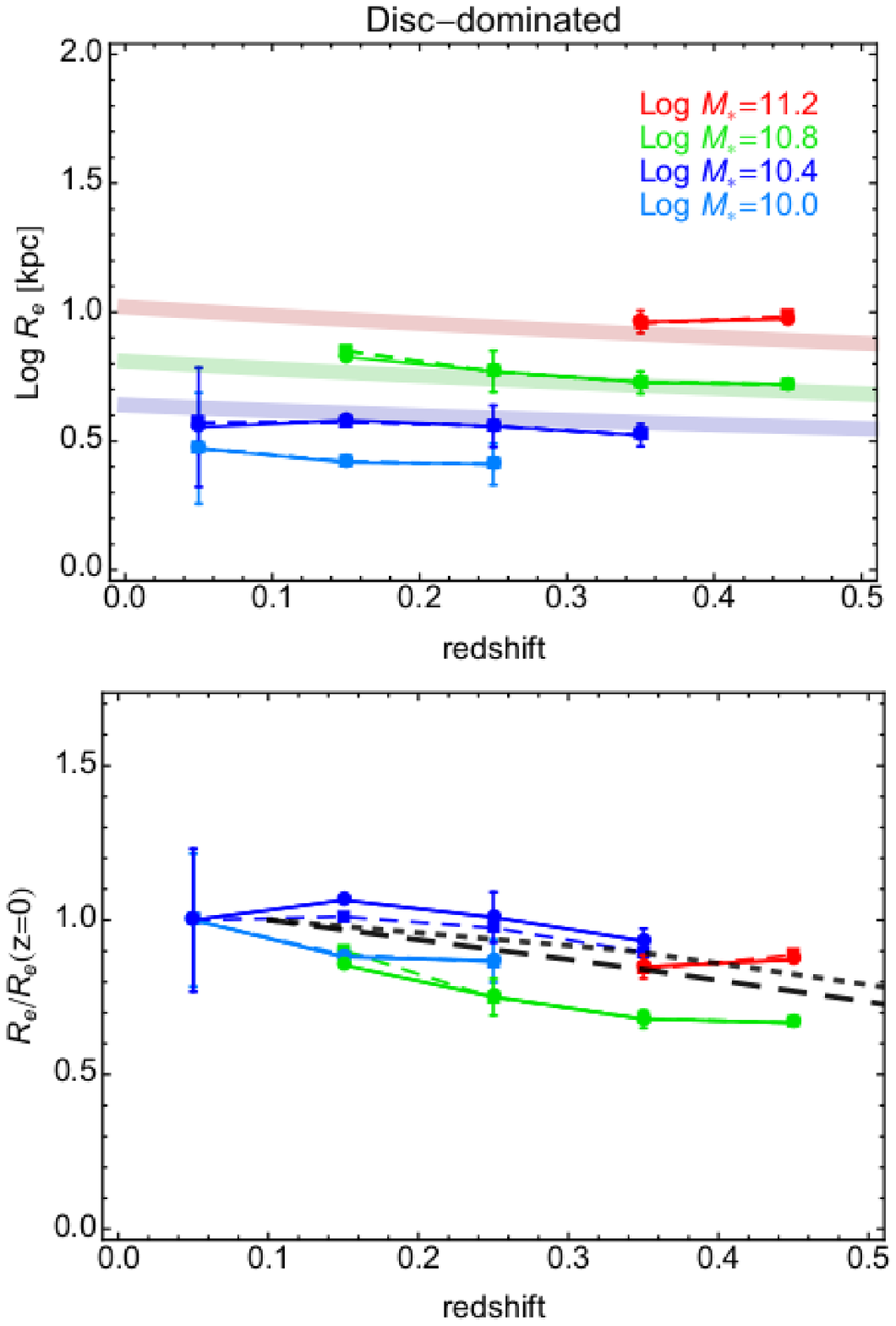}
\caption{Size vs. $z$ plots from the average size-mass parametric
fit of disc-dominated galaxies, as done for spheroids in
\Fig\ref{fig:size_mass_interc}. {\it Top}. We plot the derived
absolute intercept of the best fit relations at mass values as in
the legend. Dotted and solid lines show the results of fit from
the M+13 relations as in Eq. \ref{size_M_eq1} and HB+09 relation
as in Eq. \ref{size_M_eq2} respectively. Error bars account for
the $1\sigma$ errors in the best fit. We also overplot three
relations from HST data in vdW+14 in transparent colours,
corresponding to $\Log\ M_*/M_\odot=10.25,~10.75,~11.25$, from
bottom to the top, which nicely overlap to our measurements in
similar mass bins. {\it Bottom}. We plot the size evolution with
respect to the \Re\ at $z=0$ at different mass intercepts. The
trend of the size with redshift of disc-dominated seems constant
within the errors at all mass bins. The black dashed and dotted
lines represent the same relations for spheroids and
disc-dominated, respectively (from \citealt{Trujillo+07}).}
\label{fig:size_mass_interc_ltg}
\end{figure}

\begin{table*}
\hspace{-1cm}
\caption{$\Re=B_z (1+z)^{\beta_z}$ fit to the size-redshift
relation as derived from the average size-mass fit at different
redshifts (\Figs\ref{fig:size_mass_interc} and
\ref{fig:size_mass_interc_ltg}) and from direct fit to the
size-redshift relation in different mass bins
(\Fig\ref{size_redshift}).}
\label{tab:Re-z_etg_ltg}       
\begin{tabular}{ccccccccc}
\hline\hline
\noalign{\smallskip}
 & \multicolumn{4}{c}{Indirect from size-mass } & \multicolumn{4}{c}{Direct Fit}  \\
\hline
 & \multicolumn{2}{c}{spheroids} & \multicolumn{2}{c}{disc-dominated galaxies} & \multicolumn{2}{c}{spheroids} & \multicolumn{2}{c}{disc-dominated galaxies} \\
\noalign{\smallskip}\hline\noalign{\smallskip}
$\Log\ M_*$ & $\Log\ B_z$ & $\beta_z$& $\Log\ B_z$ & $\beta_z$& $\Log\ B_z$ & $\beta_z$& $\Log\ B_z$ & $\beta_z$\\
\noalign{\smallskip}\hline\noalign{\smallskip}
 10.0 & -- & --                            & $ 0.48\pm0.02  $ & $ -0.8\pm0.4$      & -- & --                   & $0.44\pm0.04$ & $-0.3\pm0.5$\\
 10.4 & $0.48\pm0.04  $& $ -1.4\pm0.6$ &     $0.57\pm0.02$ & $-0.3\pm0.3$&     $ 0.35\pm0.04 $& $0.4\pm0.3$     & $0.60\pm0.01$ & $ -0.7\pm0.1$\\
 10.8 & $0.75\pm0.04 $& $-1.6\pm0.5 $&     $0.88\pm0.02$ & $-1.1\pm0.2$ &    $0.67\pm0.06$& $-0.6\pm0.4 $   & $0.92\pm0.11$ & $-1.4\pm0.9$\\
 11.2 & $1.03\pm0.03 $& $ -1.5\pm0.3$ &     $ 0.91\pm0.04$ & $ 0.4\pm0.2$    &$1.11\pm0.03$& $-2.0\pm0.2$    & $0.87\pm0.03$ & $ 0.4\pm0.2$\\
 11.6 & $1.53\pm0.02 $ & $-2.0\pm0.3$ &     -- & --                          & $1.41\pm0.02 $ &$-1.5\pm0.2$        & -- & -- \\
\noalign{\smallskip}\hline\hline\noalign{\smallskip}
\end{tabular}
\end{table*}

\begin{figure*}
\includegraphics[scale=0.63]{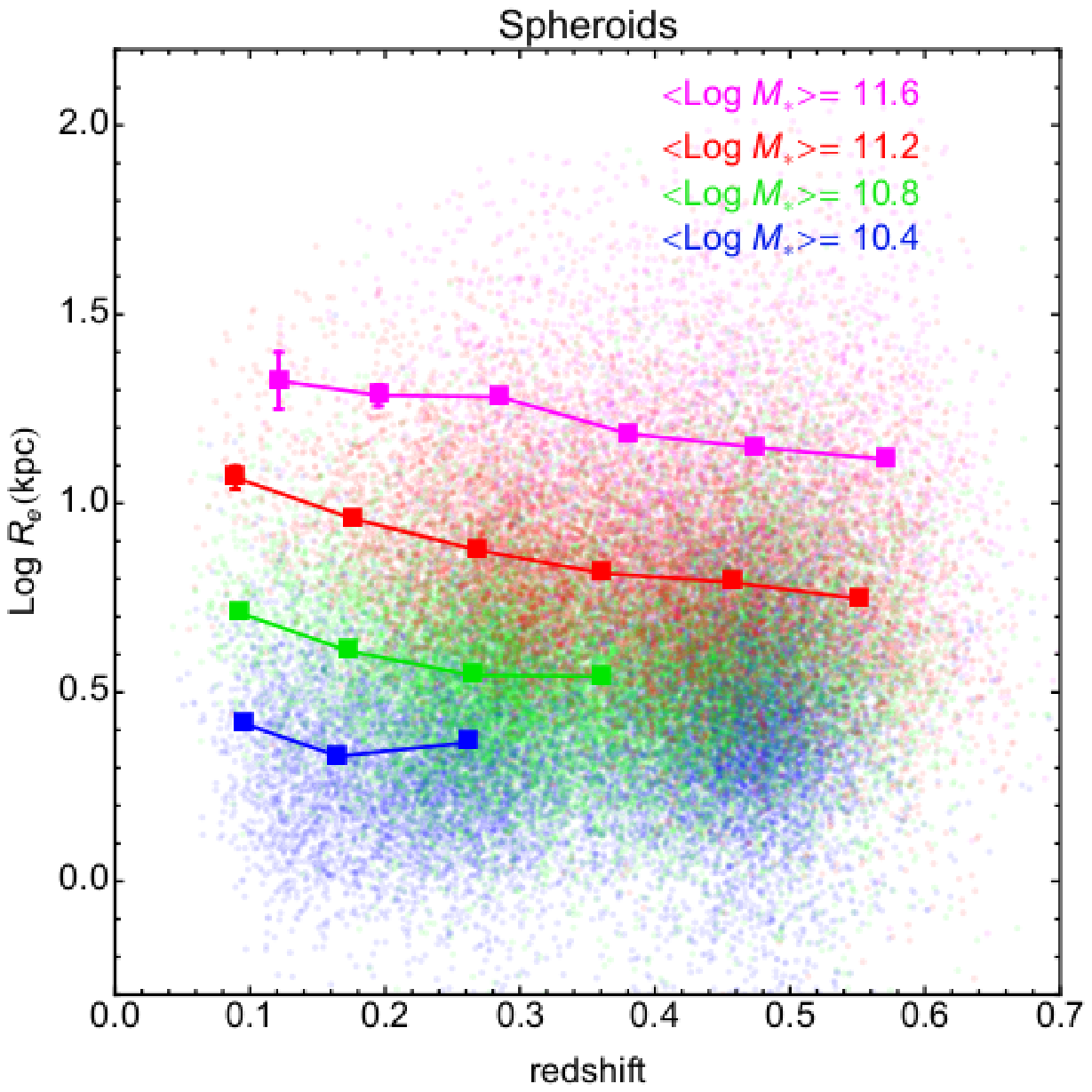}
\includegraphics[scale=0.63]{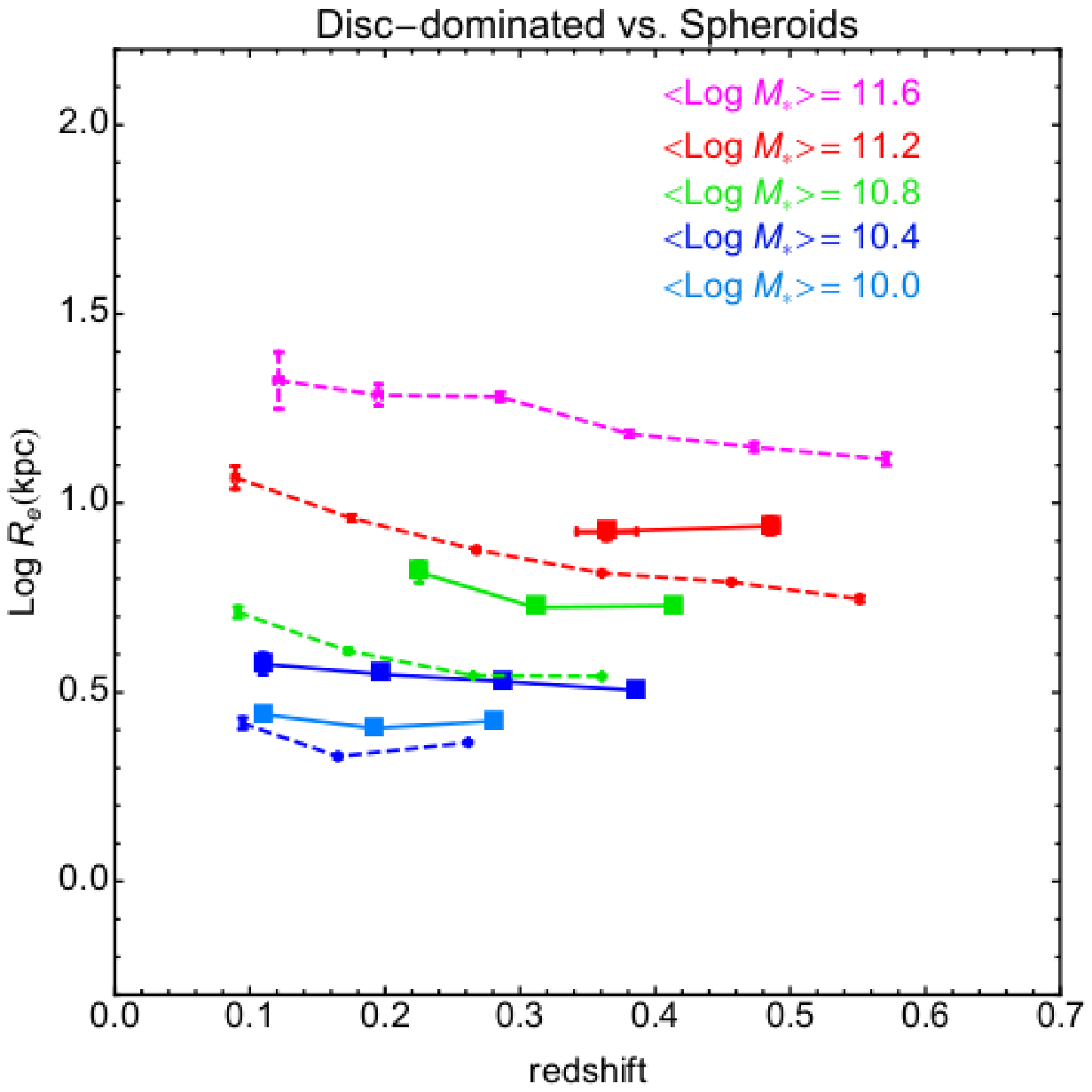}
\caption{Size evolution with respect to z for spheroids (left) and
disc-dominated galaxies (right) based on $r$ band data. In the
left panel we show the spheroids average relation as full squares
connected by solid lines against individual galaxy values colour
coded according to their average mass in the mass bins. In the
right panel disc-dominated values are given as squares connected
by solid lines and spheroids are also reported as dotted lines as
comparison. In each case mean for redshift are given and standard
deviation of mean in size is given as the error bars as filled
squares.} \label{size_redshift}
\end{figure*}

\subsubsection{disc-dominated}\label{sec:size_mass_ltg}
The mass size relation of disc-dominated systems is shown in
\Fig\ref{fig:size_mass_ltg} as open symbols and compared with the
ones of spheroids from \Fig\ref{fig:size_mass1}. In all panels we
show again the local relation, by M+13, but here represented as a
shaded area which reproduces the larger spanning of their
inferences, depending on the different selections made (LTG,
$n<2.5$, blue samples, etc.). Our $z\sim0$ results (top left) are
again very well consistent with literature, and we can see a
change in the overall slope at $\Log\ M_*/M_\odot<9$ which is not
reported in previous data.
%
%
In all other redshift bins, we see that the size--mass data tend to
tilt with respect to the local relation, around a fixed mass scale
($\Log\ M_*/M_\odot\sim10.5$).

In our sample, disc-dominated galaxies have always larger sizes
than spheroids at masses $\lsim 10^{11.0}M_\odot$, consistently
with what is found in previous literature (e.g. vdW+14), while for
higher masses we do not have a significant sample of
disc-dominated galaxies (see also the $B_z$ values in
\Tab\ref{tab:Re-z_etg_ltg} which are larger for the discs with
respect to the passive galaxies at $\Log\ M_*/M_\odot>10.8$) and
we cannot exclude that spheroids might have larger sizes at that
mass range, e.g., vdW+14. We expect to investigate more this issue
with the next KiDS data release.

We finally see that disc-dominated galaxies do not show a clear
trend with redshift as clearly seen for spheroids. In fact, in
\Fig\ref{fig:size_mass_ltg} the average relations at higher $z$ do
not deviate significantly from the one at $z=0$ (shaded region) as
observed in the case of spheroids.

As done for spheroids in \Sec\ref{sec:size_mass_etg}, we have
quantified the dependence on the redshift by fitting the $\Re-M_*$
relations at the different redshifts and determining the intercept
at different mass values. In this case we have used only the
double power law formula (\Eq\ref{size_M_eq1}), since the data do
not show any signature of the inversion of their trend at low
masses. The results are shown in
\Fig\ref{fig:size_mass_interc_ltg}, for the highest mass bins for
which the sample is complete at $z<0.5$. disc-dominated sizes show
a flat trend with redshits (see also the best fit slopes in
\Tab\ref{tab:Re-z_etg_ltg}), much flatter that the spheroids. This
is consistent with the results from vdW+14, also shown as thick
shaded lines, using the same intercept approach.

In the same \Fig\ref{fig:size_mass_interc_ltg} (bottom panel) we
have also estimated the trend with redshift of the size normalized
to the local value and our results seem to have a trend which is
spread in normalization but consistent with the ones obtained by,
e.g., \cite{Trujillo+07}. The new evidence from the KiDS sample is
that galaxies in the lower mass bins have a trend which is similar
to the one of the most massive bins (if we exclude the very
massive one, which is incomplete at lower redshift), as also
quantified in \Tab\ref{tab:Re-z_etg_ltg}. Overall the
disc-dominated systems show shallower trends than spheroids in all
mass bins.

\subsection{Spheroids and disc-dominated size evolution parametric fit}\label{sec:Re-z_fit1}
In this section we offer a complementary analysis of the size
evolution by directly deriving the $\Re-z$ relation in different
mass bins. Being this inference independent of any fitting
formula, it provides a more unbiased estimate of the actual
dependence of the size from the redshift, once the the mass
incompleteness in each redshift bin has been taken into account.

The results for the spheroids and disc-dominated galaxies are
compared in \Fig\ref{size_redshift}, where we show the average
$\Re-z$ dependence in different mass bins, following the mass
grouping and colour code adopted in the previous section. For the
spheroids, we also show the individual values with the same colour
code to better evaluate the spread of the relation. For
disc-dominated galaxies we have omitted individual values because,
being their relative normalization in the different mass bins
smaller than the spheroids case (see average values, in the right
panel, closer to each other wrt spheroids) and the scatter almost
the same of the one of spheroids, it was too crowded to appreciate
any difference among the different mass bins. We have performed
also for these average estimates the $\Re=B_z (1+z)^{\beta_z}$
fit, with best-fitting parameters being reported in
\Tab\ref{tab:Re-z_etg_ltg}.

Both the average values and the parametric fit show the same
features discussed for the size-redshift obtained for the
``indirect'' relations in the previous Section. Namely, the
spheroids show steeper decreasing trends with $z$ for mass bins
$\Log\ M_*\gsim 10.5$ while they almost flatten out at lower
masses. disc-dominated galaxies show shallower slopes (see
\Tab\ref{tab:Re-z_etg_ltg}) than spheroids and, at masses $\Log\
M_*\lsim 11.0$, they show larger sizes than the spheroids (see the
comparison between spheroids and disc-dominated galaxies in
\Fig\ref{size_redshift}, right panel). We will interpret these
different variations of the size with $z$ in the next paragraph.

Looking at the average slopes in the \Tab\ref{tab:Re-z_etg_ltg},
for the indirect fit we have a good agreement with vdW+14 (they
have found a slope of -1.48 for their ETGs, we have an average of
$-1.6\pm0.3$), while our disc-dominated systems show possibly a
shallower evolution as they find -0.75, while we have an average
slope of $-0.5\pm0.6$, but we are dominated here by the value of
the high mass which is quite uncertain being based on two points.
If we exclude that value, we obtain an average slope of
$-0.7\pm0.4$, hence consistent with the results from vdW+14.
Overall these average quantities have a large scatter due
primarily to the wide range of stellar masses covered. However, as
shown in \S\ref{sec:size_mass_etg} and \S\ref{sec:size_mass_ltg},
the consistency with vdW+14 is generally very good in the mass
bins. Similar average slopes are found for the direct fit in the
same Table.

\section{Discussion and Conclusions}\label{sec:disk_concl}
The main result of this paper is that the two main classes of
galaxies, spheroids and discs, show different relations between
size and stellar mass and size and redshift, which are well
consistent with previous literature (\citealt{Shen+03},
\citealt{Baldry+12_gama_mass}, \citealt{vanderwel+14_SM}) but
based on a sample which is much larger in the higher redshift
bins. Our sample, complete in mass down to $\Log\ M_*\lsim 9.0$ at
$z<0.2$ and down to $\Log\ M_*\gsim 10.0$ at higher$-z$, has
allowed us to highlight some features that were not clearly
assessed in previous datasets (at lower $z$, e.g. HB+09, S+03,
M+13). First, a curvature in the $\Re-M_*$, seems present at
almost all $z$-bins for both spheroids and disc-dominated
galaxies, but becomes less clear at $z>0.4$, mainly because of the
mass incompleteness. The size--mass relation of disc-dominated
galaxies also presents a knee in the relation at the very low
masses ($\Log\ M_*\lsim 9.5$) at $z<0.1$, which was not reported
in previous studies.

\begin{figure*}
\includegraphics[scale=0.66]{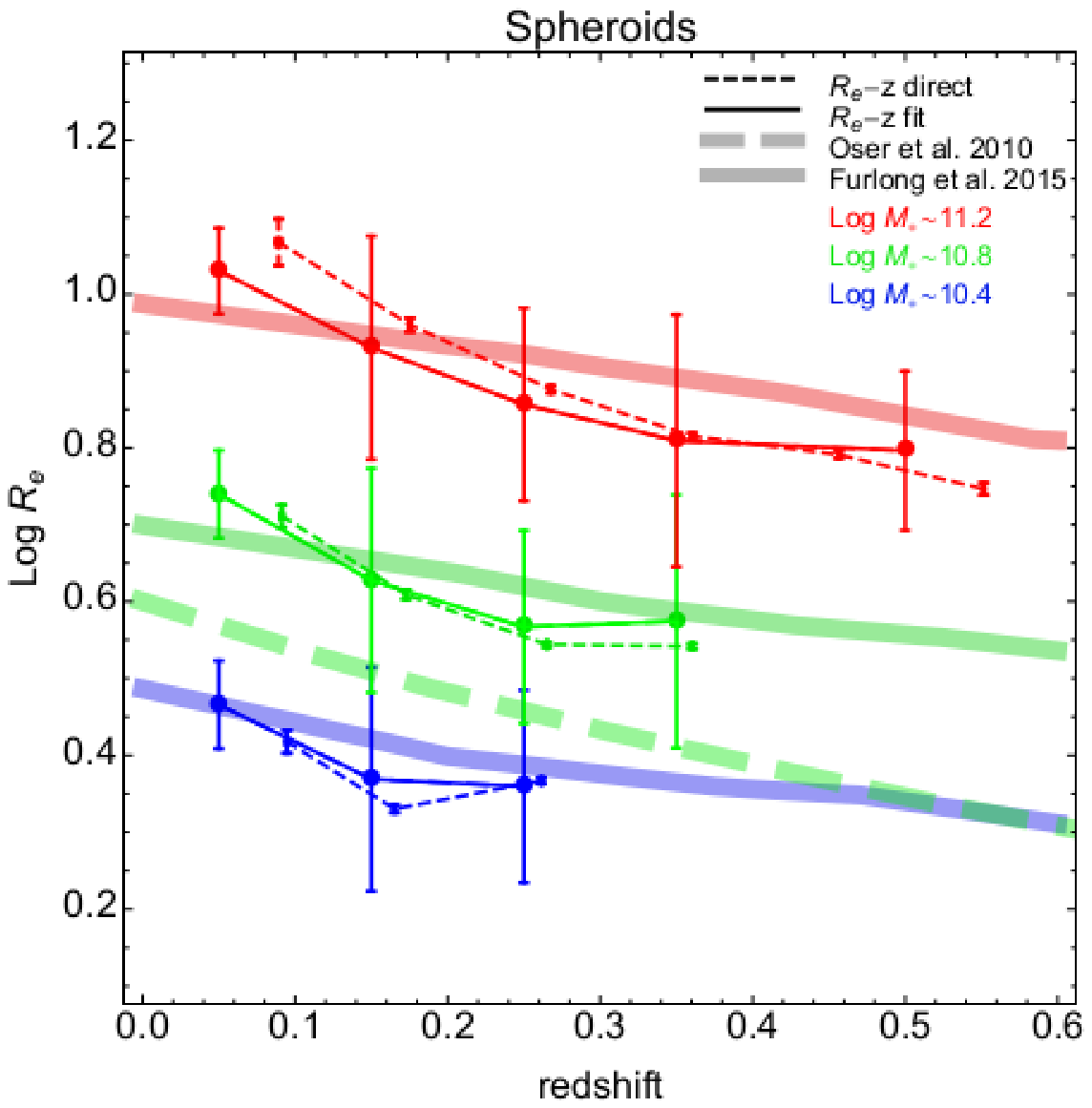}
\includegraphics[scale=0.66]{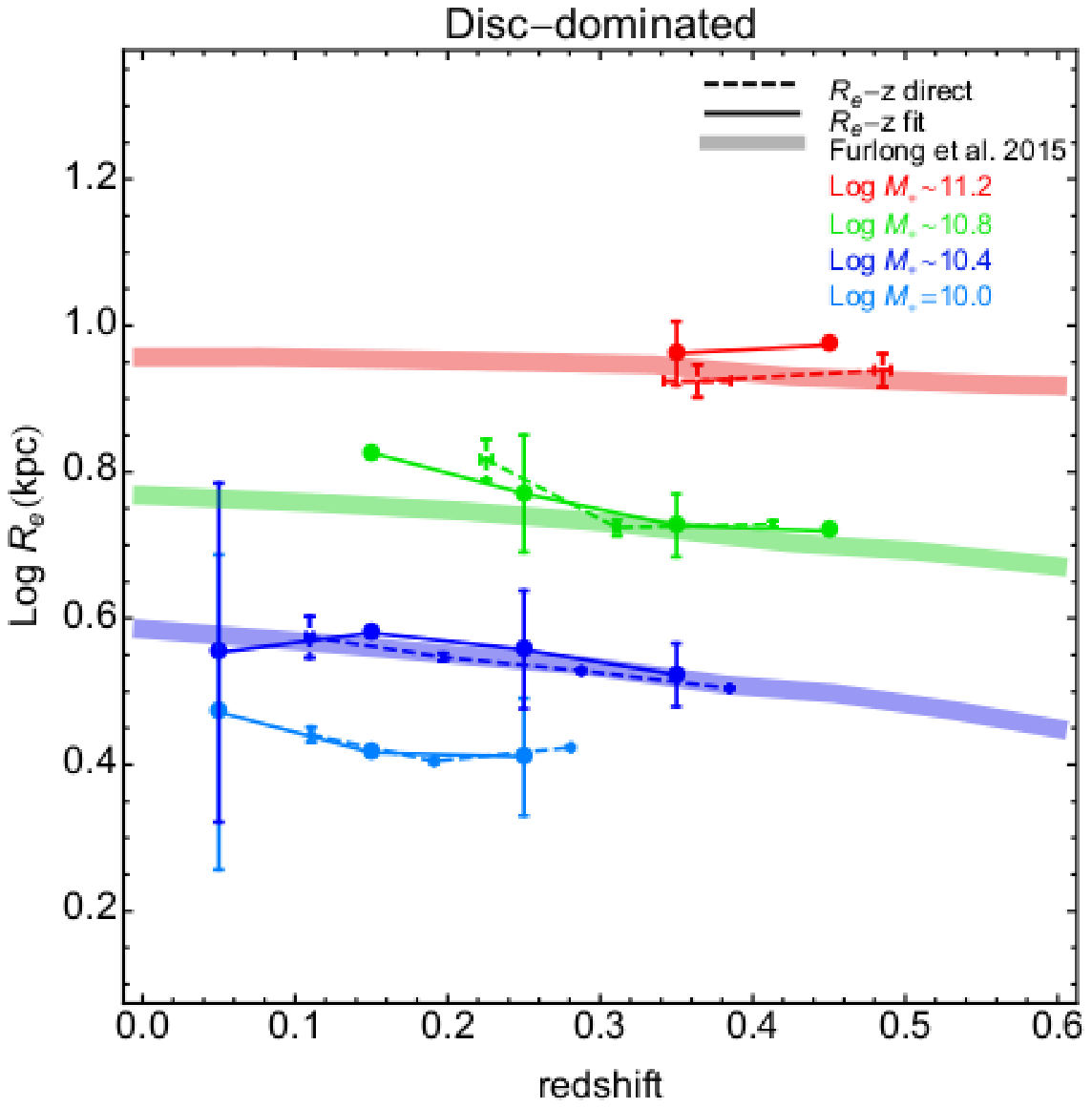}
\caption{Comparison with hydrodynamical simulations. Left panel:
spheroids size--redshift relation is compared with the EAGLE
simulations from \citet{furlong+015_SM} and with a modified
version of GADGET-2 from \citet{Oser+012_evol_massive_etg}. The
data points show the results obtained in
\Secs\ref{sec:size_mass_etg} (solid lines) and
\ref{sec:size_mass_ltg} (dashed lines) in different mass bins as
in the legends. Right panel: disc-dominated size--redshift
relation with symbols as in the left panel but with the results
from the average-size mass from \Sec\ref{sec:size_mass_ltg} (solid
lines).} \label{fig:sim_comp}
\end{figure*}

The results found for our spheroids and disc-dominated samples are
consistent with the expectation of the galaxy growth from recent
hydrodynamical simulations (\citealt{furlong+015_SM}) from the
EAGLE set-up (\citealt{Schaye+015_EAGLE}), as demonstrated in
\Fig\ref{fig:sim_comp}\footnote{We have corrected the simulation
results both a) rescaling their major axis radii as done for the
other literature, and b) linear interpolating the normalization of
their curve to the $\log M_*$ of our mass bins, to have the best
match between the data and predictions.}. Overall, the predictions
from simulations match our trends at all mass scales within
$1\sigma$, although the match of the spheroids is slightly more
discrepant with respect to the excellent agreement found for
disc-dominated galaxies, especially for the higher mass values.
However, the consistency of sizes predicted for spheroids in the
EAGLE simulations and our estimates indirectly demonstrates the
importance of feedback mechanisms to prevent the simulated systems
to collapse too much. This is a well known effect of
hydrodynamical simulations (e.g.
\citealt{Scannapieco+012_hydro_simu}) as a consequence of the
so-called angular momentum catastrophe
(\citealt{Katz_Gunn+91_GF_dynam};
\citealt{Navarro_white+94_dynamics}) consisting in a too large
angular momentum transfer into the galaxy haloes which cannot
retain the collapse of the cold gas into stars toward the galaxy
center. The effect is todays balanced by the inclusion of feedback
mechanisms in the centers, which balance the gas collapse (e.g.
\citealt{Governato+004_disk_gal};
\citealt{Sales+010_feed_simu_gal};
\citealt{Hopkins+014_feed_stellar}; etc.), but whose recipes are
still under refinement. In case an insufficient energy injection
is accounted for in simulations, the predicted sizes result to be
more compact for a given mass bin, as shown in the same Figure by
the the predictions of the $\Re-z$ for $\sim10^{11}M_\odot$
spheroids from \cite{Oser+012_evol_massive_etg} (note that they do
not provide explicitly disc-dominated predictions) with a modified
version of the parallel TreeSPH code GADGET-2
(\citealt{Springel+005_GADGET2}) and no AGN feedback. The
predicted sizes, in this case, turn out to be more than $1\sigma$
smaller the one derived in our analysis at all redshift.

The remarkable result that emerges from this comparison is that
the observed sizes are naturally explained in the context of the
galaxy assembly described in the cosmological simulations. In
particular, the size growth is interpreted in Furlong et al.
(2015) as the consequence of the accreted mass fraction since
$z=2$. The more stellar mass is accreted from sources other than
the main progenitor at a given time the more the final size of a
galaxy is found to increase. This does not take into account the
type of mergers that contribute to the size growth, but clearly
establish that size growth and accreted mass fraction are
inherently related (see their \Fig 5).

To conclude, in this paper we have demonstrated the large
potential of the KiDS dataset for the structural parameter
analysis of galaxies at least up to $z=0.6$. We have analyzed a
sample of $\sim 380,000$ galaxies with signal-to-noise ratio large
enough ($\SN_{r}>50$) to derive accurate structural parameters.
Based on mock galaxy images and performing an external comparison,
we have demonstrated that our estimates are robust. We have used
in particular the size and stellar masses to investigate the
evolution of the size-mass relation up to $z\sim0.6$ and compared
the results with hydrodynamical simulations for galaxy assembly.
The main results of our analysis can be summarized as follows:

\begin{itemize}
\item The size--mass--redshift relation show a very good agreement with
the size--mass and the size--redshift correlations obtained either
in local analyses (e.g. \citealt{Shen+03},
\citealt{Baldry+12_gama_mass}, \citealt{Mosleh+13_size_evol}) or
at higher-$z$ (e.g. \citealt{Trujillo+07},
\citealt{vanderwel+14_SM}). The size--mass relation of spheroids
shows a clear evolution of the average quantities with redshift
which we have interpreted as a consequence of the size decreases
with increasing redshift at masses larger than $\Log\
M_*/M_\odot\sim10.5$, while the evolution of the sizes for the
disc-dominated galaxies is very weak, which produces no
appreciable evolution of their size--mass relations.

\item We have derived the \Re\ vs. $z$ evolution using two approaches: 1) by
fitting the size--mass relation at different redshift bins and
then estimating the $\Re-z$ evolution along different mass
intercepts (see \Sec\ref{sec:size_mass_etg} and
\Sec\ref{sec:size_mass_ltg}) and 2) by direct fitting the measured
\Re\ vs. $z$ in different mass bins. The results of the two
methods consistently show a substantial evolution of sizes with
redshift, with spheroids having a steeper decrease of their sizes
with increasing redshifts with respect to disc-dominated galaxies.
The normalization and slope of the the \Re\ vs. $z$, parameterized
using the standard $\Re/kpc\propto(1+z)^{\beta_z}$ relation (see
\Tab\ref{tab:Re-z_etg_ltg}), are consistent with a recent analysis
using accurate HST size measurements with single S\'ersic profiles
(\citealt{vanderwel+14_SM}).

\item We have compared the data with suites of recent hydrodynamical
simulations of galaxy assembly with a full treatment of galaxy
feedback (including supernovae and AGN feedback,
\citealt{furlong+015_SM}), showing that also in this case our
results are well matched by simulations and alway consistent with
$1\sigma$ scatter of our observationally inferred $\Re-z$
relations in different mass bins for both the spheroids and
disc-dominated systems. We have also checked that simulations with
no AGN feedback (e.g. from \citealt{Oser+012_evol_massive_etg})
show a large discrepancy, showing that an insufficient feedback
recipe produces a tension with data, due to the too compact sizes
in simulated galaxies.
\end{itemize}

The large sample expected from KiDS and the image quality will
allow us to obtain unprecedented details in the evolution of the
galaxy size and mass over the cosmic time, which can be compared
with expectations from simulations. We expect to expand
considerably the analysis presented in this paper with the next
KiDS data releases, both in terms of size and depth of the sample,
as we will gather statistics toward higher redshift to confirm our
trends.

\section*{Acknowledgments}
NRN acknowledges financial support from the European Union Horizon
2020 research and innovation programme under the Marie
Sklodowska-Curie grant agreement N. 721463 to the SUNDIAL ITN
network. CT is supported through an NWO-VICI grant (project number
639.043.308). Based on data products from observations made with
ESO Telescopes at the La Silla Paranal Observatory under programme
IDs 177.A-3016, 177.A-3017 and 177.A-3018, and on data products
produced by Target/OmegaCEN, INAF-OACN, INAF-OAPD and the KiDS
production team, on behalf of the KiDS consortium. OmegaCEN and
the KiDS production team acknowledge support by NOVA and NWO-M
grants. Members of INAF-OAPD and INAF-OACN also acknowledge the
support from the Department of Physics \& Astronomy of the
University of Padova, and of the Department of Physics of Univ.
Federico II (Naples). We thank Giuseppe D'Ago for useful
discussions and feedback on the analysis performed in this paper.

\appendix
\section{Effect of the errors on the \mathinhead{$\Re-z$}{Re-Z} relations}\label{app:Re_z}

\begin{figure*}
\centering
\includegraphics[scale=0.48]{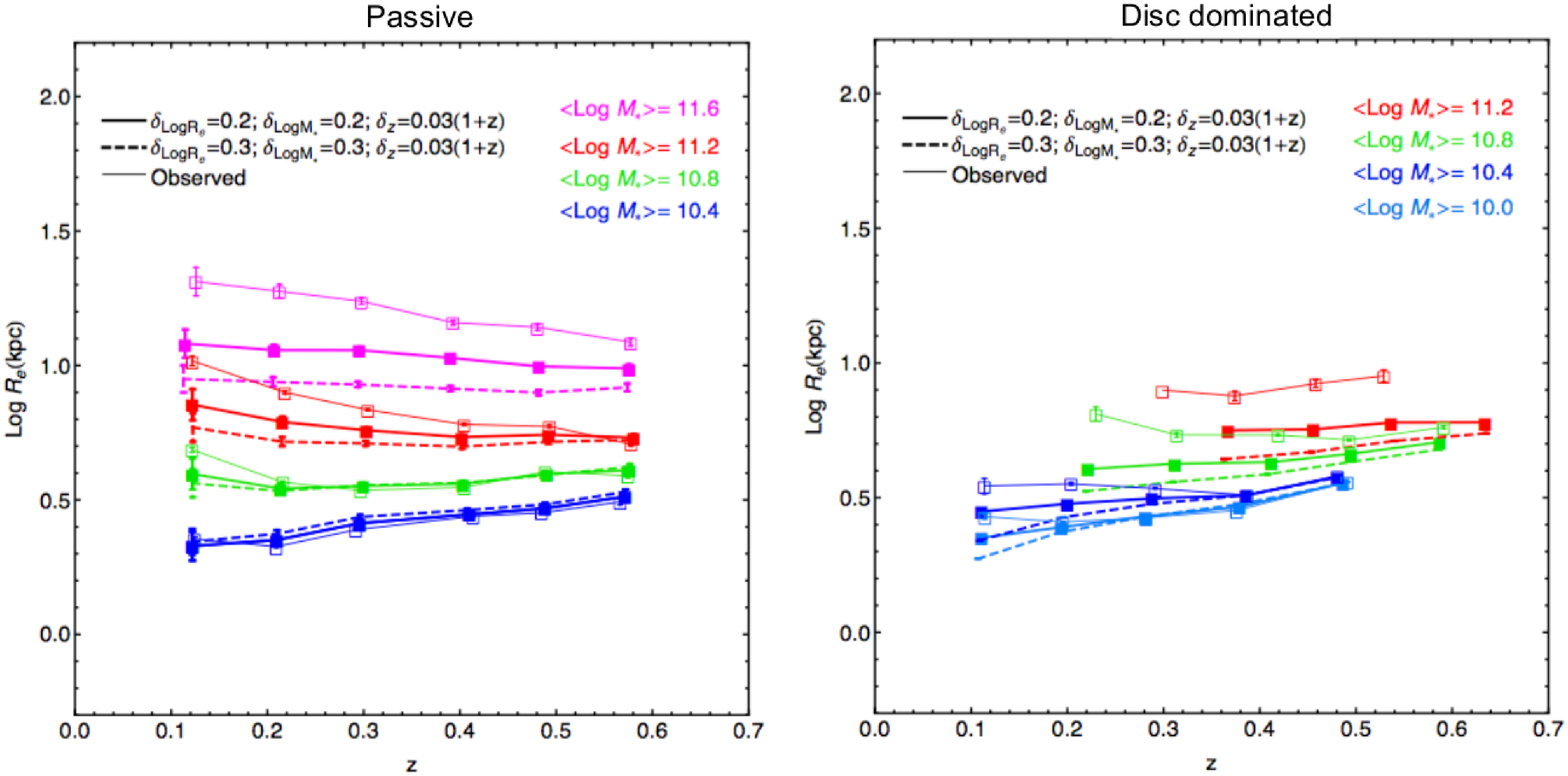}
\caption{Effect of the errors on the $\Re$-$z$ correlation as
derived in \Fig\ref{size_redshift}. Left: average trend for the
spheroid sample obtained by resampling the original sample
parameters by adding a shift from a Gaussian distribution using
average errors (solid line) and maximum errors (dashed lines) as
in the legends. The observed trend is also shown as comparison
(tiny solid line). Right: the same as in the left panel for the
disc dominated galaxies.} \label{fig:bootstrap}
\end{figure*}

We want to check the effect of the uncertainties on the different
quantities entering into the size-redshift trends discussed in
\Secs\ref{sec:size_mass_etg} and \ref{sec:size_mass_ltg}. The
trend found can indeed be affected by the intrinsic scatter of the
mass, effective radius estimate and photometric redshift. In
principle, the covariance among the individual errors might
spuriously generate a correlation from the observed quantities. On
the other way around, the observed trend can be even shallower
than the intrinsic one for the scatter due to the different
quantities that move objects from one bin to another, hence
diluting the real trends. In order to check for the presence of
these effects, and evaluate in which direction the correlations
that we have derived in \Secs\ref{sec:size_mass_etg} and
\ref{sec:size_mass_ltg} and reported in \Tab\ref{tab:Re-z_etg_ltg}
can be biased by the intrinsic scatter of the individual
parameters, we have performed a series of bootstrap experiments to
obtain random resamplings of our datasets. We have perturbed
galaxy mass, \Re\ and \zphot\ by randomly adding a offset
extracted from a Gaussian distribution having zero mean and a
constant standard deviation equal to the average error of the
different quantities (namely $\sigma_{\Log\ M_*}$, $\sigma_{\Re}$,
$\sigma_{\zphot}$ for the mass, size and \zphot\ respectively),
hence resampling the same observed relations, but adding the
effect of random errors on the individual parameters.

In \Sec\ref{sec:SED} we have mentioned that average errors on
masses are of the order of 0.2 dex (maximum errors reaching 0.3
dex), while the relative errors on \Re\ are of the order of 15 per
cent (20 per cent maximum, see e.g. \Figs\ref{simulation},
\ref{background_figs}, and \ref{spider_kids}), while the scatter
for the \zphot\ has been reported to be of the order of
$0.03(1+z_{\rm spec})$ (see \Sec\ref{sec:photo_z}). We have
re-extracted the catalog values 100 times and obtained, at every
extraction, a correlation like \Fig\ref{size_redshift}, which we
have finally averaged to obtain the average trend in each mass
bin.

\begin{figure}
\centering
\includegraphics[width=.85\columnwidth]{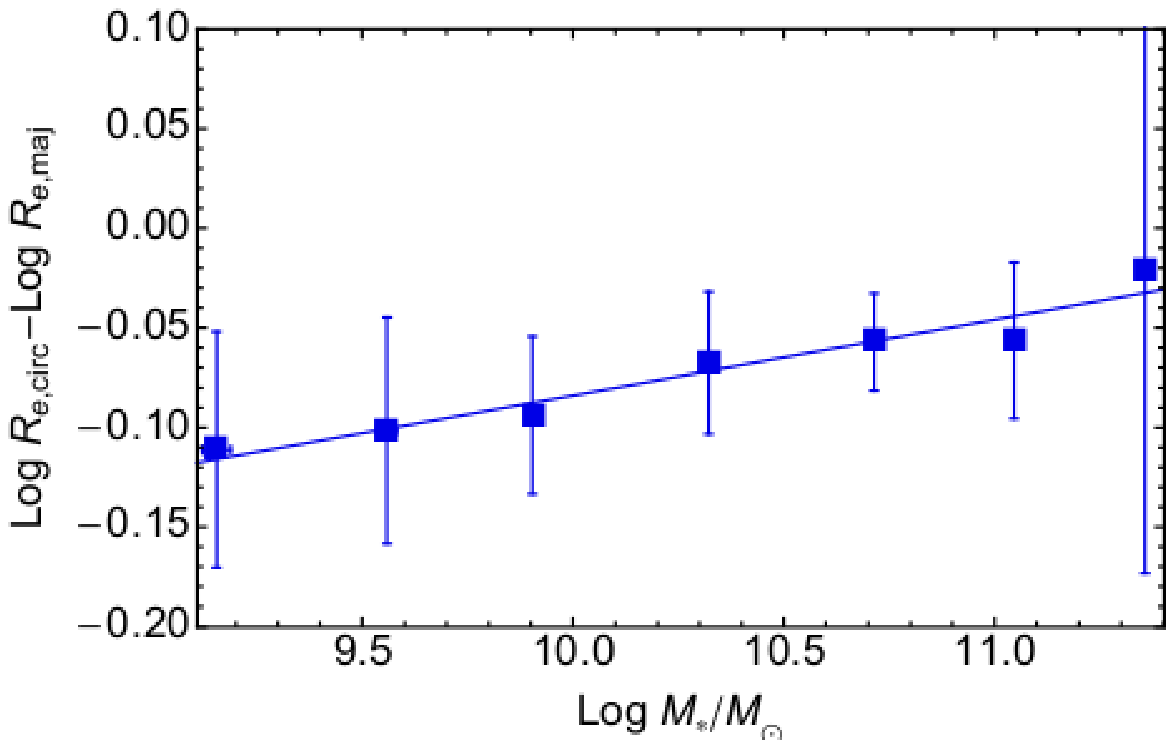}
\includegraphics[width=.85\columnwidth]{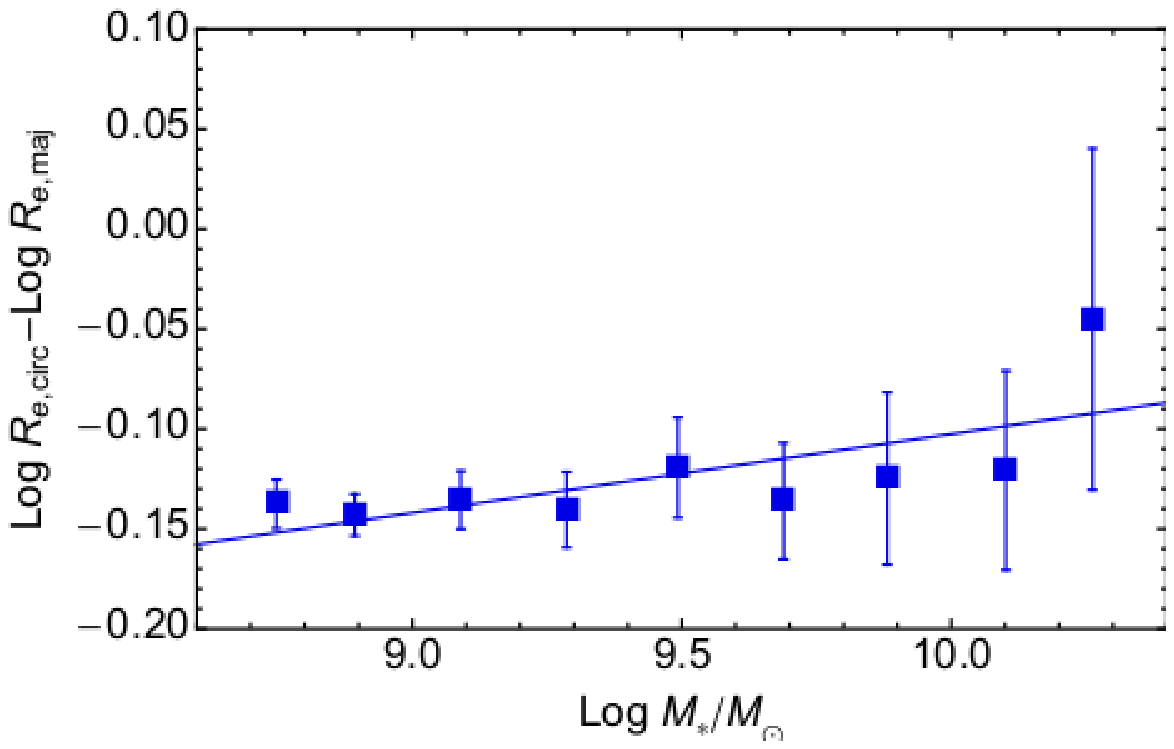}
\caption{Differences between the (log) circularized and major axis
effective radii obtained with 2DPHOT for KiDS spheroids (top) and
disc dominated (bottom) in $r$-band as a function of the stellar
mass in the lowest redshift bin as in Fig. \ref{fig:size_mass1}.
Mean differences and standard deviations are shown as data points
with error bars, together with a linear fit to the data. This
shows the statistical correction one should apply to the size-mass
relation in the case of major axis size estimates (as for B+12 and
K+12).} \label{fig:maj_circ}
\end{figure}

\begin{figure*}
\centering
\includegraphics[scale=0.7]{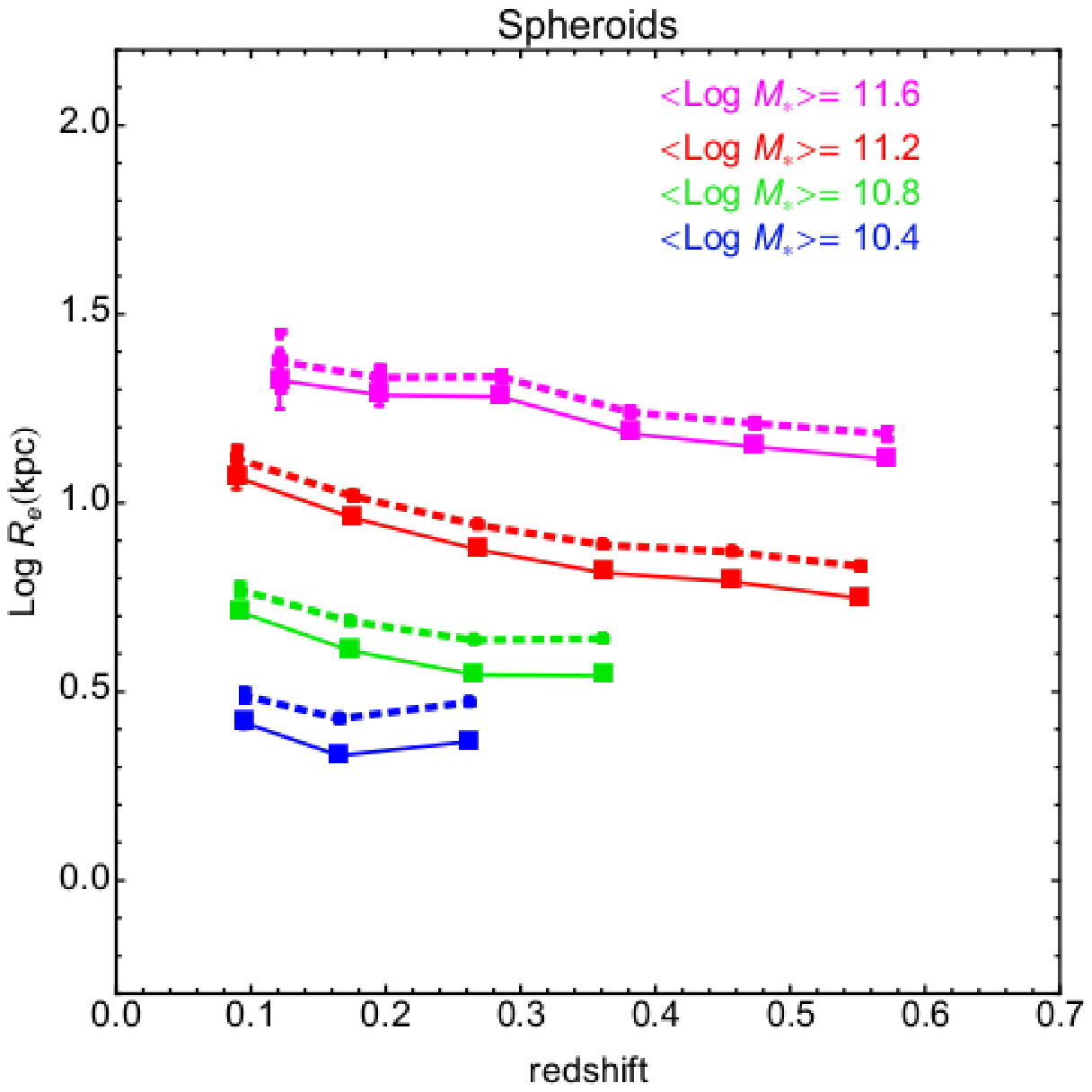}
\includegraphics[scale=0.7]{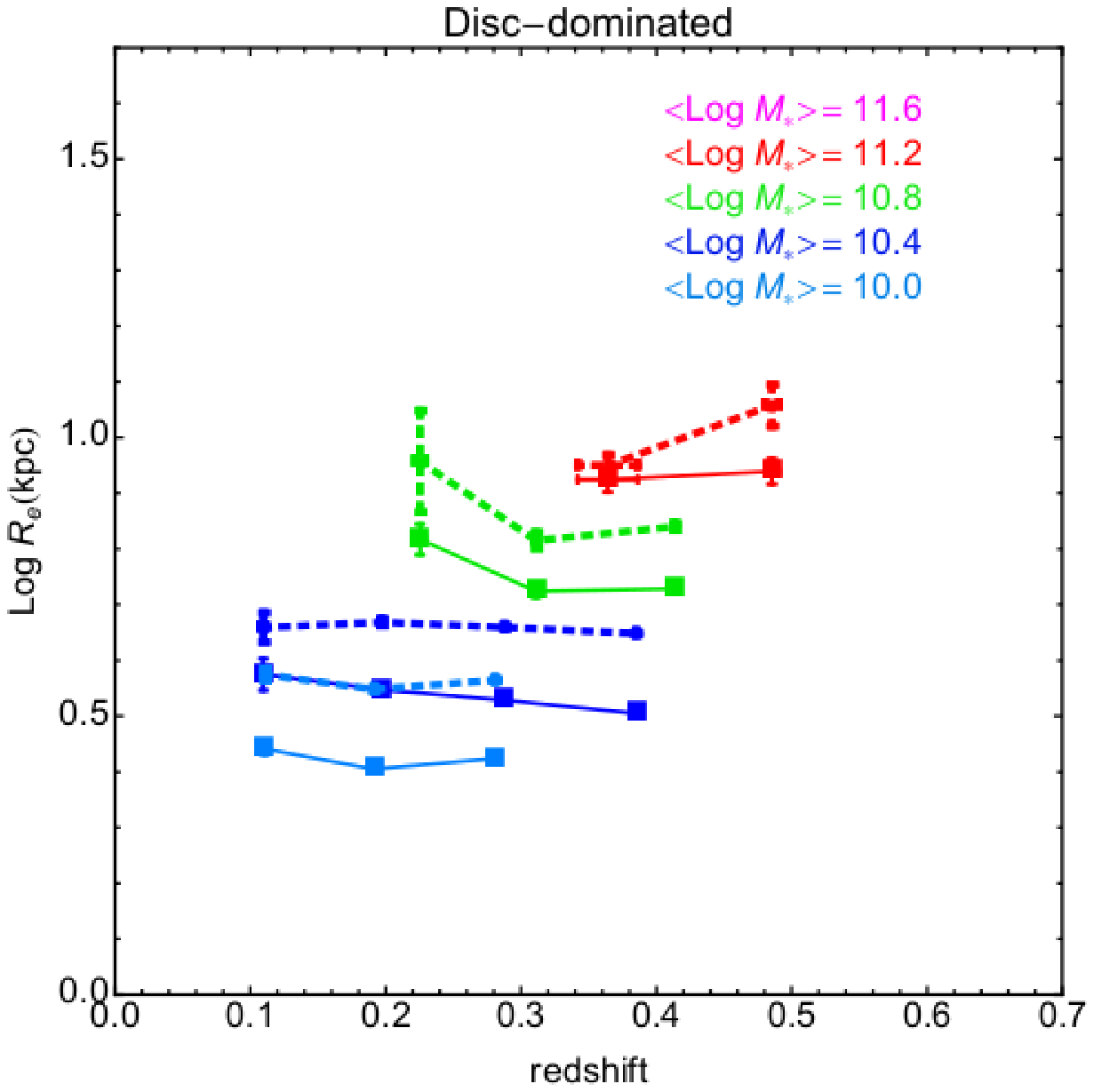}
\caption{Size-redshift correlation for the spheroids (left) and
disc dominated galaxies (right) using the circularized (solid
line) or the major axis effective radii (dashed line).}
\label{fig:maj_circ_etg_ltg}
\end{figure*}
We have also checked that the quantities that are affecting more
the trend is the mass as the scattered quantities move galaxies
from the central mass bins to the contiguous (small and larger
mass) ones, hence making all relations to converge toward the ones
of the intermediate bins, as shown from the case of maximum
errors.

In \Fig\ref{fig:bootstrap} we show the ``bootstrap'' results for
spheroids and disc-dominated galaxies obtained both for the
average errors (solid lines) and for the maximum errors (dashed
lines). We can clearly see that for the spheroids, the larger the
errors assumed the flatter is the final trend obtained. This
demonstrate that the effect of the uncertainties on the quantities
is statistically to reduce the steepness of the observed trends
(tiny lines in \Fig\ref{fig:bootstrap}) rather then to introduce a
spurious slope. The same effect is also seen for the disc-dominated
galaxies although, for the lower mass bin, we see that
errors produce a steepening of the correlation in the lower
redshift bins.

Overall, this test demonstrates that the trends discussed in
\Tab\ref{tab:Re-z_etg_ltg} are real and possibly shallower than
the ones that we had measured if we could reduce the uncertainties
on the observed quantities. The only exception is for low-mass
disc-dominated systems (i.e. $\Log\ M_*/M_\odot\lsim10.2$), that
at lower redshift ($z<0.2$) might have a steeper trend with
respect to the almost flat trend observed in
\Fig\ref{size_redshift}.

As discussed above, the major source of uncertainties in the
derived trends is the one on the mass, which we plan to reduce in
the future by adding NearInfraRed (NIR) bands in our SED
estimates.

\section{Effect of the size definition: circularized vs. major axis effective radius}
\label{app:maj_circ} In this section we want to statistical assess
the effect of the size definition on the size mass relations
discussed in \S\ref{sec:Resu}. We have seen that different
analyses have made different choices on the effective radius to
use for their relations, i.e. by using the simple major axis
radius ($R_{m}$, which does not take into account the observed
flattening of the galaxy) or the circularized one, defined as $\Re
= (b/a)^{1/2}R_{m}$ (as done in \S\ref{sec:stru_par}). Since the
axis ratio changes with the luminosity and mass of galaxies, the
ratio between the $\Re$ and $R_{m}$, which is exactly
$f=\sqrt{b/a}$, also changes with these parameters and hence one
should measure a tilt of the size-mass relation depending on
whether this is based on the use of \Re\ or $R_{m}$. We can
quantify how the ratio of the two quantities $f=\sqrt{b/a}$
changes with stellar mass for the KiDS spheroids and disc
dominated galaxies by looking at the difference of the logarithm
of these two radii in Fig. \ref{fig:maj_circ}.

In the plot we see that for spheroids $\Re/R_m$ is almost equal to one for the more massive galaxies
(which are dominated by early-type rounder systems) and then decreases for lower masses
consistently with the known anti-correlation between the flattening and the mass, i.e.
the axis ratio tends to increase for lower mass systems. This is more marked for
disc dominated systems which are intrinsically flatter at almost all masses.
If on one hand, we can use the correlation shown in Fig. \ref{fig:maj_circ} to convert the
correlations found in \S\ref{sec:Resu} for the circularized radii into the same
correlations for the major axis sizes, on the other hand we can use the same
correlation to rescale the literature results based on major axis sizes to the
circularized size they should have if they statistically have the same flattening
variation with mass as the KiDS sample. We note that, in practice, the operation
of correcting the circularized radii of the KiDS data to match the major axis
definition of other analysis (e.g. B+12 and K+12) would bring to the same conclusion
if the major axis data would be converted into the circularized ones using the inverse
ratio $1/f$, but with a different normalization, which is the approach we used for Fig.
\ref{fig:size_mass1}, when comparing different results from literature.

\begin{figure*}
\includegraphics[scale=0.7]{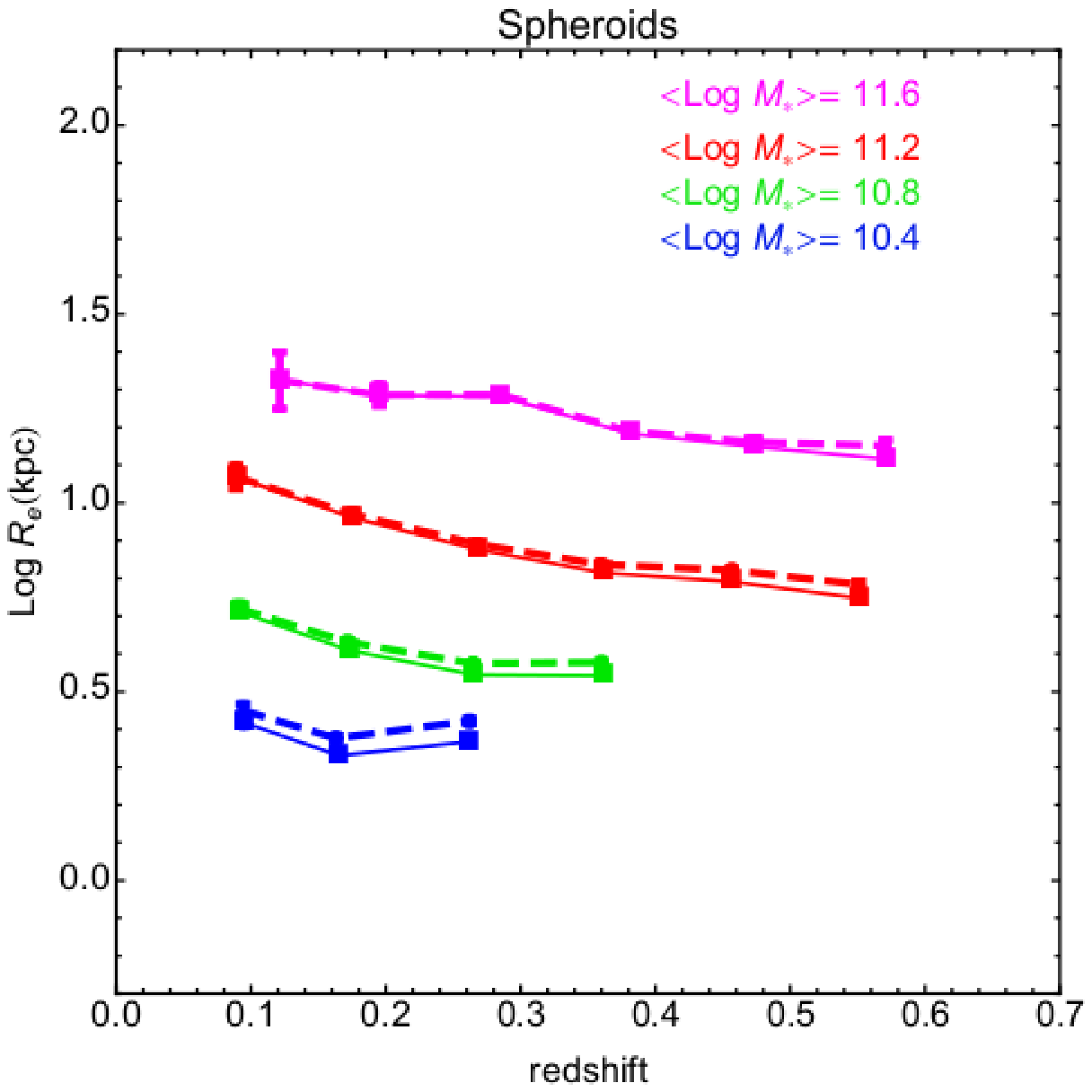}
\includegraphics[scale=0.7]{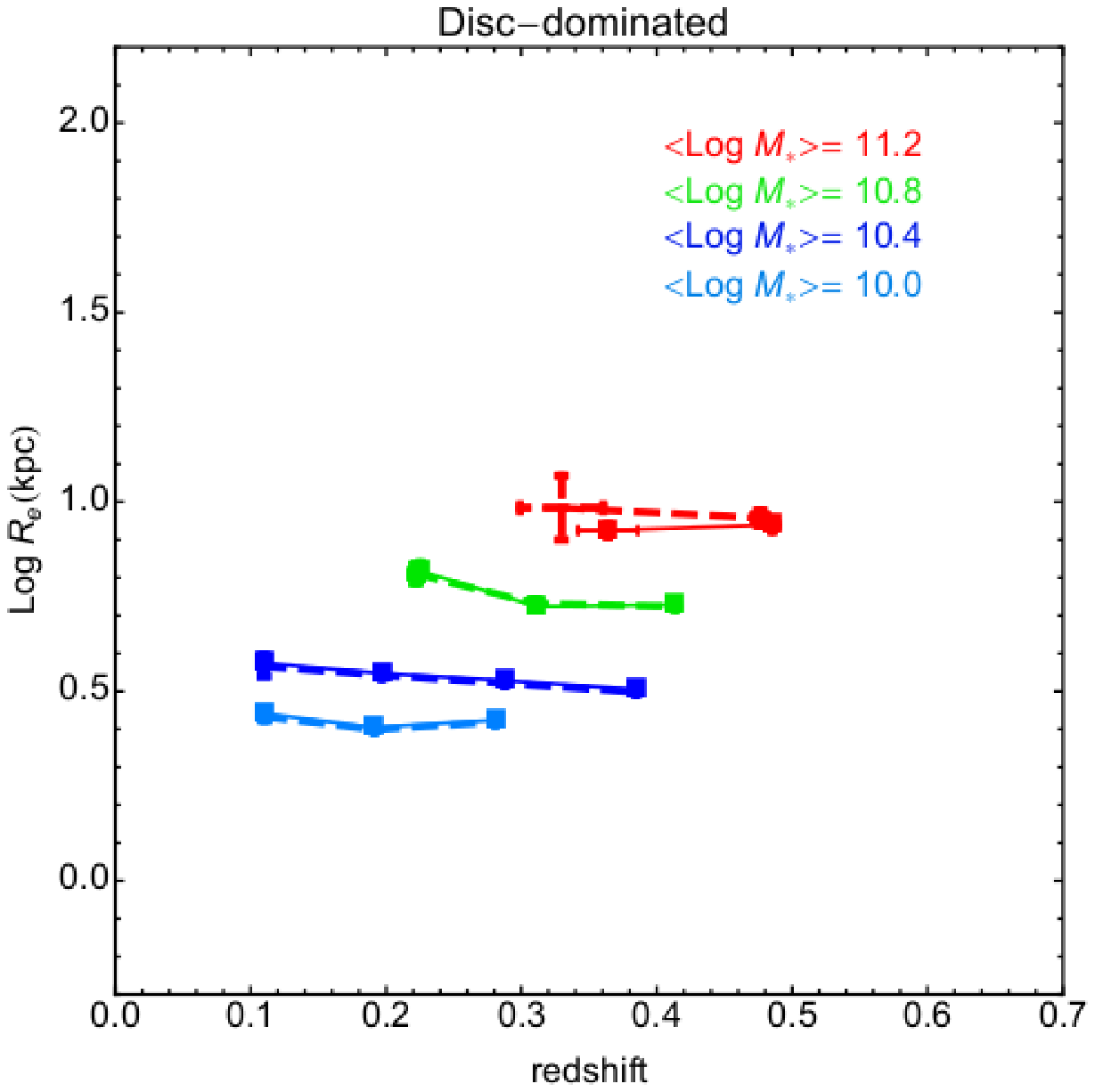}
\caption{Size--redshift correlation for the spheroids (left)
and disc dominated galaxies (right) using $n=2.5$ (solid line, see also
Fig. \ref{size_redshift}) or $n=3.5$ (dashed line).} \label{fig:n25_n35}
\end{figure*}

In order to better quantify the effect of the size definition on our results,
we compare the Size vs. $z$ relations of the spheroids and disc-dominated systems
in Fig. \ref{fig:maj_circ_etg_ltg}, obtained both for the \Re and $R_m$.
The major effect is seen in the normalization of the relations, being the $R_m$
generally larger than the \Re\ and it is more evident for disc-dominated systems
which are intrinsically more flattened. Also, there is a tendency to show a larger
difference toward higher-z, being generally all systems less round going back in time
in their evolution. Overall, the $R_m$ vs. redshift does not seem to drive to different
conclusions of the one discussed for the \Re, with spheroids showing a gain a
significant growth toward low-z and the disc-dominated systems almost no evolutions at all
all mass scales.

\section{Effect of the S\'ersic index systematics}
\label{app:sersic}
In \S\ref{sec:sdss_comp} we have found a systematic offset of the $n$-index
estimated with 2DPHOT with respect to the ones by GALFIT in the sample common to the SDSS
analysed in K+12. We have discussed that it is not possible at the moment to assess which
of the two set of inferences is unbiased. Despite our test on mock galaxies in \S\ref{sec:simul}
shows no bias in the 2DPHOT estimates, we want to quantify which would be the effect of
of a biased the spheroids/discs splitting of our sample on the final size evolution, if
our S\'ersic index are systematic overestimate of ground truth, assuming these latter
given by the ones from GALFIT on GAMA galaxies. If this is the case, then
our criterion of $n > 2.5$ to select spheroids should have possibly including a large
fraction of actual discs.
We recall here that the galaxy classification we have adopted is based on
both the S\'ersic index cut and SED classes as discussed in \S\ref{sec:SED},
and the former only partially plays a role.

To proceed with this test we decided to compare the results on the
size-redshift relation obtained with $n=2.5$ as discriminant for
spheroids/discs with the same relation obtained for $n=3.5$, which
is a more conservative value in case the 2DPHOT estimates
correspond to intrinsically smaller $n$-indexes. This is shown in
Fig. \ref{fig:n25_n35}, where we can see that the change of the
correlations for the spheroid and disc-dominated galaxies are
almost unchanged. This shows that the combination of the $n$-index
and the SED class basically prevents significant
misclassifications which might be expected in the case only
$n$-index would be used.

\section{Effect of wavelength on the \mathinhead{$\Re$}{Re}}\label{app:Re_lamda}
Galaxies have color gradients, which means that, on average, their
optical \Re\ can change as a function of the band in which they
are observed. E.g. spheroids have negative gradients (they are
redder in their center and bluer in their outskirts) which implies
that they are larger in bluer bands (see e.g.,
\citealt{Sparks_Jorgensen93}; \citealt{HB09_curv};
\citealt{LaBarbera_deCarvalho+09, SPIDER-IV}; \citealt{Roche+10};
\citealt{Tortora+10CG, Tortora+018_DM_KiDS}; \citealt{Vulcani+14};
\citealt{Beifiori+14}).

\begin{figure}
\includegraphics[scale=0.6]{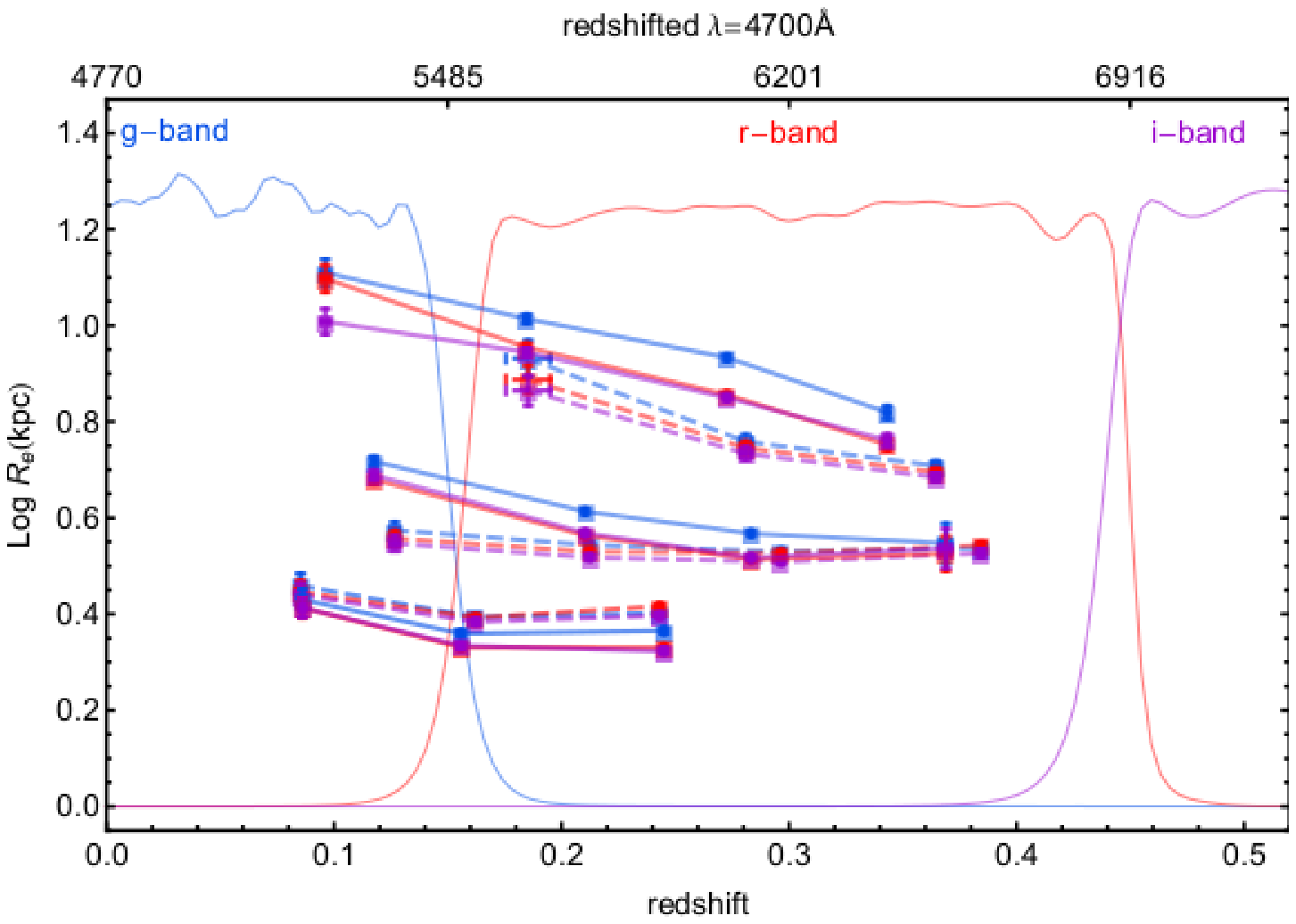}
\caption{ Size--redshift relation for the spheroids (solid lines)
in the mass bins  $\Log\ M_*/M_\odot=$10.4, 10.8, 11.2 and disc
dominated galaxies in the mass bins $\Log\ M_*/M_\odot=$10.0,
10.4, 10.8 (dashed lines) in $g$ (blue), $r$ (red) and $i$
(purple) bands.} The top axis reports the wavelength of the
$g$-band central $\lambda_g=4770$\AA\ redshifted according to the
corresponding $z$ on the bottom axis. The filter response in
arbitrary units are shown with the wavelength consistent with the
top axis. See text for more details. \label{fig:wav_effect}
\end{figure}

Going to higher redshifts, the blue part of galaxy SEDs are
redshifted into redder bands, hence the $r$-band which we have
used as reference band for our analysis, covers different rest
frame wavelengths in different galaxies, depending on each galaxy
redshift. If the colour gradients persist at these epochs, this
implies that the observed frame $r$-band of a high-$z$ galaxy,
which corresponds to a $g$-band rest frame, is intrinsically
larger than the $r$-band of a lower-$z$, which is closer to the
$r$-band rest frame itself (see \citealt{Vulcani+14}). E.g., the
mean wavelengths of OmegaCAM filters are $\lambda_{g,r,i} =
\{4770, 6231, 7625\}$ \AA , i.e. a galaxy at $z=0$ observed in
$g$-band at $\lambda_g=4770$\AA\ is redshifted to
$\lambda_r=6287$\AA\ for $z=0.32$, and to the lower limit of the
$r$-band wavelength range $\lambda_{r,min}\sim5200$\AA\ for
$z\sim0.1$ (see Fig. \ref{fig:wav_effect}).

This implies that if we use $r$-band as a reference for galaxies
at lower (e.g. $z\lsim0.15$) and higher redshift (e.g.
$z\gsim0.15$), in case the ones at higher-$z$ have negative colour
gradients, these might look larger than the ones at lower-$z$
because their emission in $r$-band is dominated by the rest-frame
$g$-band. The opposite will happen for galaxies at higher-$z$ with
positive gradient.

To quantify this effect we could proceed in two ways. First, we
can check whether there is an observed difference between the
average \Re\ estimated in bins of redshift and mass in the
different bands. Second, we could compute the rest frame \Re\ at
some reference wavelength by linearly interpolating the \Re\
values obtained in different bands and estimate the evolution with
redshift of this latter. We will show that the second approach is
not convenient to apply at this stage of the project since we are
lacking of completeness and sample size, to have robust
inferences. We limit here to demonstrate that the adoption of the
deeper $r$-band is not expected to affect the main results of our
paper.

Looking at the $\Re-z$ in the other bands, it is important to
check the mass dependence because color gradients have been
observed to change with mass (\citealt{LaBarbera_deCarvalho+09,
SPIDER-IV}; \citealt{Tortora+10CG, Tortora+018_DM_KiDS}). In order
to do that we needed to use a mass complete sample of galaxies for
which we have measurements of the \Re\ in the $gri$ bands
simultaneously (we have excluded the $u$-band since this would
have reduced the sample too much, see below) and selected also in
this case the ones with highest \SN\ $(>50)$ and best
$\chi^2~(<1.3)$. Since the depth and completeness mass in the $g$-
and $i$-band are lower than the $r$-band because of the survey
strategy, the  final sample of galaxies  available for this test
is almost one third of the one found for $r$-band (see
\S\ref{sec:etg_ltg}), i.e. $\sim70$k galaxies.

In Fig. \ref{fig:wav_effect} (left) we show the average $\Re-z$ of
the selected sample of galaxies in $gri$-bands for the spheroids
(solid lines) and disc-dominated (dashed lines). In particular, we
show the relations obtained for different mass bins (bottom lines
with smaller \Re\ to the top), which are $\Log\ M_*/M_\odot=$10.4,
10.8, 11.2 for spheroids and $\Log\ M_*/M_\odot=$10.0, 10.4, 10.8
for disc-dominated galaxies. In the same figure we show on the top
axis the wavelength of the $g$-band central $\lambda_g=4770$\AA\
redshifted according to the corresponding $z$ on the bottom axis.
We also overplot the filter response in arbitrary units with the
wavelength distributed according to the top axis. This allows us
to visualize how the $g$-band rest frame is shifted into other
bands at any redshift, and compare this with the \Re\ inferences
in the different bands. We can see that the adoption of the
$r$-band as a reference filter for our analysis is motivated by
the fact that this covers the largest part of the redshift window
of our sample (i.e. $z=0.1$ to $0.5$). If one would fairly compare
the galaxy sizes in the same rest frame wavelength range, than the
$g$-band estimates would work approximately until $z\sim0.15$ and
the $r$-band between $z\sim0.15$ and $z\sim0.45$, while $i$-band
should be used at $z\gsim0.45$.

Overall, we see that the disc-dominated galaxies show almost no
differences in the \Re\ estimates in all bands and for the lower
mass bin, they look almost undistinguishable. This suggests that
discs have almost no colour gradients or possibly mild negative
one. This latter is more evident for the larger mass bin at low-z,
which goes along the direction of previous finding in local
samples (e.g., \citealt{Tortora+10CG}). Spheroids show negative
gradients at almost all mass (and increasing with it) and
redshift, except possibly for the $g-r$ of the most massive
systems at low-$z$, where the $g$- and $r$-band estimates almost
coincide. On the other hand the $r$- and $i$-band average
estimates are almost identical at all masses and redshift except
for the latter case. The spheroids negative gradients are also
larger for more massive systems.

The discussion on the trends of the colour gradients is beyond the
scopes of this paper, and there will be forthcoming analysis
dedicated to that, here we are interested on the effect of these
gradients on the main conclusions of our paper. As anticipated,
the presence of negative gradients would imply that, while the
measurement of the $r$-band at $0.15\lsim z \lsim 0.45$ are a good
representation of the rest frame $g$-band, at lower $z$ it is the
$g$-band \Re\ the one to use. But these latter are systematically
larger than the ones from $r$-band and hence the slope that we
would measure should be steeper that the one obtained using the
$r$-band at all redshift. Note that even if the $i$-band would
represent the ideal band to cover the $g$-band rest frame at
$z>0.45$, being $r$ and $i$-band estimates almost identical, we do
not expect that the use of $r$-band has caused any effect.

In this perspective the only big change we could expect by using
proper rest frame sizes would be a steepening of the correlation
with $z$ toward the lower redshift. Since the current sample of
galaxies analysed in the three $gri$ bands is limited in number,
mass and redshift (see Fig. \ref{fig:wav_effect}), we reserve this
analysis to next datasets from the subsequent data releases. E.g.,
we expect to collect up to 500k galaxies with $gri$ sizes in a
larger photo-z range (using the updated estimates from
\citealt{Bilicki+017_photoz_arxiiv}) with the upcoming KiDS data
release 4 based on 900 deg$^2$ of the survey.

\bibliographystyle{mnras}
\bibliography{myrefs}

\begin{thebibliography}{}
\makeatletter
\relax
\def\mn@urlcharsother{\let\do\@makeother \do\$\do\&\do\#\do\^\do\_\do\%\do\~}
\def\mn@doi{\begingroup\mn@urlcharsother \@ifnextchar [ {\mn@doi@}
  {\mn@doi@[]}}
\def\mn@doi@[#1]#2{\def\@tempa{#1}\ifx\@tempa\@empty \href
  {http://dx.doi.org/#2} {doi:#2}\else \href {http://dx.doi.org/#2} {#1}\fi
  \endgroup}
\def\mn@eprint#1#2{\mn@eprint@#1:#2::\@nil}
\def\mn@eprint@arXiv#1{\href {http://arxiv.org/abs/#1} {{\tt arXiv:#1}}}
\def\mn@eprint@dblp#1{\href {http://dblp.uni-trier.de/rec/bibtex/#1.xml}
  {dblp:#1}}
\def\mn@eprint@#1:#2:#3:#4\@nil{\def\@tempa {#1}\def\@tempb {#2}\def\@tempc
  {#3}\ifx \@tempc \@empty \let \@tempc \@tempb \let \@tempb \@tempa \fi \ifx
  \@tempb \@empty \def\@tempb {arXiv}\fi \@ifundefined
  {mn@eprint@\@tempb}{\@tempb:\@tempc}{\expandafter \expandafter \csname
  mn@eprint@\@tempb\endcsname \expandafter{\@tempc}}}

\bibitem[\protect\citeauthoryear{{Ahn} et~al.,}{{Ahn}
  et~al.}{2012}]{Ahn+12_SDSS_DR9}
{Ahn} C.~P.,  et~al., 2012, \mn@doi [\apjs] {10.1088/0067-0049/203/2/21}, \href
  {http://adsabs.harvard.edu/abs/2012ApJS..203...21A} {203, 21}

\bibitem[\protect\citeauthoryear{{Arnouts}, {Cristiani}, {Moscardini},
  {Matarrese}, {Lucchin}, {Fontana}  \& {Giallongo}}{{Arnouts}
  et~al.}{1999}]{Arnouts+99}
{Arnouts} S.,  {Cristiani} S.,  {Moscardini} L.,  {Matarrese} S.,  {Lucchin}
  F.,  {Fontana} A.,   {Giallongo} E.,  1999, \mn@doi [\mnras]
  {10.1046/j.1365-8711.1999.02978.x}, \href
  {http://adsabs.harvard.edu/abs/1999MNRAS.310..540A} {310, 540}

\bibitem[\protect\citeauthoryear{{Arnouts} et~al.,}{{Arnouts}
  et~al.}{2001}]{Arnouts+01_gal_counts}
{Arnouts} S.,  et~al., 2001, \mn@doi [\aap] {10.1051/0004-6361:20011341}, \href
  {http://adsabs.harvard.edu/abs/2001A%26A...379..740A} {379, 740}

\bibitem[\protect\citeauthoryear{{Baldry} et~al.,}{{Baldry}
  et~al.}{2012}]{Baldry+12_gama_mass}
{Baldry} I.~K.,  et~al., 2012, \mn@doi [\mnras]
  {10.1111/j.1365-2966.2012.20340.x}, \href
  {http://adsabs.harvard.edu/abs/2012MNRAS.421..621B} {421, 621}

\bibitem[\protect\citeauthoryear{{Beifiori} et~al.,}{{Beifiori}
  et~al.}{2014}]{Beifiori+14}
{Beifiori} A.,  et~al., 2014, preprint (\mn@eprint {arXiv} {1405.1431})

\bibitem[\protect\citeauthoryear{{Bernardi} et~al.,}{{Bernardi}
  et~al.}{2003}]{Bernardi+03_Mg2}
{Bernardi} M.,  et~al., 2003, \mn@doi [\aj] {10.1086/367795}, \href
  {http://adsabs.harvard.edu/abs/2003AJ....125.1882B} {125, 1882}

\bibitem[\protect\citeauthoryear{{Bertin} \& {Arnouts}}{{Bertin} \&
  {Arnouts}}{1996}]{Bertin_Arnouts+96}
{Bertin} E.,  {Arnouts} S.,  1996, \mn@doi [\aaps] {10.1051/aas:1996164}, \href
  {http://adsabs.harvard.edu/abs/1996A%26AS..117..393B} {117, 393}

\bibitem[\protect\citeauthoryear{{Bezanson}, {van Dokkum}, {Tal}, {Marchesini},
  {Kriek}, {Franx}  \& {Coppi}}{{Bezanson}
  et~al.}{2009}]{Bezanson+09_minor_merge}
{Bezanson} R.,  {van Dokkum} P.~G.,  {Tal} T.,  {Marchesini} D.,  {Kriek} M.,
  {Franx} M.,   {Coppi} P.,  2009, \mn@doi [\apj]
  {10.1088/0004-637X/697/2/1290}, \href
  {http://adsabs.harvard.edu/abs/2009ApJ...697.1290B} {697, 1290}

\bibitem[\protect\citeauthoryear{{Bilicki} et~al.,}{{Bilicki}
  et~al.}{2017}]{Bilicki+017_photoz_arxiiv}
{Bilicki} M.,  et~al., 2017, ArXiv:1709.04205

\bibitem[\protect\citeauthoryear{{Bower}, {Lucey}  \& {Ellis}}{{Bower}
  et~al.}{1992}]{Bower+92_CM}
{Bower} R.~G.,  {Lucey} J.~R.,   {Ellis} R.~S.,  1992, \mn@doi [\mnras]
  {10.1093/mnras/254.4.601}, \href
  {http://adsabs.harvard.edu/abs/1992MNRAS.254..601B} {254, 601}

\bibitem[\protect\citeauthoryear{{Brescia}, {Cavuoti}, {D'Abrusco}, {Longo}  \&
  {Mercurio}}{{Brescia} et~al.}{2013}]{Brescia+13}
{Brescia} M.,  {Cavuoti} S.,  {D'Abrusco} R.,  {Longo} G.,   {Mercurio} A.,
  2013, \mn@doi [\apj] {10.1088/0004-637X/772/2/140}, \href
  {http://adsabs.harvard.edu/abs/2013ApJ...772..140B} {772, 140}

\bibitem[\protect\citeauthoryear{{Brescia}, {Cavuoti}, {Longo}  \& {De
  Stefano}}{{Brescia} et~al.}{2014}]{Brescia+14}
{Brescia} M.,  {Cavuoti} S.,  {Longo} G.,   {De Stefano} V.,  2014, \mn@doi
  [\aap] {10.1051/0004-6361/201424383}, \href
  {http://adsabs.harvard.edu/abs/2014A%26A...568A.126B} {568, A126}

\bibitem[\protect\citeauthoryear{{Bruzual} \& {Charlot}}{{Bruzual} \&
  {Charlot}}{2003}]{BC03}
{Bruzual} G.,  {Charlot} S.,  2003, \mn@doi [\mnras]
  {10.1046/j.1365-8711.2003.06897.x}, \href
  {http://adsabs.harvard.edu/abs/2003MNRAS.344.1000B} {344, 1000}

\bibitem[\protect\citeauthoryear{{Capaccioli} \& {Schipani}}{{Capaccioli} \&
  {Schipani}}{2011}]{Capaccioli_Schipani11}
{Capaccioli} M.,  {Schipani} P.,  2011, The Messenger, \href
  {http://adsabs.harvard.edu/abs/2011Msngr.146....2C} {146, 2}

\bibitem[\protect\citeauthoryear{{Capaccioli}, {Caon}  \&
  {D'Onofrio}}{{Capaccioli} et~al.}{1992}]{Capaccioli+92a}
{Capaccioli} M.,  {Caon} N.,   {D'Onofrio} M.,  1992, \mnras, \href
  {http://adsabs.harvard.edu/abs/1992MNRAS.259..323C} {259, 323}

\bibitem[\protect\citeauthoryear{{Capak} et~al.,}{{Capak}
  et~al.}{2004}]{Capak+04_gal_counts}
{Capak} P.,  et~al., 2004, \mn@doi [\aj] {10.1086/380611}, \href
  {http://adsabs.harvard.edu/abs/2004AJ....127..180C} {127, 180}

\bibitem[\protect\citeauthoryear{{Cappellari} et~al.,}{{Cappellari}
  et~al.}{2012}]{Cappellari+12}
{Cappellari} M.,  et~al., 2012, \mn@doi [\nat] {10.1038/nature10972}, \href
  {http://adsabs.harvard.edu/abs/2012Natur.484..485C} {484, 485}

\bibitem[\protect\citeauthoryear{{Cardone}, {Del Popolo}, {Tortora}  \&
  {Napolitano}}{{Cardone} et~al.}{2011}]{Cardone+11SIM}
{Cardone} V.~F.,  {Del Popolo} A.,  {Tortora} C.,   {Napolitano} N.~R.,  2011,
  \mn@doi [\mnras] {10.1111/j.1365-2966.2011.19162.x}, \href
  {http://adsabs.harvard.edu/abs/2011MNRAS.416.1822C} {416, 1822}

\bibitem[\protect\citeauthoryear{{Cassata} et~al.,}{{Cassata}
  et~al.}{2010}]{Cassata+10}
{Cassata} P.,  et~al., 2010, \mn@doi [\apjl] {10.1088/2041-8205/714/1/L79},
  \href {http://adsabs.harvard.edu/abs/2010ApJ...714L..79C} {714, L79}

\bibitem[\protect\citeauthoryear{{Cavuoti}, {Brescia}, {De Stefano}  \&
  {Longo}}{{Cavuoti} et~al.}{2015a}]{Cavuoti+15_photoz}
{Cavuoti} S.,  {Brescia} M.,  {De Stefano} V.,   {Longo} G.,  2015a, \mn@doi
  [Experimental Astronomy] {10.1007/s10686-015-9443-4}, \href
  {http://adsabs.harvard.edu/abs/2015ExA....39...45C} {39, 45}

\bibitem[\protect\citeauthoryear{{Cavuoti} et~al.,}{{Cavuoti}
  et~al.}{2015b}]{Cavuoti+15_KIDS_I}
{Cavuoti} S.,  et~al., 2015b, \mn@doi [\mnras] {10.1093/mnras/stv1496}, \href
  {http://adsabs.harvard.edu/abs/2015MNRAS.452.3100C} {452, 3100}

\bibitem[\protect\citeauthoryear{{Cavuoti} et~al.,}{{Cavuoti}
  et~al.}{2017}]{Cavuoti+17_KiDS}
{Cavuoti} S.,  et~al., 2017, \mn@doi [\mnras] {10.1093/mnras/stw3208}, \href
  {http://adsabs.harvard.edu/abs/2017MNRAS.466.2039C} {466, 2039}

\bibitem[\protect\citeauthoryear{{Chabrier}}{{Chabrier}}{2001}]{Chabrier01}
{Chabrier} G.,  2001, \mn@doi [\apj] {10.1086/321401}, \href
  {http://adsabs.harvard.edu/abs/2001ApJ...554.1274C} {554, 1274}

\bibitem[\protect\citeauthoryear{{Coleman}, {Wu}  \& {Weedman}}{{Coleman}
  et~al.}{1980}]{CWW80}
{Coleman} G.~D.,  {Wu} C.-C.,   {Weedman} D.~W.,  1980, \mn@doi [\apjs]
  {10.1086/190674}, \href {http://adsabs.harvard.edu/abs/1980ApJS...43..393C}
  {43, 393}

\bibitem[\protect\citeauthoryear{{Colless} et~al.,}{{Colless}
  et~al.}{2001}]{Colless+001_2dfGRS}
{Colless} M.,  et~al., 2001, \mn@doi [\mnras]
  {10.1046/j.1365-8711.2001.04902.x}, \href
  {http://adsabs.harvard.edu/abs/2001MNRAS.328.1039C} {328, 1039}

\bibitem[\protect\citeauthoryear{{Conroy} \& {van Dokkum}}{{Conroy} \& {van
  Dokkum}}{2012}]{Conroy_vanDokkum12b}
{Conroy} C.,  {van Dokkum} P.~G.,  2012, \mn@doi [\apj]
  {10.1088/0004-637X/760/1/71}, \href
  {http://adsabs.harvard.edu/abs/2012ApJ...760...71C} {760, 71}

\bibitem[\protect\citeauthoryear{{Courteau}, {Dutton}, {van den Bosch},
  {MacArthur}, {Dekel}, {McIntosh}  \& {Dale}}{{Courteau}
  et~al.}{2007}]{Courteau+007}
{Courteau} S.,  {Dutton} A.~A.,  {van den Bosch} F.~C.,  {MacArthur} L.~A.,
  {Dekel} A.,  {McIntosh} D.~H.,   {Dale} D.~A.,  2007, \mn@doi [\apj]
  {10.1086/522193}, \href {http://adsabs.harvard.edu/abs/2007ApJ...671..203C}
  {671, 203}

\bibitem[\protect\citeauthoryear{{D'Onofrio}, {Capaccioli}, {Zaggia}  \&
  {Caon}}{{D'Onofrio} et~al.}{1997}]{DOnofrio+97_distance}
{D'Onofrio} M.,  {Capaccioli} M.,  {Zaggia} S.~R.,   {Caon} N.,  1997, \mn@doi
  [\mnras] {10.1093/mnras/289.4.847}, \href
  {http://adsabs.harvard.edu/abs/1997MNRAS.289..847D} {289, 847}

\bibitem[\protect\citeauthoryear{{Daddi} et~al.,}{{Daddi}
  et~al.}{2005}]{Daddi+05}
{Daddi} E.,  et~al., 2005, \mn@doi [\apj] {10.1086/430104}, \href
  {http://adsabs.harvard.edu/abs/2005ApJ...626..680D} {626, 680}

\bibitem[\protect\citeauthoryear{{Damjanov}, {Geller}, {Zahid}  \&
  {Hwang}}{{Damjanov} et~al.}{2015}]{Damjanov+15_compacts}
{Damjanov} I.,  {Geller} M.~J.,  {Zahid} H.~J.,   {Hwang} H.~S.,  2015, \mn@doi
  [\apj] {10.1088/0004-637X/806/2/158}, \href
  {http://adsabs.harvard.edu/abs/2015ApJ...806..158D} {806, 158}

\bibitem[\protect\citeauthoryear{{Davis} et~al.,}{{Davis}
  et~al.}{2003}]{Davis+003_DEEP2}
{Davis} M.,  et~al., 2003, in {Guhathakurta} P.,  ed.,  \procspie Vol. 4834,
  Discoveries and Research Prospects from 6- to 10-Meter-Class Telescopes II.
  pp 161--172 (\mn@eprint {} {astro-ph/0209419}), \mn@doi{10.1117/12.457897}

\bibitem[\protect\citeauthoryear{{Dressler}, {Lynden-Bell}, {Burstein},
  {Davies}, {Faber}, {Terlevich}  \& {Wegner}}{{Dressler}
  et~al.}{1987}]{Dressler+87_FP}
{Dressler} A.,  {Lynden-Bell} D.,  {Burstein} D.,  {Davies} R.~L.,  {Faber}
  S.~M.,  {Terlevich} R.,   {Wegner} G.,  1987, \mn@doi [\apj]
  {10.1086/164947}, \href {http://adsabs.harvard.edu/abs/1987ApJ...313...42D}
  {313, 42}

\bibitem[\protect\citeauthoryear{{Driver} et~al.,}{{Driver}
  et~al.}{2011}]{Driver+11_GAMA}
{Driver} S.~P.,  et~al., 2011, \mn@doi [\mnras]
  {10.1111/j.1365-2966.2010.18188.x}, \href
  {http://adsabs.harvard.edu/abs/2011MNRAS.413..971D} {413, 971}

\bibitem[\protect\citeauthoryear{{Dutton}, {van den Bosch}, {Dekel}  \&
  {Courteau}}{{Dutton} et~al.}{2007}]{Dutton+07}
{Dutton} A.~A.,  {van den Bosch} F.~C.,  {Dekel} A.,   {Courteau} S.,  2007,
  \mn@doi [\apj] {10.1086/509314}, \href
  {http://adsabs.harvard.edu/abs/2007ApJ...654...27D} {654, 27}

\bibitem[\protect\citeauthoryear{{Faber} \& {Jackson}}{{Faber} \&
  {Jackson}}{1976}]{Faber_Jackson+76}
{Faber} S.~M.,  {Jackson} R.~E.,  1976, \mn@doi [\apj] {10.1086/154215}, \href
  {http://adsabs.harvard.edu/abs/1976ApJ...204..668F} {204, 668}

\bibitem[\protect\citeauthoryear{{Fan}, {Lapi}, {De Zotti}  \& {Danese}}{{Fan}
  et~al.}{2008}]{Fan+08}
{Fan} L.,  {Lapi} A.,  {De Zotti} G.,   {Danese} L.,  2008, \mn@doi [\apjl]
  {10.1086/595784}, \href {http://adsabs.harvard.edu/abs/2008ApJ...689L.101F}
  {689, L101}

\bibitem[\protect\citeauthoryear{{Ferrarese} \& {Merritt}}{{Ferrarese} \&
  {Merritt}}{2000}]{Ferrarese_Merritt+00_BH}
{Ferrarese} L.,  {Merritt} D.,  2000, \mn@doi [\apjl] {10.1086/312838}, \href
  {http://adsabs.harvard.edu/abs/2000ApJ...539L...9F} {539, L9}

\bibitem[\protect\citeauthoryear{{Franx}, {van Dokkum}, {F{\"o}rster
  Schreiber}, {Wuyts}, {Labb{\'e}}  \& {Toft}}{{Franx}
  et~al.}{2008}]{Franx+08_surf_dens}
{Franx} M.,  {van Dokkum} P.~G.,  {F{\"o}rster Schreiber} N.~M.,  {Wuyts} S.,
  {Labb{\'e}} I.,   {Toft} S.,  2008, \mn@doi [\apj] {10.1086/592431}, \href
  {http://adsabs.harvard.edu/abs/2008ApJ...688..770F} {688, 770}

\bibitem[\protect\citeauthoryear{{Furlong} et~al.,}{{Furlong}
  et~al.}{2015}]{furlong+015_SM}
{Furlong} M.,  et~al., 2015, \mn@doi [\mnras] {10.1093/mnras/stv852}, \href
  {http://adsabs.harvard.edu/abs/2015MNRAS.450.4486F} {450, 4486}

\bibitem[\protect\citeauthoryear{{Gebhardt} et~al.,}{{Gebhardt}
  et~al.}{2000}]{Gebhardt+00_BH_mass_velo}
{Gebhardt} K.,  et~al., 2000, \mn@doi [\apjl] {10.1086/312840}, \href
  {http://adsabs.harvard.edu/abs/2000ApJ...539L..13G} {539, L13}

\bibitem[\protect\citeauthoryear{{Governato} et~al.,}{{Governato}
  et~al.}{2004}]{Governato+004_disk_gal}
{Governato} F.,  et~al., 2004, \mn@doi [\apj] {10.1086/383516}, \href
  {http://adsabs.harvard.edu/abs/2004ApJ...607..688G} {607, 688}

\bibitem[\protect\citeauthoryear{{Guo} \& {White}}{{Guo} \&
  {White}}{2008}]{Guo_White+08_merge}
{Guo} Q.,  {White} S.~D.~M.,  2008, \mn@doi [\mnras]
  {10.1111/j.1365-2966.2007.12619.x}, \href
  {http://adsabs.harvard.edu/abs/2008MNRAS.384....2G} {384, 2}

\bibitem[\protect\citeauthoryear{{Guzman}, {Lucey}, {Carter}  \&
  {Terlevich}}{{Guzman} et~al.}{1992}]{Guzman+92_Mg2}
{Guzman} R.,  {Lucey} J.~R.,  {Carter} D.,   {Terlevich} R.~J.,  1992, \mn@doi
  [\mnras] {10.1093/mnras/257.2.187}, \href
  {http://adsabs.harvard.edu/abs/1992MNRAS.257..187G} {257, 187}

\bibitem[\protect\citeauthoryear{{Hildebrandt} et~al.,}{{Hildebrandt}
  et~al.}{2017}]{Hildebrandt+017}
{Hildebrandt} H.,  et~al., 2017, \mn@doi [\mnras] {10.1093/mnras/stw2805},
  \href {http://adsabs.harvard.edu/abs/2017MNRAS.465.1454H} {465, 1454}

\bibitem[\protect\citeauthoryear{{Hilz}, {Naab}  \& {Ostriker}}{{Hilz}
  et~al.}{2013}]{Hilz+13}
{Hilz} M.,  {Naab} T.,   {Ostriker} J.~P.,  2013, \mn@doi [\mnras]
  {10.1093/mnras/sts501}, \href
  {http://adsabs.harvard.edu/abs/2013MNRAS.429.2924H} {429, 2924}

\bibitem[\protect\citeauthoryear{{Hopkins}, {Lauer}, {Cox}, {Hernquist}  \&
  {Kormendy}}{{Hopkins} et~al.}{2009}]{Hopkins+09_DELGN_III}
{Hopkins} P.~F.,  {Lauer} T.~R.,  {Cox} T.~J.,  {Hernquist} L.,   {Kormendy}
  J.,  2009, \mn@doi [\apjs] {10.1088/0067-0049/181/2/486}, \href
  {http://adsabs.harvard.edu/abs/2009ApJS..181..486H} {181, 486}

\bibitem[\protect\citeauthoryear{{Hopkins}, {Kere{\v s}}, {O{\~n}orbe},
  {Faucher-Gigu{\`e}re}, {Quataert}, {Murray}  \& {Bullock}}{{Hopkins}
  et~al.}{2014}]{Hopkins+014_feed_stellar}
{Hopkins} P.~F.,  {Kere{\v s}} D.,  {O{\~n}orbe} J.,  {Faucher-Gigu{\`e}re}
  C.-A.,  {Quataert} E.,  {Murray} N.,   {Bullock} J.~S.,  2014, \mn@doi
  [\mnras] {10.1093/mnras/stu1738}, \href
  {http://adsabs.harvard.edu/abs/2014MNRAS.445..581H} {445, 581}

\bibitem[\protect\citeauthoryear{{Huertas-Company} et~al.,}{{Huertas-Company}
  et~al.}{2013}]{Huertas+013_SM}
{Huertas-Company} M.,  et~al., 2013, \mn@doi [\mnras] {10.1093/mnras/sts150},
  \href {http://adsabs.harvard.edu/abs/2013MNRAS.428.1715H} {428, 1715}

\bibitem[\protect\citeauthoryear{{Hyde} \& {Bernardi}}{{Hyde} \&
  {Bernardi}}{2009}]{HB09_curv}
{Hyde} J.~B.,  {Bernardi} M.,  2009, \mn@doi [\mnras]
  {10.1111/j.1365-2966.2009.14445.x}, \href
  {http://adsabs.harvard.edu/abs/2009MNRAS.394.1978H} {394, 1978}

\bibitem[\protect\citeauthoryear{{Ilbert} et~al.,}{{Ilbert}
  et~al.}{2006}]{Ilbert+06}
{Ilbert} O.,  et~al., 2006, \mn@doi [\aap] {10.1051/0004-6361:20065138}, \href
  {http://adsabs.harvard.edu/abs/2006A%26A...457..841I} {457, 841}

\bibitem[\protect\citeauthoryear{{Iodice} et~al.,}{{Iodice}
  et~al.}{2016}]{Iodice+16}
{Iodice} E.,  et~al., 2016, \mn@doi [\apj] {10.3847/0004-637X/820/1/42}, \href
  {http://adsabs.harvard.edu/abs/2016ApJ...820...42I} {820, 42}

\bibitem[\protect\citeauthoryear{{Kashikawa} et~al.,}{{Kashikawa}
  et~al.}{2004}]{Kashikawa+04_gal_counts}
{Kashikawa} N.,  et~al., 2004, \mn@doi [\pasj] {10.1093/pasj/56.6.1011}, \href
  {http://adsabs.harvard.edu/abs/2004PASJ...56.1011K} {56, 1011}

\bibitem[\protect\citeauthoryear{{Katz} \& {Gunn}}{{Katz} \&
  {Gunn}}{1991}]{Katz_Gunn+91_GF_dynam}
{Katz} N.,  {Gunn} J.~E.,  1991, \mn@doi [\apj] {10.1086/170367}, \href
  {http://adsabs.harvard.edu/abs/1991ApJ...377..365K} {377, 365}

\bibitem[\protect\citeauthoryear{{Kauffmann}}{{Kauffmann}}{1996}]{Kauffmann+96_merging}
{Kauffmann} G.,  1996, \mn@doi [\mnras] {10.1093/mnras/281.2.487}, \href
  {http://adsabs.harvard.edu/abs/1996MNRAS.281..487K} {281, 487}

\bibitem[\protect\citeauthoryear{{Kauffmann} et~al.,}{{Kauffmann}
  et~al.}{2003}]{Kauffmann+03}
{Kauffmann} G.,  et~al., 2003, \mn@doi [\mnras]
  {10.1046/j.1365-8711.2003.06292.x}, \href
  {http://adsabs.harvard.edu/abs/2003MNRAS.341...54K} {341, 54}

\bibitem[\protect\citeauthoryear{{Kelvin} et~al.,}{{Kelvin}
  et~al.}{2012}]{Kelvin+012}
{Kelvin} L.~S.,  et~al., 2012, \mn@doi [\mnras]
  {10.1111/j.1365-2966.2012.20355.x}, \href
  {http://adsabs.harvard.edu/abs/2012MNRAS.421.1007K} {421, 1007}

\bibitem[\protect\citeauthoryear{{Kelvin} et~al.,}{{Kelvin}
  et~al.}{2014}]{Kelvin+014}
{Kelvin} L.~S.,  et~al., 2014, \mn@doi [\mnras] {10.1093/mnras/stu1507}, \href
  {http://adsabs.harvard.edu/abs/2014MNRAS.444.1647K} {444, 1647}

\bibitem[\protect\citeauthoryear{{Kinney}, {Calzetti}, {Bohlin}, {McQuade},
  {Storchi-Bergmann}  \& {Schmitt}}{{Kinney} et~al.}{1996}]{Kinney+96}
{Kinney} A.~L.,  {Calzetti} D.,  {Bohlin} R.~C.,  {McQuade} K.,
  {Storchi-Bergmann} T.,   {Schmitt} H.~R.,  1996, \mn@doi [\apj]
  {10.1086/177583}, \href {http://adsabs.harvard.edu/abs/1996ApJ...467...38K}
  {467, 38}

\bibitem[\protect\citeauthoryear{{Koekemoer} et~al.,}{{Koekemoer}
  et~al.}{2011}]{CANDELS+11}
{Koekemoer} A.~M.,  et~al., 2011, \mn@doi [\apjs] {10.1088/0067-0049/197/2/36},
  \href {http://adsabs.harvard.edu/abs/2011ApJS..197...36K} {197, 36}

\bibitem[\protect\citeauthoryear{{Komatsu} et~al.,}{{Komatsu}
  et~al.}{2011}]{Komatsu+11_WMAP7}
{Komatsu} E.,  et~al., 2011, \mn@doi [\apjs] {10.1088/0067-0049/192/2/18},
  \href {http://adsabs.harvard.edu/abs/2011ApJS..192...18K} {192, 18}

\bibitem[\protect\citeauthoryear{{Kormendy}}{{Kormendy}}{1977}]{Kormendy+77}
{Kormendy} J.,  1977, \mn@doi [\apj] {10.1086/155687}, \href
  {http://adsabs.harvard.edu/abs/1977ApJ...218..333K} {218, 333}

\bibitem[\protect\citeauthoryear{{Kormendy}, {Fisher}, {Cornell}  \&
  {Bender}}{{Kormendy} et~al.}{2009}]{Kormendy+09}
{Kormendy} J.,  {Fisher} D.~B.,  {Cornell} M.~E.,   {Bender} R.,  2009, \mn@doi
  [\apjs] {10.1088/0067-0049/182/1/216}, \href
  {http://adsabs.harvard.edu/abs/2009ApJS..182..216K} {182, 216}

\bibitem[\protect\citeauthoryear{{Kuijken} et~al.,}{{Kuijken}
  et~al.}{2015}]{Kuijken+15_weak_lensing_kids}
{Kuijken} K.,  et~al., 2015, \mn@doi [\mnras] {10.1093/mnras/stv2140}, \href
  {http://adsabs.harvard.edu/abs/2015MNRAS.454.3500K} {454, 3500}

\bibitem[\protect\citeauthoryear{{La Barbera} \& {de Carvalho}}{{La Barbera} \&
  {de Carvalho}}{2009}]{LaBarbera_deCarvalho+09}
{La Barbera} F.,  {de Carvalho} R.~R.,  2009, \mn@doi [\apjl]
  {10.1088/0004-637X/699/2/L76}, \href
  {http://adsabs.harvard.edu/abs/2009ApJ...699L..76L} {699, L76}

\bibitem[\protect\citeauthoryear{{La Barbera}, {Busarello}, {Merluzzi},
  {Massarotti}  \& {Capaccioli}}{{La Barbera}
  et~al.}{2002}]{LaBarbera+02_2dphot}
{La Barbera} F.,  {Busarello} G.,  {Merluzzi} P.,  {Massarotti} M.,
  {Capaccioli} M.,  2002, \mn@doi [\apj] {10.1086/340021}, \href
  {http://adsabs.harvard.edu/abs/2002ApJ...571..790L} {571, 790}

\bibitem[\protect\citeauthoryear{{La Barbera}, {de Carvalho}, {Kohl-Moreira},
  {Gal}, {Soares-Santos}, {Capaccioli}, {Santos}  \& {Sant'anna}}{{La Barbera}
  et~al.}{2008}]{LaBarbera_08_2DPHOT}
{La Barbera} F.,  {de Carvalho} R.~R.,  {Kohl-Moreira} J.~L.,  {Gal} R.~R.,
  {Soares-Santos} M.,  {Capaccioli} M.,  {Santos} R.,   {Sant'anna} N.,  2008,
  \mn@doi [\pasp] {10.1086/588614}, \href
  {http://adsabs.harvard.edu/abs/2008PASP..120..681L} {120, 681}

\bibitem[\protect\citeauthoryear{{La Barbera}, {De Carvalho}, {De La Rosa},
  {Gal}, {Swindle}  \& {Lopes}}{{La Barbera} et~al.}{2010a}]{SPIDER-IV}
{La Barbera} F.,  {De Carvalho} R.~R.,  {De La Rosa} I.~G.,  {Gal} R.~R.,
  {Swindle} R.,   {Lopes} P.~A.~A.,  2010a, \mn@doi [\aj]
  {10.1088/0004-6256/140/5/1528}, \href
  {http://adsabs.harvard.edu/abs/2010AJ....140.1528L} {140, 1528}

\bibitem[\protect\citeauthoryear{{La Barbera}, {de Carvalho}, {de La Rosa},
  {Lopes}, {Kohl-Moreira}  \& {Capelato}}{{La Barbera}
  et~al.}{2010b}]{SPIDER-I}
{La Barbera} F.,  {de Carvalho} R.~R.,  {de La Rosa} I.~G.,  {Lopes} P.~A.~A.,
  {Kohl-Moreira} J.~L.,   {Capelato} H.~V.,  2010b, \mn@doi [\mnras]
  {10.1111/j.1365-2966.2010.16850.x}, \href
  {http://adsabs.harvard.edu/abs/2010MNRAS.408.1313L} {408, 1313}

\bibitem[\protect\citeauthoryear{{La Barbera}, {Ferreras}, {de Carvalho},
  {Lopes}, {Pasquali}, {de la Rosa}  \& {De Lucia}}{{La Barbera}
  et~al.}{2011}]{LaBarbera+11_CG}
{La Barbera} F.,  {Ferreras} I.,  {de Carvalho} R.~R.,  {Lopes} P.~A.~A.,
  {Pasquali} A.,  {de la Rosa} I.~G.,   {De Lucia} G.,  2011, \mn@doi [\apjl]
  {10.1088/2041-8205/740/2/L41}, \href
  {http://adsabs.harvard.edu/abs/2011ApJ...740L..41L} {740, L41}

\bibitem[\protect\citeauthoryear{{La Barbera}, {Ferreras}, {Vazdekis}, {de la
  Rosa}, {de Carvalho}, {Trevisan}, {Falc{\'o}n-Barroso}  \&
  {Ricciardelli}}{{La Barbera} et~al.}{2013}]{LaBarbera+13_SPIDERVIII_IMF}
{La Barbera} F.,  {Ferreras} I.,  {Vazdekis} A.,  {de la Rosa} I.~G.,  {de
  Carvalho} R.~R.,  {Trevisan} M.,  {Falc{\'o}n-Barroso} J.,   {Ricciardelli}
  E.,  2013, \mn@doi [\mnras] {10.1093/mnras/stt943}, \href
  {http://adsabs.harvard.edu/abs/2013MNRAS.433.3017L} {433, 3017}

\bibitem[\protect\citeauthoryear{{Lange} et~al.,}{{Lange}
  et~al.}{2015}]{Lange+015}
{Lange} R.,  et~al., 2015, \mn@doi [\mnras] {10.1093/mnras/stu2467}, \href
  {http://adsabs.harvard.edu/abs/2015MNRAS.447.2603L} {447, 2603}

\bibitem[\protect\citeauthoryear{{Lelli}, {McGaugh}  \& {Schombert}}{{Lelli}
  et~al.}{2016}]{Lelli+016}
{Lelli} F.,  {McGaugh} S.~S.,   {Schombert} J.~M.,  2016, \mn@doi [\apjl]
  {10.3847/2041-8205/816/1/L14}, \href
  {http://adsabs.harvard.edu/abs/2016ApJ...816L..14L} {816, L14}

\bibitem[\protect\citeauthoryear{{Magorrian} et~al.,}{{Magorrian}
  et~al.}{1998}]{Magorrian+98_BH}
{Magorrian} J.,  et~al., 1998, \mn@doi [\aj] {10.1086/300353}, \href
  {http://adsabs.harvard.edu/abs/1998AJ....115.2285M} {115, 2285}

\bibitem[\protect\citeauthoryear{{McCracken} et~al.,}{{McCracken}
  et~al.}{2003}]{McCracken+03_gal_counts}
{McCracken} H.~J.,  et~al., 2003, \mn@doi [\aap] {10.1051/0004-6361:20031081},
  \href {http://adsabs.harvard.edu/abs/2003A%26A...410...17M} {410, 17}

\bibitem[\protect\citeauthoryear{{Mo}, {Mao}  \& {White}}{{Mo}
  et~al.}{1998}]{Mo+1998}
{Mo} H.~J.,  {Mao} S.,   {White} S.~D.~M.,  1998, \mn@doi [\mnras]
  {10.1046/j.1365-8711.1998.01227.x}, \href
  {http://adsabs.harvard.edu/abs/1998MNRAS.295..319M} {295, 319}

\bibitem[\protect\citeauthoryear{{Mosleh}, {Williams}  \& {Franx}}{{Mosleh}
  et~al.}{2013}]{Mosleh+13_size_evol}
{Mosleh} M.,  {Williams} R.~J.,   {Franx} M.,  2013, \mn@doi [\apj]
  {10.1088/0004-637X/777/2/117}, \href
  {http://adsabs.harvard.edu/abs/2013ApJ...777..117M} {777, 117}

\bibitem[\protect\citeauthoryear{{Moster}, {Somerville}, {Maulbetsch}, {van den
  Bosch}, {Macci{\`o}}, {Naab}  \& {Oser}}{{Moster} et~al.}{2010}]{Moster+10}
{Moster} B.~P.,  {Somerville} R.~S.,  {Maulbetsch} C.,  {van den Bosch} F.~C.,
  {Macci{\`o}} A.~V.,  {Naab} T.,   {Oser} L.,  2010, \mn@doi [\apj]
  {10.1088/0004-637X/710/2/903}, \href
  {http://adsabs.harvard.edu/abs/2010ApJ...710..903M} {710, 903}

\bibitem[\protect\citeauthoryear{{Muzzin}, {van Dokkum}, {Franx}, {Marchesini},
  {Kriek}  \& {Labb{\'e}}}{{Muzzin} et~al.}{2009}]{Muzzin+09_Mass_of_gal}
{Muzzin} A.,  {van Dokkum} P.,  {Franx} M.,  {Marchesini} D.,  {Kriek} M.,
  {Labb{\'e}} I.,  2009, \mn@doi [\apjl] {10.1088/0004-637X/706/1/L188}, \href
  {http://adsabs.harvard.edu/abs/2009ApJ...706L.188M} {706, L188}

\bibitem[\protect\citeauthoryear{{Naab}, {Johansson}  \& {Ostriker}}{{Naab}
  et~al.}{2009}]{Naab+09}
{Naab} T.,  {Johansson} P.~H.,   {Ostriker} J.~P.,  2009, \mn@doi [\apjl]
  {10.1088/0004-637X/699/2/L178}, \href
  {http://adsabs.harvard.edu/abs/2009ApJ...699L.178N} {699, L178}

\bibitem[\protect\citeauthoryear{{Napolitano} et~al.,}{{Napolitano}
  et~al.}{2016}]{Napolitano+15_proc_lensing_KiDS}
{Napolitano} N.~R.,  et~al., 2016, \mn@doi [The Universe of Digital Sky
  Surveys] {10.1007/978-3-319-19330-4_20}, \href
  {http://adsabs.harvard.edu/abs/2016ASSP...42..129N} {42, 129}

\bibitem[\protect\citeauthoryear{{Navarro} \& {White}}{{Navarro} \&
  {White}}{1994}]{Navarro_white+94_dynamics}
{Navarro} J.~F.,  {White} S.~D.~M.,  1994, \mn@doi [\mnras]
  {10.1093/mnras/267.2.401}, \href
  {http://adsabs.harvard.edu/abs/1994MNRAS.267..401N} {267, 401}

\bibitem[\protect\citeauthoryear{{Oser}, {Naab}, {Ostriker}  \&
  {Johansson}}{{Oser} et~al.}{2012}]{Oser+012_evol_massive_etg}
{Oser} L.,  {Naab} T.,  {Ostriker} J.~P.,   {Johansson} P.~H.,  2012, \mn@doi
  [\apj] {10.1088/0004-637X/744/1/63}, \href
  {http://adsabs.harvard.edu/abs/2012ApJ...744...63O} {744, 63}

\bibitem[\protect\citeauthoryear{{Peng}, {Ho}, {Impey}  \& {Rix}}{{Peng}
  et~al.}{2002}]{Peng+002_GALFIT}
{Peng} C.~Y.,  {Ho} L.~C.,  {Impey} C.~D.,   {Rix} H.-W.,  2002, \mn@doi [\aj]
  {10.1086/340952}, \href {http://adsabs.harvard.edu/abs/2002AJ....124..266P}
  {124, 266}

\bibitem[\protect\citeauthoryear{{Petrillo} et~al.,}{{Petrillo}
  et~al.}{2017}]{Petrillo+17_CNN}
{Petrillo} C.~E.,  et~al., 2017, \mn@doi [\mnras] {10.1093/mnras/stx2052},
  \href {http://adsabs.harvard.edu/abs/2017MNRAS.472.1129P} {472, 1129}

\bibitem[\protect\citeauthoryear{{Radovich} et~al.,}{{Radovich}
  et~al.}{2017}]{Radovich+17_KiDS}
{Radovich} M.,  et~al., 2017, \mn@doi [\aap] {10.1051/0004-6361/201629353},
  \href {http://adsabs.harvard.edu/abs/2017A%26A...598A.107R} {598, A107}

\bibitem[\protect\citeauthoryear{{Ravindranath} et~al.,}{{Ravindranath}
  et~al.}{2002}]{Ravindranath+02}
{Ravindranath} S.,  et~al., 2002, in American Astronomical Society Meeting
  Abstracts. p.~1099

\bibitem[\protect\citeauthoryear{{Roche}, {Bernardi}  \& {Hyde}}{{Roche}
  et~al.}{2010}]{Roche+10}
{Roche} N.,  {Bernardi} M.,   {Hyde} J.,  2010, \mn@doi [\mnras]
  {10.1111/j.1365-2966.2010.16976.x}, \href
  {http://adsabs.harvard.edu/abs/2010MNRAS.407.1231R} {407, 1231}

\bibitem[\protect\citeauthoryear{{Rykoff}, {Rozo}  \& {Keisler}}{{Rykoff}
  et~al.}{2015}]{Rykoff+015}
{Rykoff} E.~S.,  {Rozo} E.,   {Keisler} R.,  2015, preprint (\mn@eprint {arXiv}
  {1509.00870})

\bibitem[\protect\citeauthoryear{{Saglia} et~al.,}{{Saglia}
  et~al.}{2010}]{Saglia+10}
{Saglia} R.~P.,  et~al., 2010, \mn@doi [\aap] {10.1051/0004-6361/201014703},
  \href {http://adsabs.harvard.edu/abs/2010A%26A...524A...6S} {524, A6}

\bibitem[\protect\citeauthoryear{{Sales}, {Navarro}, {Schaye}, {Dalla Vecchia},
  {Springel}  \& {Booth}}{{Sales} et~al.}{2010}]{Sales+010_feed_simu_gal}
{Sales} L.~V.,  {Navarro} J.~F.,  {Schaye} J.,  {Dalla Vecchia} C.,  {Springel}
  V.,   {Booth} C.~M.,  2010, \mn@doi [\mnras]
  {10.1111/j.1365-2966.2010.17391.x}, \href
  {http://adsabs.harvard.edu/abs/2010MNRAS.409.1541S} {409, 1541}

\bibitem[\protect\citeauthoryear{{Scannapieco} \& {Athanassoula}}{{Scannapieco}
  \& {Athanassoula}}{2012}]{Scannapieco+012_hydro_simu}
{Scannapieco} C.,  {Athanassoula} E.,  2012, \mn@doi [\mnras]
  {10.1111/j.1745-3933.2012.01291.x}, \href
  {http://adsabs.harvard.edu/abs/2012MNRAS.425L..10S} {425, L10}

\bibitem[\protect\citeauthoryear{{Schaye} et~al.,}{{Schaye}
  et~al.}{2015}]{Schaye+015_EAGLE}
{Schaye} J.,  et~al., 2015, \mn@doi [\mnras] {10.1093/mnras/stu2058}, \href
  {http://adsabs.harvard.edu/abs/2015MNRAS.446..521S} {446, 521}

\bibitem[\protect\citeauthoryear{{Schlafly} \& {Finkbeiner}}{{Schlafly} \&
  {Finkbeiner}}{2011}]{Schlafly_Finkbeiner11}
{Schlafly} E.~F.,  {Finkbeiner} D.~P.,  2011, \mn@doi [\apj]
  {10.1088/0004-637X/737/2/103}, \href
  {http://adsabs.harvard.edu/abs/2011ApJ...737..103S} {737, 103}

\bibitem[\protect\citeauthoryear{{Shen}, {Mo}, {White}, {Blanton}, {Kauffmann},
  {Voges}, {Brinkmann}  \& {Csabai}}{{Shen} et~al.}{2003}]{Shen+03}
{Shen} S.,  {Mo} H.~J.,  {White} S.~D.~M.,  {Blanton} M.~R.,  {Kauffmann} G.,
  {Voges} W.,  {Brinkmann} J.,   {Csabai} I.,  2003, \mn@doi [\mnras]
  {10.1046/j.1365-8711.2003.06740.x}, 343, 978

\bibitem[\protect\citeauthoryear{{Sparks} \& {Jorgensen}}{{Sparks} \&
  {Jorgensen}}{1993}]{Sparks_Jorgensen93}
{Sparks} W.~B.,  {Jorgensen} I.,  1993, \mn@doi [\aj] {10.1086/116552}, 105,
  1753

\bibitem[\protect\citeauthoryear{{Spiniello} et~al.,}{{Spiniello}
  et~al.}{2018}]{Spiniello+18_quads}
{Spiniello} C.,  et~al., 2018, preprint (\mn@eprint {arXiv} {1805.12436})

\bibitem[\protect\citeauthoryear{{Springel}}{{Springel}}{2005}]{Springel+005_GADGET2}
{Springel} V.,  2005, \mn@doi [\mnras] {10.1111/j.1365-2966.2005.09655.x}, 364,
  1105

\bibitem[\protect\citeauthoryear{{Strateva} et~al.,}{{Strateva}
  et~al.}{2001}]{Strateva+01}
{Strateva} I.,  et~al., 2001, \mn@doi [\aj] {10.1086/323301}, 122, 1861

\bibitem[\protect\citeauthoryear{{Szomoru} et~al.,}{{Szomoru}
  et~al.}{2010}]{Szomoru+10}
{Szomoru} D.,  et~al., 2010, \mn@doi [\apjl] {10.1088/2041-8205/714/2/L244},
  \href {http://adsabs.harvard.edu/abs/2010ApJ...714L.244S} {714, L244}

\bibitem[\protect\citeauthoryear{{Tortora}, {Napolitano}, {Romanowsky},
  {Capaccioli}  \& {Covone}}{{Tortora} et~al.}{2009}]{Tortora+09}
{Tortora} C.,  {Napolitano} N.~R.,  {Romanowsky} A.~J.,  {Capaccioli} M.,
  {Covone} G.,  2009, \mn@doi [\mnras] {10.1111/j.1365-2966.2009.14789.x}, 396,
  1132

\bibitem[\protect\citeauthoryear{{Tortora}, {Napolitano}, {Cardone},
  {Capaccioli}, {Jetzer}  \& {Molinaro}}{{Tortora} et~al.}{2010}]{Tortora+10CG}
{Tortora} C.,  {Napolitano} N.~R.,  {Cardone} V.~F.,  {Capaccioli} M.,
  {Jetzer} P.,   {Molinaro} R.,  2010, \mn@doi [\mnras]
  {10.1111/j.1365-2966.2010.16938.x}, \href
  {http://adsabs.harvard.edu/abs/2010MNRAS.407..144T} {407, 144}

\bibitem[\protect\citeauthoryear{{Tortora}, {La Barbera}, {Napolitano}, {de
  Carvalho}  \& {Romanowsky}}{{Tortora} et~al.}{2012}]{SPIDER-VI}
{Tortora} C.,  {La Barbera} F.,  {Napolitano} N.~R.,  {de Carvalho} R.~R.,
  {Romanowsky} A.~J.,  2012, \mn@doi [\mnras]
  {10.1111/j.1365-2966.2012.21506.x}, \href
  {http://adsabs.harvard.edu/abs/2012MNRAS.425..577T} {425, 577}

\bibitem[\protect\citeauthoryear{{Tortora}, {Romanowsky}  \&
  {Napolitano}}{{Tortora} et~al.}{2013}]{Tortora+13_SPIDER_IMF}
{Tortora} C.,  {Romanowsky} A.~J.,   {Napolitano} N.~R.,  2013, \mn@doi [\apj]
  {10.1088/0004-637X/765/1/8}, \href
  {http://adsabs.harvard.edu/abs/2013ApJ...765....8T} {765, 8}

\bibitem[\protect\citeauthoryear{{Tortora}, {Romanowsky}, {Cardone},
  {Napolitano}  \& {Jetzer}}{{Tortora} et~al.}{2014a}]{Tortora+14_MOND}
{Tortora} C.,  {Romanowsky} A.~J.,  {Cardone} V.~F.,  {Napolitano} N.~R.,
  {Jetzer} P.,  2014a, \mn@doi [\mnras] {10.1093/mnrasl/slt155}, \href
  {http://adsabs.harvard.edu/abs/2014MNRAS.438L..46T} {438, L46}

\bibitem[\protect\citeauthoryear{{Tortora}, {La Barbera}, {Napolitano},
  {Romanowsky}, {Ferreras}  \& {de Carvalho}}{{Tortora}
  et~al.}{2014b}]{Tortora+14_DMslope}
{Tortora} C.,  {La Barbera} F.,  {Napolitano} N.~R.,  {Romanowsky} A.~J.,
  {Ferreras} I.,   {de Carvalho} R.~R.,  2014b, \mn@doi [\mnras]
  {10.1093/mnras/stu1616}, \href
  {http://adsabs.harvard.edu/abs/2014MNRAS.445..115T} {445, 115}

\bibitem[\protect\citeauthoryear{{Tortora}, {Napolitano}, {Saglia},
  {Romanowsky}, {Covone}  \& {Capaccioli}}{{Tortora}
  et~al.}{2014c}]{Tortora+14_DMevol}
{Tortora} C.,  {Napolitano} N.~R.,  {Saglia} R.~P.,  {Romanowsky} A.~J.,
  {Covone} G.,   {Capaccioli} M.,  2014c, \mn@doi [\mnras]
  {10.1093/mnras/stu1712}, \href
  {http://adsabs.harvard.edu/abs/2014MNRAS.445..162T} {445, 162}

\bibitem[\protect\citeauthoryear{{Tortora} et~al.,}{{Tortora}
  et~al.}{2016}]{Tortora+16_compact_KiDS}
{Tortora} C.,  et~al., 2016, \mn@doi [\mnras] {10.1093/mnras/stw184}, 457, 2845

\bibitem[\protect\citeauthoryear{{Tortora} et~al.,}{{Tortora}
  et~al.}{2018a}]{Tortora+18_UCMG}
{Tortora} C.,  et~al., 2018a, preprint (\mn@eprint {arXiv} {1806.01307})

\bibitem[\protect\citeauthoryear{{Tortora}, {Napolitano}, {Roy}, {Radovich},
  {Getman}, {Koopmans}, {Verdoes Kleijn}  \& {Kuijken}}{{Tortora}
  et~al.}{2018b}]{Tortora+18}
{Tortora} C.,  {Napolitano} N.~R.,  {Roy} N.,  {Radovich} M.,  {Getman} F.,
  {Koopmans} L.~V.~E.,  {Verdoes Kleijn} G.~A.,   {Kuijken} K.~H.,  2018b,
  \mn@doi [\mnras] {10.1093/mnras/stx2390}, \href
  {http://adsabs.harvard.edu/abs/2018MNRAS.473..969T} {473, 969}

\bibitem[\protect\citeauthoryear{{Tortora}, {Napolitano}, {Roy}, {Radovich},
  {Getman}, {Koopmans}, {Verdoes Kleijn}  \& {Kuijken}}{{Tortora}
  et~al.}{2018c}]{Tortora+018_DM_KiDS}
{Tortora} C.,  {Napolitano} N.~R.,  {Roy} N.,  {Radovich} M.,  {Getman} F.,
  {Koopmans} L.~V.~E.,  {Verdoes Kleijn} G.~A.,   {Kuijken} K.~H.,  2018c,
  \mn@doi [\mnras] {10.1093/mnras/stx2390}, \href
  {http://adsabs.harvard.edu/abs/2018MNRAS.473..969T} {473, 969}

\bibitem[\protect\citeauthoryear{{Tremaine} et~al.,}{{Tremaine}
  et~al.}{2002}]{Tremaine+02_BH_velo}
{Tremaine} S.,  et~al., 2002, \mn@doi [\apj] {10.1086/341002}, \href
  {http://adsabs.harvard.edu/abs/2002ApJ...574..740T} {574, 740}

\bibitem[\protect\citeauthoryear{{Treu}, {Auger}, {Koopmans}, {Gavazzi},
  {Marshall}  \& {Bolton}}{{Treu} et~al.}{2010}]{Treu+10}
{Treu} T.,  {Auger} M.~W.,  {Koopmans} L.~V.~E.,  {Gavazzi} R.,  {Marshall}
  P.~J.,   {Bolton} A.~S.,  2010, \mn@doi [\apj]
  {10.1088/0004-637X/709/2/1195}, \href
  {http://adsabs.harvard.edu/abs/2010ApJ...709.1195T} {709, 1195}

\bibitem[\protect\citeauthoryear{{Trujillo} et~al.,}{{Trujillo}
  et~al.}{2006}]{Trujillo+06}
{Trujillo} I.,  et~al., 2006, \mn@doi [\apj] {10.1086/506464}, \href
  {http://adsabs.harvard.edu/abs/2006ApJ...650...18T} {650, 18}

\bibitem[\protect\citeauthoryear{{Trujillo}, {Conselice}, {Bundy}, {Cooper},
  {Eisenhardt}  \& {Ellis}}{{Trujillo} et~al.}{2007}]{Trujillo+07}
{Trujillo} I.,  {Conselice} C.~J.,  {Bundy} K.,  {Cooper} M.~C.,  {Eisenhardt}
  P.,   {Ellis} R.~S.,  2007, \mn@doi [\mnras]
  {10.1111/j.1365-2966.2007.12388.x}, \href
  {http://adsabs.harvard.edu/abs/2007MNRAS.382..109T} {382, 109}

\bibitem[\protect\citeauthoryear{{Trujillo}, {Ferreras}  \& {de La
  Rosa}}{{Trujillo} et~al.}{2011}]{Trujillo+11}
{Trujillo} I.,  {Ferreras} I.,   {de La Rosa} I.~G.,  2011, \mn@doi [\mnras]
  {10.1111/j.1365-2966.2011.19017.x}, \href
  {http://adsabs.harvard.edu/abs/2011MNRAS.415.3903T} {415, 3903}

\bibitem[\protect\citeauthoryear{{Trujillo}, {Carrasco}  \&
  {Ferr{\'e}-Mateu}}{{Trujillo} et~al.}{2012}]{Trujillo+12_compacts}
{Trujillo} I.,  {Carrasco} E.~R.,   {Ferr{\'e}-Mateu} A.,  2012, \mn@doi [\apj]
  {10.1088/0004-637X/751/1/45}, \href
  {http://adsabs.harvard.edu/abs/2012ApJ...751...45T} {751, 45}

\bibitem[\protect\citeauthoryear{{Tully} \& {Fisher}}{{Tully} \&
  {Fisher}}{1977}]{TF77}
{Tully} R.~B.,  {Fisher} J.~R.,  1977, \aap, \href
  {http://adsabs.harvard.edu/abs/1977A%26A....54..661T} {54, 661}

\bibitem[\protect\citeauthoryear{{Vulcani} et~al.,}{{Vulcani}
  et~al.}{2014}]{Vulcani+14}
{Vulcani} B.,  et~al., 2014, \mn@doi [\mnras] {10.1093/mnras/stu632}, \href
  {http://adsabs.harvard.edu/abs/2014MNRAS.441.1340V} {441, 1340}

\bibitem[\protect\citeauthoryear{{Yasuda} et~al.,}{{Yasuda}
  et~al.}{2001}]{Yasuda+01_gal_counts}
{Yasuda} N.,  et~al., 2001, \mn@doi [\aj] {10.1086/322093}, \href
  {http://adsabs.harvard.edu/abs/2001AJ....122.1104Y} {122, 1104}

\bibitem[\protect\citeauthoryear{{York} et~al.,}{{York}
  et~al.}{2000}]{York+000_SDSS_tech}
{York} D.~G.,  et~al., 2000, \mn@doi [\aj] {10.1086/301513}, \href
  {http://adsabs.harvard.edu/abs/2000AJ....120.1579Y} {120, 1579}

\bibitem[\protect\citeauthoryear{{de Jong}}{{de Jong}}{2008}]{deJong+08_PSF}
{de Jong} R.~S.,  2008, \mn@doi [\mnras] {10.1111/j.1365-2966.2008.13505.x},
  \href {http://adsabs.harvard.edu/abs/2008MNRAS.388.1521D} {388, 1521}

\bibitem[\protect\citeauthoryear{{de Jong} et~al.,}{{de Jong}
  et~al.}{2015}]{deJong+15_KiDS_paperI}
{de Jong} J.~T.~A.,  et~al., 2015, \mn@doi [\aap]
  {10.1051/0004-6361/201526601}, \href
  {http://adsabs.harvard.edu/abs/2015A%26A...582A..62D} {582, A62}

\bibitem[\protect\citeauthoryear{{de Jong} et~al.,}{{de Jong}
  et~al.}{2017}]{deJong+017_KIDS_DR3}
{de Jong} J.~T.~A.,  et~al., 2017, \mn@doi [\aap]
  {10.1051/0004-6361/201730747}, \href
  {http://adsabs.harvard.edu/abs/2017A%26A...604A.134D} {604, A134}

\bibitem[\protect\citeauthoryear{{de Zeeuw}}{{de
  Zeeuw}}{2001}]{deZeeuw+01_Blk_HOLe}
{de Zeeuw} T.,  2001, in {Kaper} L.,  {Heuvel} E.~P.~J.~V.~D.,   {Woudt} P.~A.,
   eds, Black Holes in Binaries and Galactic Nuclei. p.~78 (\mn@eprint {}
  {astro-ph/0009249}), \mn@doi{10.1007/10720995_12}

\bibitem[\protect\citeauthoryear{{de la Rosa}, {La Barbera}, {Ferreras},
  {S{\'a}nchez Almeida}, {Dalla Vecchia}, {Mart{\'{\i}}nez-Valpuesta}  \&
  {Stringer}}{{de la Rosa} et~al.}{2016}]{delarosa+16_highz_compact}
{de la Rosa} I.~G.,  {La Barbera} F.,  {Ferreras} I.,  {S{\'a}nchez Almeida}
  J.,  {Dalla Vecchia} C.,  {Mart{\'{\i}}nez-Valpuesta} I.,   {Stringer} M.,
  2016, \mn@doi [\mnras] {10.1093/mnras/stw130}, \href
  {http://adsabs.harvard.edu/abs/2016MNRAS.457.1916D} {457, 1916}

\bibitem[\protect\citeauthoryear{{van Dokkum} et~al.,}{{van Dokkum}
  et~al.}{2008}]{vandokkum+09_compactness}
{van Dokkum} P.~G.,  et~al., 2008, \mn@doi [\apjl] {10.1086/587874}, \href
  {http://adsabs.harvard.edu/abs/2008ApJ...677L...5V} {677, L5}

\bibitem[\protect\citeauthoryear{{van Dokkum} et~al.,}{{van Dokkum}
  et~al.}{2010}]{vanDokkum+10}
{van Dokkum} P.~G.,  et~al., 2010, \mn@doi [\apj]
  {10.1088/0004-637X/709/2/1018}, \href
  {http://adsabs.harvard.edu/abs/2010ApJ...709.1018V} {709, 1018}

\bibitem[\protect\citeauthoryear{{van der Wel}, {Bell}, {van den Bosch},
  {Gallazzi}  \& {Rix}}{{van der Wel} et~al.}{2009}]{vanderwel+09_size_mass}
{van der Wel} A.,  {Bell} E.~F.,  {van den Bosch} F.~C.,  {Gallazzi} A.,
  {Rix} H.-W.,  2009, \mn@doi [\apj] {10.1088/0004-637X/698/2/1232}, \href
  {http://adsabs.harvard.edu/abs/2009ApJ...698.1232V} {698, 1232}

\bibitem[\protect\citeauthoryear{{van der Wel} et~al.,}{{van der Wel}
  et~al.}{2014}]{vanderwel+14_SM}
{van der Wel} A.,  et~al., 2014, \mn@doi [\apj] {10.1088/0004-637X/788/1/28},
  \href {http://adsabs.harvard.edu/abs/2014ApJ...788...28V} {788, 28}

\makeatother
\end{thebibliography}
\bsp	
\label{lastpage}
\end{document}